\documentclass[12pt]{article} 
\pdfoutput=1
\usepackage{amsmath,amssymb,amsfonts}
\usepackage{psfrag}
\usepackage{color}
\definecolor{darkblue}{rgb}{0.1,0.1,.7}
\usepackage[colorlinks, linkcolor=darkblue, citecolor=darkblue, urlcolor=darkblue, linktocpage]{hyperref} 
\usepackage[square, comma, compress,numbers]{natbib}
\usepackage[]{amsmath}
\usepackage[]{graphicx}
\usepackage[]{latexsym}
\usepackage[utf8]{inputenc}
\usepackage{geometry}
\usepackage{tikz}
\usepackage{amscd}
\usepackage[all,cmtip]{xy}
\usepackage{mathrsfs}
\usepackage{makecell} 
\setcellgapes{2pt}
\usepackage{booktabs}  
\usepackage{bbold}
\usepackage{subfigure}
\usepackage[margin=10pt,font=small,labelfont=bf]{caption}
\usepackage{dsdshorthand}
\usepackage{simplewick}
\usepackage{changepage}
\usepackage[multiple]{footmisc}
\usepackage[utf8]{inputenc}
\usepackage{multirow}

\numberwithin{equation}{section}

\newcommand{\beq}{\begin{equation}}
\newcommand{\eeq}{\end{equation}}

\renewcommand{\be}{\begin{eqnarray}}
\renewcommand{\ee}{\end{eqnarray}}
\newcommand{\bea}{\begin{eqnarray}}
\newcommand{\eea}{\end{eqnarray}}

\newcommand{\Co}{\mathcal{C}}
\newcommand{\co}{C}

\def\Xint#1{\mathchoice
   {\XXint\displaystyle\textstyle{#1}}%
   {\XXint\textstyle\scriptstyle{#1}}%
   {\XXint\scriptstyle\scriptscriptstyle{#1}}%
   {\XXint\scriptscriptstyle\scriptscriptstyle{#1}}%
   \!\int}
\def\XXint#1#2#3{{\setbox0=\hbox{$#1{#2#3}{\int}$}
     \vcenter{\hbox{$#2#3$}}\kern-.5\wd0}}

\def\dashint{\Xint-}

\newcommand{\remarkjc}[1]{{\renewcommand{\bfdefault}{b}{\color[RGB]{0,0,150}{\textbf{#1}}}}}
\providecommand{\remarkjc}[1]{\ignorespaces}

\begin{document}

\vspace*{-.12in}
\thispagestyle{empty}
\vspace{.2in}

{\Large
\begin{center}
{\bf Hexagons and Correlators in the Fishnet Theory}
\end{center}
}
\vspace{.2in}
\begin{center}
{
Benjamin~Basso$^{a}$,
Jo\~ao~Caetano$^{a,b,c}$, 
Thiago~Fleury$^{a,d}$
}
\\
\vspace{.2in} 
$^a${\it \footnotesize  Laboratoire de Physique Th\'eorique de l'\'Ecole Normale Sup\'erieure, CNRS,
Universit\'e PSL, Sorbonne Universit\'es, Universit\'e Pierre et Marie Curie,
24 rue Lhomond, 75005 Paris, France}\\
$^b${\it \footnotesize C. N. Yang Institute for Theoretical Physics, SUNY, Stony Brook, NY 11794-3840, USA}\\
$^c${\it \footnotesize Simons Center for Geometry and Physics, SUNY, Stony Brook, NY 11794-3636, USA} \\
$^d${\it \footnotesize  International Institute of Physics, Federal University of Rio Grande do Norte,
Campus Universit\'ario, Lagoa Nova, Natal, RN 59078-970, Brazil}\\
$^d${\it \footnotesize  Instituto de F\'isica Te\'orica, UNESP - Univ. Estadual Paulista, ICTP South American Institute for
Fundamental Research, Rua Dr. Bento Teobaldo Ferraz 271, 01140-070, S\~ao Paulo, SP, Brazil
}
\end{center}

\vspace{.2in}

\begin{abstract}

We investigate the hexagon formalism in the planar 4d conformal fishnet theory. 
This theory arises from  $\mathcal{N}=4$ SYM by a deformation that preserves both conformal symmetry and integrability.
Based on this relation, we obtain the hexagon form factors for a large class of states, including the BMN vacuum, some excited states, and the Lagrangian density. We apply these form factors to the computation of several correlators and match the results with direct Feynman diagrammatic calculations. We also study the renormalisation of the hexagon form factor expansion for a family of diagonal structure constants and test the procedure at higher orders through comparison with a known universal formula for the Lagrangian insertion.

\end{abstract}

\newpage

\renewcommand{\baselinestretch}{0.8}\normalsize
\tableofcontents
\renewcommand{\baselinestretch}{1.0}\normalsize

\newpage

\section{Introduction}

The conformal fishnet theory \cite{Gurdogan:2015csr,Caetano:2016ydc,Kazakov:2018qez} may well be the simplest interacting CFT in higher dimensions that is integrable in the planar limit. Defined as the extreme limit of a twisted version \cite{Leigh:1995ep,Lunin:2005jy,Frolov:2005dj,Beisert:2005if} of the 4d maximally supersymmetric Yang-Mills theory ($\mathcal{N}=4$ SYM), the theory is minimalistic, but still highly nontrivial. It counts only two complex scalar fields and a single quartic coupling,
\beq\label{dil}
\mathcal{L}_{\textrm{int}} = g^2 \textrm{tr}\, \phi_{1}^{\dagger}\phi_{2}^{\dagger}\phi_{1}\phi_{2}\, ,
\eeq
with the fields filling $N\times N$ matrices. It depends, in the planar limit $N\rightarrow \infty$, on a single marginal coupling $g^2$, much like $\mathcal{N}=4$ SYM, if not that here double-trace deformations must be switched on and finely adjusted to maintain criticality \cite{Grabner:2017pgm,Sieg:2016vap}. The theory lacks unitarity but serves nonetheless as a natural stage for a broad family of perfectly meaningful conformal Feynman integrals, the fishnet graphs. These diagrams host one of the first observed manifestations of integrability in higher dimensions~\cite{Zamolodchikov:1980mb} and, although very special, they give us a hint at the remarkable mathematical structures that underlie Feynman integrals in general, see e.g.~\cite{Chicherin:2017cns,Chicherin:2017frs,Isaev:2003tk,Gromov:2017cja,Basso:2017jwq,Derkachov:2018rot,Bourjaily:2018ycu,Gromov:2018hut,Coronado:2018cxj,Coronado:2018ypq}. They also form an irreducible subset of the conformal integrals needed to span correlators and amplitudes in general perturbative CFTs, and in $\mathcal{N}=4$ SYM in particular, see e.g.~\cite{Bourjaily:2018ycu,Coronado:2018cxj,Coronado:2018ypq}.

The integrability of the fishnet theory is not as mysterious as in its supersymmetric parent. It traces back to the properties of the quartic coupling and links directly to the dynamics of non-compact conformal spin chains \cite{Zamolodchikov:1980mb,Gromov:2017cja,Chicherin:2012yn}. Fishnet theories, in general, offer a natural setting for discussing the integrability of these non-compact magnets, in a field theoretical language, and expressing their remarkable properties, at the Feynman diagrammatic level. They are also intimately tied to integrable non-compact sigma models \cite{Basso:2018agi}, in the graph thermodynamic limit \cite{Zamolodchikov:1980mb}, offering new perspectives on the problem of their quantization. Last but not least, fishnet theories form a laboratory for experimenting the techniques put forward for computing correlation functions and scattering amplitudes at finite coupling in more sophisticated integrable theories, like $\mathcal{N}=4$ SYM, see e.g.~\cite{Basso:2013vsa,Basso:2015zoa,Bajnok:2015hla,Fleury:2016ykk,Eden:2016xvg,Fleury:2017eph,Bargheer:2017nne,Eden:2017ozn,Ben-Israel:2018ckc,Bargheer:2018jvq}.

In this paper, we will apply one of these techniques - the hexagon factorisation - to the correlation functions and Feynman integrals of the fishnet theory. The method was first developed for computing structure constants in $\mathcal{N}=4$ SYM \cite{Basso:2015zoa} and was later on upgraded to encompass higher-point functions \cite{Fleury:2016ykk,Eden:2016xvg} and non-planar corrections \cite{Bargheer:2017nne,Eden:2017ozn}. Although the hexagon framework has been fairly tested, see e.g.~\cite{Basso:2015eqa,Eden:2015ija,Jiang:2015bvm,Jiang:2016dsr,Jiang:2016ulr,Caetano:2016keh,Basso:2017khq,Eden:2018vug,Chicherin:2018avq,Coronado:2018cxj,Coronado:2018ypq},
it is still far from being a well-oiled machinery and remains limited in some of its applications. The problem is partly due to the nature of the approach, which builds on a form-factor decomposition and requires that complicated sums and integrals over all the magnonic states be taken to non-perturbatively recover the original observable. Progress with the hexagon formalism is also hindered by the need of renormalising the divergences that show up at wrapping orders~\cite{Basso:2017muf}, when the magnons can circulate around a (non-protected) local operator. To date, no systematic removal of these divergences is known and it is challenging to push the hexagon strategy to higher loops in $\mathcal{N}=4$ SYM, even for the simplest structure constant, with one non-protected and two half-BPS operators, see \cite{Goncalves:2016vir,Eden:2016aqo,Georgoudis:2017meq,Chicherin:2018avq} for the state of the art on the field theory side.

The fishnet theory appears as an interesting playground to address these issues. For instance, the simplest structure constants of the fishnet theory are all about wrapping corrections, exposing the problem in its minimal form. Moreover, the ingredients entering the integrability framework acquire a direct diagrammatic meaning in the fishnet theory, a feature which helps testing their correctness. We will substantially benefit from this graphical intuition, in this paper. It will allow us, for instance, to fill a gap in the hexagon approach and incorporate the ``dilaton" (\ref{dil}) in its dictionary. Interest in this operator stems from its relation to the coupling dependence of the Green functions. Its insertion in a pair of conjugated operators, for instance, is fixed in terms of the spectral data \cite{Costa:2010rz}, offering a mean of testing the ability of the hexagon method at encoding the scaling dimensions of the theory.

The main outcome of this paper is a proposal for a large class of hexagon form factors of the fishnet theory, applicable to a variety of states, including the BMN vacuum, in the SYM terminology. Our formulae can be understood as a projection to the fishnet theory of the conjectures pushed forward for the SYM theory. We will subject them to a series of tests, by means of comparison with diagrammatic computations in the fishnet theory, and will obtain, on the way, a few predictions for a certain class of three-point Feynman integrals.

Finally, we will test the hexagons' aptitude at reproducing the scaling dimension of the BMN vacuum by considering diagonal structure constants with a Lagrangian insertion. To this end, we will generalise the renormalisation procedure put forward in \cite{Basso:2017muf} and derive, in a particular regime, an all order representation using the Leclair-Mussardo formula \cite{Leclair:1999ys}. We will verify the renormalised expansion so-obtained up to NNLO by a comparison with the Thermodynamical Bethe Ansatz (TBA) equations.

The paper is structured as follows. In Section \ref{Sect2}, we briefly recap the ingredients entering the hexagon program and detail the approach we shall follow to obtain their counterparts in the fishnet theory. In Section \ref{Sect3}, we perform several classic tests of our hexagon form factors through the computation of correlators, including some with excited states. In Section \ref{Sect4}, we discuss more advanced applications to a family of diagonal structure constants, mostly focusing on the Lagrangian insertion and its higher-charge siblings. We conclude in Section~\ref{Sect5}.  The details omitted in the main text are presented in several Appendices.

\section{Hexagons}\label{Sect2}

In this paper, we will analyse planar correlators in the fishnet theory using the hexagon factorisation. The prototype is the three-point function between a conjugate pair of BMN vacua and a third operator. The former are vacuum states in the spin-chain picture and can be chosen as
\beq\label{one-two}
\mathcal{O}_{1} = \textrm{tr}\, \phi_{1}^{L_{1}}\, , \qquad \mathcal{O}_{2} = \textrm{tr}\, \phi_{1}^{\dagger L_{2}}\, ,
\eeq
where the traces run over the color degrees of freedom; they have minimal dimensions $\Delta_{1,2}$ given their $U(1)$ charges, i.e., spin-chain lengths $L_{1,2}$. The third operator is designed such as to permit contractions with both operators in the pair. In $\mathcal{N} =4$ SYM, we can pick yet another BMN vacuum, by rotating the fields in (\ref{one-two}) using an $SO(6)$ transformation, and work with e.g.
\beq\label{reservoir}
\phi'_{1} = \phi_{1}+\phi^{\dagger}_{1} + \phi_{i}-\phi_{i}^{\dagger} \qquad \Rightarrow \qquad \mathcal{O}_{3} = \textrm{tr}\, (\phi'_{1})^{L_{3}}\, ,
\eeq
where $\phi_{i\neq 1}$ is a complex scalar field, charged under a different Cartan generator. This choice underlies the SYM hexagon framework and the third operator built in this manner is the reservoir in the terminology of \cite{Basso:2015zoa}. As well known, the structure constant for three BMN operators is protected in the SYM theory and given to all orders by its tree level expression.

\begin{figure}
\begin{center}
\includegraphics[scale=0.42]{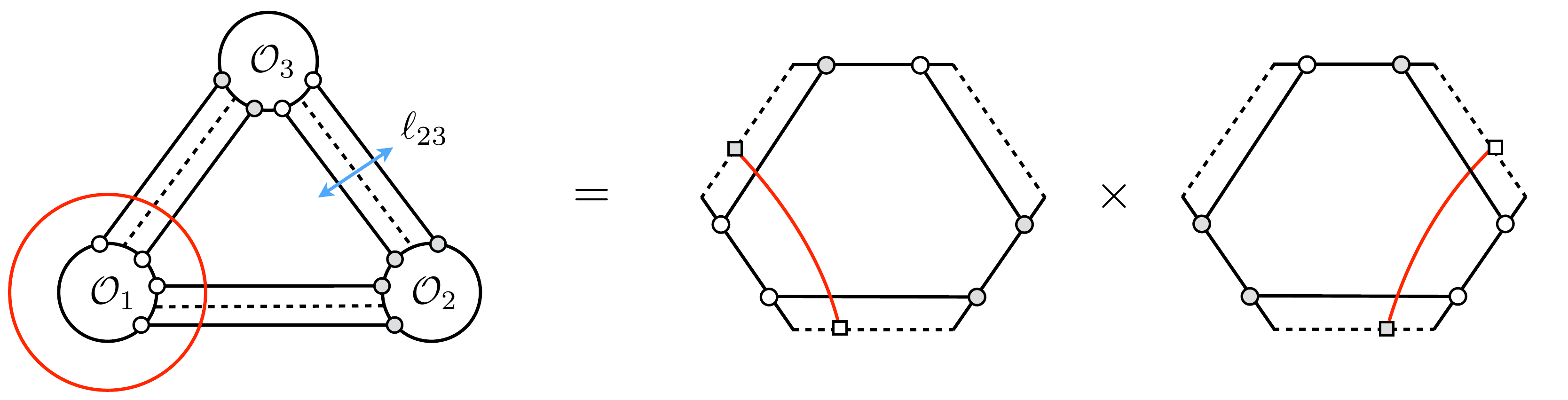}
\end{center}
\caption{Wheeled Feynman diagram contributing to a fishnet structure constant. White and grey dots represent the fields $\phi_{1}$ and $\phi_{1}^{\dagger}$ in the operators. Black and red lines represent propagators for $\phi_{1}$ and $\phi_{2}$, respectively.  The bridge length $\ell_{ij}$ counts the number of black propagators along each edge. Cutting along the three edges, as shown here in dashed lines, splits the Feynman diagram into two hexagons and cuts open the wheel.}\label{fC123} 
\end{figure}

In the fishnet theory, it is not possible to take the third operator in the form (\ref{reservoir}), since the above mixture is not an eigenstate of the dilatation operator, due to lack of symmetry. In fact, it is \textit{generically} not possible to have the three operators appearing on an equal footing, in the fishnet theory, since no BMN vacuum appears in the OPE of $\mathcal{O}_{1}$ and $\mathcal{O}_{2}$, barring extremal processes.%
\footnote{Extremal processes are found when the length of the third operator obeys $L_{3} = \pm (L_{1}-L_{2})$, a condition which permits the third operator to be a vacuum state. However, the associated structure constant is expected to vanish if the admixtures of double-trace operators are take into account.}
Instead, the operators entering this OPE look like domain walls of $\phi_{1}$ and $\phi_{1}^{\dagger}$, and the simplest choice of third operator corresponds to
\beq\label{third}
\mathcal{O}_{3} = \textrm{tr}\, \mathcal{\phi}^{\dagger \ell_{13}}_{1} \mathcal{\phi}^{\ell_{23}}_{1} \, ,
\eeq
where the splitting lengths, a.k.a bridge lengths, $\ell_{ij} = \ell_{ji}$ determine the pairing of fields in the BMN pair (\ref{one-two}), see figure \ref{fC123}, and are such that $\ell_{13}-\ell_{23} = L_{1}-L_{2}$, for charge conservation.

Interestingly, the domain-wall operator (\ref{third}) is protected in the fishnet theory, as long as $\ell_{13}, \ell_{23}\neq 0$; its anomalous dimension $\gamma_{3} = 0$, in the planar limit. It belongs to a broader family of protected states, which includes, in particular, the Lagrangian density~(\ref{dil}), as discussed in Subsection \ref{charged}. On the contrary, the BMN operators (\ref{one-two}), which are half-BPS in the SYM theory, receive anomalous dimensions in the fishnet theory, in lack of supersymmetry. Their anomalous dimensions are induced by the so-called wheel graphs \cite{Gurdogan:2015csr,Broadhurst:1985vq} which feature loops of the second complex scalar $\phi_{2}$ around the operators,
\beq
\gamma_{1,2} = \Delta_{1,2}-L_{1,2} = O(g^{2L_{1,2}})\, ,
\eeq
Every wheel costs $L_{1,2}$ powers of $g^2$ and thus the RHS above runs in integer powers of $g^{2L_{1,2}}$.

Assembling our three operators together, we obtain the vacuum structure constant
\beq\label{prototype}
C^{\bullet \circ \bullet}_{132} = \mathcal{h}\textrm{tr}\, \{\phi_{1}^{\ell_{21}}\phi_{1}^{\ell_{13}}\}(0)\,  \textrm{tr}\, \{\mathcal{\phi}_{1}^{\dagger \ell_{13}}\mathcal{\phi}^{\ell_{32}}_{1}\}(1)\,  \textrm{tr}\, \{\phi_{1}^{\dagger \ell_{32}}\phi_{1}^{\dagger \ell_{21}}\}(\infty)\mathcal{i}\, ,
\eeq
where, to prepare the ground for the hexagons, we parameterized all the operators in terms of the bridge lengths, with $\ell_{12} = L_{1}-\ell_{13} = L_{2}-\ell_{23}$; the latter count the numbers of $\mathcal{h}\phi_{1}\phi_{1}^{\dagger}\mathcal{i}$'s in each bridge, as shown in figure \ref{fC123}. Similar structure constants were discussed recently in \cite{Grabner:2017pgm,Gromov:2018hut}; see also \cite{Cavaglia:2018lxi} for a related set-up. The graphs contributing to (\ref{prototype}) are simply obtained by bringing together the wheels dressing each BMN operator; the third operator brings nothing in this respect. Altogether, they generate a double expansion in integer powers of $g^{2L_{1}}$ and $g^{2L_{2}}$, and, accordingly, the structure constant reads
\beq\label{C123pre}
C^{\bullet \circ \bullet}_{132} = \sqrt{L_{1}L_{2}}(1+\mathcal{O}(g^{2L_{1}})+\mathcal{O}(g^{2L_{2}}))\, ,
\eeq
for canonically normalised operators and after removal of the color factor $\sim 1/N$.

Traditionally, in the spin-chain picture, the $\phi_{2}$'s are seen as magnons propagating on top of the lattice defined by the $\phi_{1}$'s \cite{Minahan:2002ve}. The magnons circulating along the wheels are made of the same wood but are not attached to a specific operator. They are the so-called mirror magnons, which live between two locally BMN operators and account for the virtual particles winding around them~\cite{Ambjorn:2005wa,Bajnok:2008bm}. They are classified according to the little group of the two boundary operators: each magnon is then labelled with a momentum $p$, or a rapidity $u = p/2$, for dilatation $r\partial/\partial r = ip(u)$, and a pair of equal spins $(\tfrac{1}{2}(a-1), \tfrac{1}{2}(a-1))$, with $a=1, 2, \ldots\,$, for Lorentz rotations $\sim O(4)$, see Subsection~\ref{fishnet-hex}.

For illustration, a magnon inserted between $\mathcal{O}_{1}$ and $\mathcal{O}_{2}$, sitting at respectively $0$ and $\infty$, is given, in the fishnet theory, as a plane wave along the radial direction,
\beq\label{states}
|\phi_{2}(u)\mathcal{i}_{0\infty} = \phi^{\#}_{1}(0)\cdot \int\limits_{0}^{\infty} dr \, r^{ip(u)} \phi_{2}(r) \cdot \phi_{1}^{\dagger \#}(\infty)\, ,
\eeq
dropping the orbital part and associated spin labels, for simplicity. (An analogous picture is used to add excitations in the background of a null polygonal Wilson loop, in the form of insertions along its edges \cite{Belitsky:2011nn,Basso:2013aha,Belitsky:2014rba,Belitsky:2016fce}.) A generic Bethe state is obtained by concatenating magnons, $|\phi_{2}(\textbf{u})\mathcal{i}_{0\infty} = |\phi_{2}(u_{1}) \ldots \phi_{2}(u_{n})\mathcal{i}_{0\infty}$, and can be cast in the form (\ref{states}) by smearing $n$ insertions within a suitable wave function $\psi_{\textbf{u}}(\{r_{i}\})$. An essential property of the Bethe states, which determines their wave functions, is that they diagonalise the quartic interactions contained inside the bridge. Namely, the bridge $ij$ should be transparent to a Bethe state in the associated frame,
\beq\label{bridgeij}
\textrm{bridge}_{ij} \cdot |\phi_{2}(\textbf{u})\mathcal{i}_{ij} = e^{-E(\textbf{u}) \ell_{ij}} |\phi_{2}(\textbf{u})\mathcal{i}_{ij}\, ,
\eeq
up to an overall factor, controlled by the energy of the state, $E(\textbf{u}) = \sum_{i}E_{a_{i}}(u_{i})$. The embedding of the fishnet theory inside $\mathcal{N}=4$ SYM dictates that
\beq\label{Ea}
E_{a}(u) = -\log{g^2/(u^2+\tfrac{a^2}{4})}
\eeq
for the individual energy of a magnon in the wave $|p(u), a\mathcal{i}$, and, as expected, the transport of the state across the bridge results in $n\times \ell_{ij}$ powers of the coupling constant.

The idea underlying the hexagon factorization is to liberate the mirror magnons by opening up the traces in (\ref{prototype}) and cutting along the bridges. In the process, every wheel is cut open twice and the end-points so produced are mapped to mirror magnons sitting along the edges of two hexagons, see figure \ref{fC123}.
\begin{figure}
\begin{center}
\includegraphics[scale=0.45]{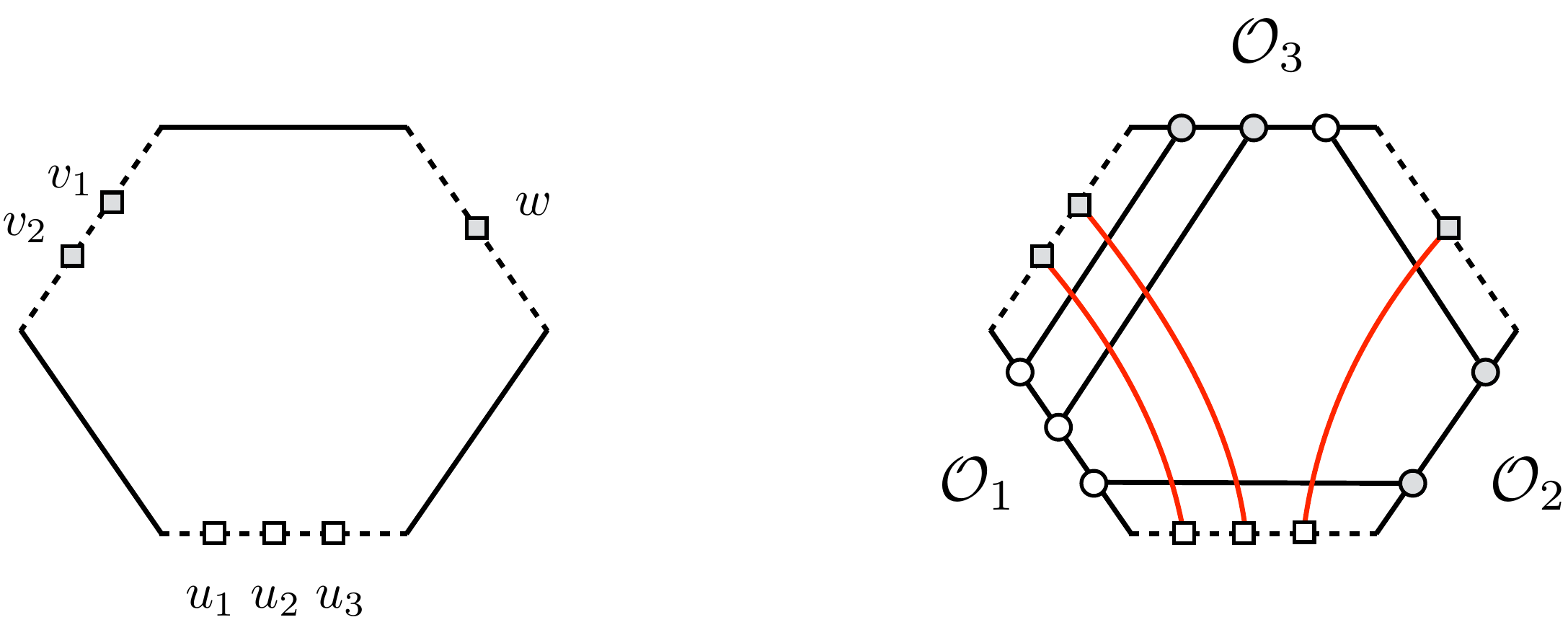}
\end{center} 
\caption{Hexagon form factor with magnons on the mirror edges and its fishnet counterpart. The quartic interactions are pushed to the boundary and absorbed inside the bridge factors. The hexagon form factor captures the splitting of the magnons' wave function according to the pattern of free (red) propagators.}\label{fig:mp1} 
\end{figure}
The hexagon form factors measure the overlaps between the three Bethe states in the three mirror cuts, as shown in figure \ref{fig:mp1},
\beq\label{Hpreamble}
H(\textbf{u}, \textbf{v}, \textbf{w}) = \, _{13\otimes 32}\mathcal{h} \phi_{2}(\textbf{v})^{\dagger} \otimes \phi_{2}(\textbf{w})^{\dagger}|\phi_{2}(\textbf{u})\mathcal{i}_{12}\, .
\eeq
In the basis of Bethe states, the effect of the bridges boils down to inserting the energy factors (\ref{bridgeij}) and, as a result, the structure constant is given, schematically, as \cite{Basso:2015zoa}
\beq\label{C123}
C^{\bullet\circ\bullet}_{132}/C_{132}^{\textrm{tree}} = \sum_{\textbf{u}, \textbf{v}, \textbf{w}} e^{-E(\textbf{u})\ell_{12}-E(\textbf{v})\ell_{13}-E(\textbf{w})\ell_{32}} \times  |H(\textbf{u}, \textbf{v}, \textbf{w})|^2 \, , 
\eeq
where each sum runs over a complete basis of states on the associated mirror cut. This expansion is readily seen to reproduce the structure of the perturbative series in (\ref{C123pre}), after taking into account that the number of magnons is conserved, for the processes under consideration, $|\textbf{u}| = |\textbf{v}|+|\textbf{w}|$, and that the hexagon form factors are coupling independent, in the fishnet theory, for properly normalised Bethe states.

In the following, we derive the expression for $H$, starting from the conjecture put forward in the SYM theory. Prior to move to this technical analysis, let us comment on a qualitative aspect of the hexagons in the fishnet theory. As should be clear from figure \ref{fig:mp1}, all the physics is pushed to the boundary, where the field theory interactions reside, and only the free propagators stay inside. The hexagons are seemingly made out of thin air, and, as for the tree-level pentagon OPE \cite{Basso:2013aha,Belitsky:2014rba,Belitsky:2016fce} or the tailoring procedure~\cite{Escobedo:2010xs}, the analysis boils down to studying free propagators. (The relation between free propagators and hexagons will be made more precise in Section \ref{Sect3}.) The analysis stays nontrivial, since the propagators must be convoluted with the mirror wave functions~$\psi$ in the relevant frames. These wave functions are not known in general; constructing them explicitly, using e.g.~the Schr\"odinger equation (\ref{bridgeij}), is demanding and evaluating their overlaps (\ref{Hpreamble}) even more. The hexagon bootstrap bypasses this difficulty by focusing on their asymptotic behaviours, which are controlled by the S matrix, but it entails a certain amount of guesswork too. It would be interesting to place the formalism on firm ground, using ``microscopic'' methods for building the wave functions. The corresponding problem for null polygonal Wilson loops was solved, for instance, in \cite{Belitsky:2014rba,Belitsky:2016fce} using the $SL(2)$ Baxter operator and its supersymmetric cousins, and progress was made recently with correlators in the 2d fishnet theory using an $SL(2, \mathbb{C})$ version of the formalism \cite{Derkachov:2018rot}. A generalisation to $SL(4)$ appears to be needed for the correlators of the 4d fishnet theory.

\subsection{SYM hexagon}

The SYM theory has many more fields than the fishnet theory but also many more symmetries. Its magnons come in more flavours but can all be packed together inside short irreducible representations of the BMN symmetry group $SU(2|2)^2$, or, to be precise, of a suitable extension thereof \cite{Beisert:2006qh}. In particular, the lightest magnons fill a bi-fundamental (16-dimensional) representation,
\beq
\chi_{A\dot{A}}(u) = \chi_{A} \otimes \chi_{\dot{A}}(u) \, ,
\eeq
with $\chi_{A} \in (\varphi_{a=1,2}\, |\, \psi_{\alpha=1,2})$ a quartet of bosonic$|$fermionic fields and with the rapidity $u$ labelling the energy $E(u)$ and momentum $p(u)$. Heavier magnons are obtained by binding $a$ fundamental magnons together \cite{Dorey:2006dq}, in the appropriate channel, and fill $(4a)^2$-dimensional irreps, with $a=1,2,3...$. In the following, we will drop the bound state label, keeping in mind that formulae for bound states entail fusing those for the elementary magnons.

\begin{figure}[t]
\begin{center}
\includegraphics[scale=0.45]{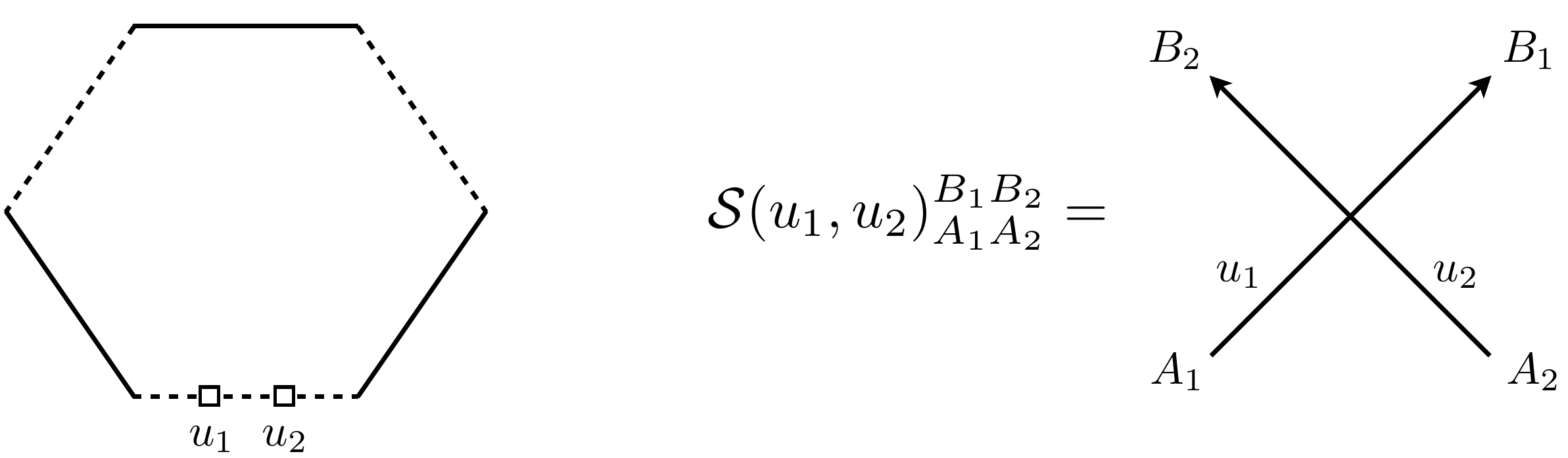}
\end{center}
\caption{Two-magnon hexagon form factor and its matrix part. A pair of magnons on a mirror edge is absorbed by the hexagon. The module of the amplitude is controlled by the abelian factor $h(u_{1}, u_{2})$. The matrix part accounts for the contraction of the magnons' left and right indices. Raising the right indices with the conjugation matrix, we can write it as the matrix element of the fundamental S matrix $\mathcal{S}$ shown in the right panel.}\label{fmp} 
\end{figure}

Hexagon processes in the SYM theory are also richer than their fishnet counterparts, as they capture more graphs. In particular, the SYM hexagon can absorb or produce magnons. The simplest form factor quantifies this effect and comes with an ordered set of magnons $\textbf{u} = \{u_{1}, u_{2}, \ldots\}$ along a single given edge, as shown in figure \ref{fmp}.%
\footnote{Note that, in this paper, we work with the anti-clockwise ordering, when drawing magnon sets along the contour of the hexagon.} It can be written formally as
\beq\label{hexff}
h_{A_1 \dot{A}_1, A_2 \dot{A}_2, \ldots} (\textbf{u}) = \mathcal{h}\mathfrak{h}|\chi_{A_{1}\dot{A}_{1}}(u_{1}), \chi_{A_{2}\dot{A}_{2}}(u_{2}), \ldots \mathcal{i} \otimes |0 \mathcal{i} \otimes |0 \mathcal{i}\, ,
\eeq
where the bra represents the hexagon vertex and the kets the states on its edges. Reshuffling magnons in a state follows from the action of the S matrix and translates into a constraint on the form factor (\ref{hexff}). The latter is a universal axiom known as the Watson relation. E.g., for two magnons, it requires that
\beq\label{perm}
h_{A_1 \dot{A}_1, A_{2}\dot{A}_{2}} (u_{1}, u_{2}) = \mathbb{S}(u_{1}, u_{2})_{A_{1}\dot{A}_{1}, A_{2}\dot{A}_{2}}^{B_{1}\dot{B}_{1}, B_{2}\dot{B}_{2}} \, h_{B_2 \dot{B}_2, B_{1}\dot{B}_{1}} (u_{2}, u_{1})\, ,
\eeq
with implicit sums over the $B$'s, and with~\cite{Beisert:2006qh,Arutyunov:2006yd,Arutyunov:2006iu}
\beq\label{Ssym}
\mathbb{S}(u_{1}, u_{2})_{A_{1}\dot{A}_{1}, A_{2}\dot{A}_{2}}^{B_{1}\dot{B}_{1}, B_{2}\dot{B}_{2}}  = (-1)^{\textbf{f}} S(u_{1}, u_{2})\mathcal{S}(u_{1}, u_{2})_{A_{1}A_{2}}^{B_{1}B_{2}}\mathcal{S}(u_{1}, u_{2})_{\dot{A}_{1}\dot{A}_{2}}^{\dot{B}_{1}\dot{B}_{2}}\, ,
\eeq
the 2-magnon S matrix, with $S$ the abelian factor, $\mathcal{S}$ its left/right component, and $\textbf{f} = f_{\dot{A}_{1}}f_{A_{2}}+f_{\dot{B}_{2}}f_{B_{1}}$ a grading factor for the left-right scattering, with $f_{A}$ the fermion number of $\chi_A$, etc.

The factorised ansatz put forward in \cite{Basso:2015zoa} expresses the form factor (\ref{hexff}) as a square root of the S matrix, obtained by dropping the right $\mathcal{S}$ matrix and mapping the right magnons' components to outgoing particles. More precisely, it casts it into the form
\beq\label{hansatz}
h_{A_1 \dot{A}_1, \ldots} (\textbf{u}) = h_{<}(\textbf{u})\times \mathcal{M}_{A_1 \dot{A}_1, \ldots} \, , \qquad h_{<}(\textbf{u}, \textbf{u}) = \prod_{i<j} h(u_{i}, u_{j})\,,
\eeq
where $h(u, v)$ is an explicitly known function, called dynamical or abelian factor, fulfilling $h(u, v)/h(v, u) = S(u, v)$, and with the matrix part $\mathcal{M}$ given by
\beq\label{simple}
\mathcal{M}(u_{1}, u_{2}, \ldots )_{A_{1}\dot{A}_{1}, A_{2}\dot{A}_{2}, \ldots} = (-1)^{\mathfrak{f}}\, \mathcal{S}_{123\ldots}(u_{1}, u_{2}, \ldots )_{A_{1}A_{2}\ldots }^{B_{1}B_{2}\ldots } \, \Co_{B_{1}\dot{A}_{1}}\Co_{B_{2}\dot{A}_{2}}\ldots \, ,
\eeq
where
\beq
\mathcal{S}_{123...} = \mathcal{S}_{<}(\textbf{u}, \textbf{u})= \ldots\mathcal{S}_{23}\mathcal{S}_{13}\mathcal{S}_{12}
\eeq
is the factorised many-body $\mathcal{S}$ matrix. $\Co_{AB}$ is a fixed conjugation matrix, $\Co_{AB} = \epsilon_{ab} | i\epsilon_{\alpha\beta}$, with $\epsilon_{12} = -\epsilon_{21} = 1$, needed to cross the right indices, and $\mathfrak{f} = \sum_{i>j}f_{A_{i}}f_{\dot{A}_{j}}$ is a grading factor for the reshuffling of the left and right components in the state. Note that one could also raise the right indices in (\ref{simple}) using the inverse matrix $\Co^{\dot{A} B}$, defined as $\Co^{A B} \Co_{B C} =\delta^{A}_{C}$, and write $\mathcal{M}$ as a standard $\mathcal{S}$ matrix element. The explicit expressions for the components of $\mathcal{S}$, to be used later on, can be read out from \cite{Caetano:2016keh}. The ansatz (\ref{hansatz}) is the simplest tensor one can write that is invariant w.r.t.~the diagonal subgroup of symmetries $SU(2|2)_{D}\subset SU(2|2)^2$ preserved by the hexagon. In fact, the diagonal symmetry fixes the solution uniquely, up to the abelian factor, for two magnons \cite{Basso:2015zoa}. Also, the Watson relation is easily seen to be satisfied, thanks to the double copy structure of the full S matrix (\ref{Ssym}) and the fundamental properties of $\mathcal{S}$, i.e., Yang-Baxter relation, unitarity, etc.

For our investigation, cf.~earlier discussion, the magnons should lie in the mirror kinematics. The latter is usually reached by transporting magnons using the mirror (90$^{\circ}$) rotation $\gamma: u \rightarrow u^{\gamma}$ starting from the spin chain kinematics. To avoid cluttering our formulae, we shall drop the upper-scripts referring to this mirror move and place ourselves on the mirror sheet from the onset. To handle this kinematics properly, we shall adopt the string worldsheet normalisation and work in the so-called string frame~\cite{Arutyunov:2006iu}.

\begin{figure}[t]
\begin{center}
\includegraphics[scale=0.45]{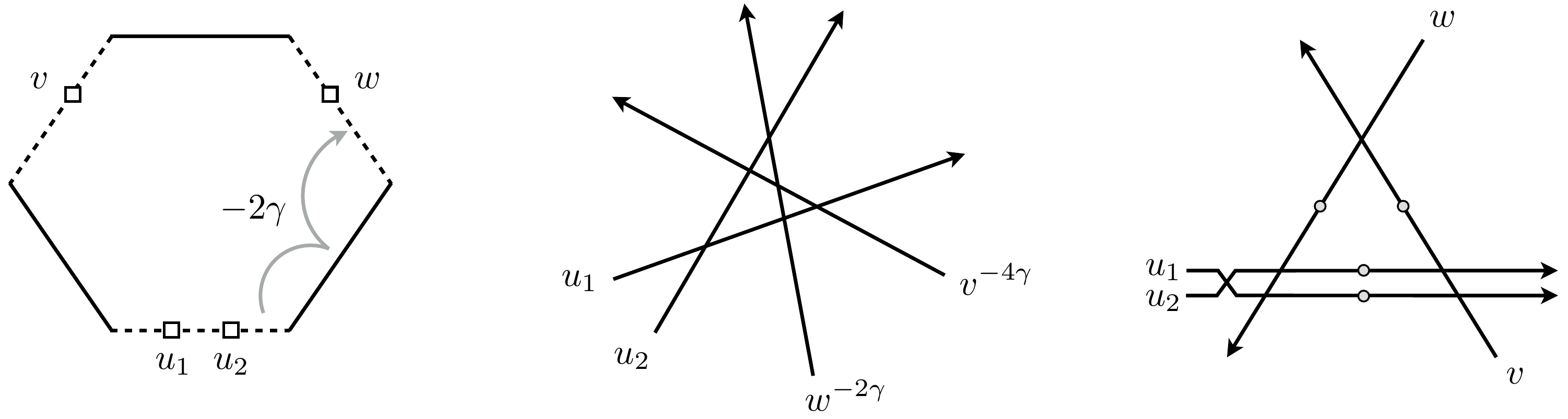}
\end{center}
\caption{A generic hexagon form factor. The magnons are distributed on the three mirror edges as in the leftmost panel. In the middle panel, we have the standard representation of the matrix part, obtained by analytically continuing rapidities to the crossed and doubly crossed kinematics. In the rightmost panel, we show an alternative representation where all rapidities are set back to the same kinematics 
using the crossing properties of $h$ and $\mathcal{S}$. This operation flips the orientation of the $\textbf{w}$ lines and makes the cyclic symmetry manifest. The price to pay for this re-organisation is a grading of the sums over intermediate states in the loops, represented by the dots.}\label{mp0} 
\end{figure}

More importantly, the magnons should be more evenly distributed on the top and bottom edges of the hexagon, as in e.g.~figure \ref{mp0}, since the magnons to be considered will be charged w.r.t.~the diagonal subgroup. These more generic form factors,
\beq\label{hgeneric}
h_{A_1 \dot{A}_1, \ldots;  \, A_2 \dot{A}_2, \ldots; \, A_3 \dot{A}_3, \ldots}  (\textbf{u}, \textbf{v}, \textbf{w}) = \mathcal{h}\mathfrak{h}|\chi_{A_{1}\dot{A}_{1}}(u_{1})\ldots \mathcal{i}  \otimes |\chi_{A_{3}\dot{A}_{3}}(w_{1})\ldots \mathcal{i} \otimes |\chi_{A_{2}\dot{A}_{2}}(v_{1})\ldots \mathcal{i} \, ,
\eeq
can be obtained by implementing mirror moves, or crossing transformations \cite{Janik:2006dc}, on the magnons in (\ref{hansatz}), following the rules spelled out in the appendices of Refs.~\cite{Basso:2015zoa} and \cite{Caetano:2016keh}. Performing these manipulations gives the form factor (\ref{hgeneric}) as a $\mathcal{S}$ matrix element with arguments $\textbf{u}, \textbf{w}^{-2\gamma}, \textbf{v}^{-4\gamma}$; see middle panel in \ref{mp0}. One can massage this expression and obtain a cyclic symmetric representation with all the arguments lying on the same kinematical sheet. To do so, one simply makes use of the crossing properties of $h$ and $\mathcal{S}$. More precisely, one needs, see \cite{Basso:2015zoa,Arutyunov:2006iu,Janik:2006dc},   
\begin{equation}
h(u^{2 \gamma}, v^{2 \gamma}) = h(u,v) \, , \quad
\quad 
h(u^{4 \gamma}, v) =\frac{1}{h(v,u)} \, , 
\end{equation}
together with
\begin{equation}
\begin{aligned}
h(w,v) h(w^{-2\gamma}, v^{-4\gamma}) \mathcal{S}(w^{-2\gamma}, v^{- 4\gamma})_{AB}^{CD} &= \Co_{AE} \, \mathcal{S}(v, w)^{DE}_{BF}
\, \Co^{FC} \, ,  \\ 
& \\
h(u, w)h(u,w^{-2\gamma})
\mathcal{S}(u, w^{-2\gamma})_{A B}^{CD}  &= \Co^{DE} \, \mathcal{S}_{EA}^{FC}(w,u) \, \Co_{FB} \, ,   
\end{aligned}
\end{equation}
and 
\begin{equation}
\mathcal{S}(u, v^{ 4 \gamma})_{AB}^{CD} = (-1)^{f_D} \mathcal{S}_{AB}^{CD}(u,v) (-1)^{f_B} \, , \quad   
\mathcal{S}(u^{4 \gamma}, v)_{AB}^{CD} = (-1)^{f_C} \mathcal{S}_{AB}^{CD}(u,v) (-1)^{f_A} \, . 
\end{equation} 
These relations are used, graphically, to flip the orientation of the $\textbf{w}$ lines (as well as to undo the $-4\gamma$ move of the $\textbf{v}$'s). Assembling all pieces together, we get the cyclic representation
\beq
h(\textbf{u}, \textbf{v}, \textbf{w})_{A_1 \dot{A}_1, \ldots} = \frac{h_{<}(\textbf{u}, \textbf{u})h_{<}(\textbf{v}, \textbf{v})h_{<}(\textbf{w}, \textbf{w})}{h(\textbf{u}, \textbf{w})h(\textbf{w}, \textbf{v})h(\textbf{v}, \textbf{u})} \times \mathcal{M}_{A_1 \dot{A}_1, \ldots}  \, ,
\label{eq:hexagonsamesheet} 
\eeq
where the matrix part is illustrated in the right panel of figure \ref{mp0} on a particular example. The matrix part is easy to spell out for a single magnon on each edge and reads
\beq\label{3M}
\mathcal{M}(u, v, w)_{A_{1}\dot{A}_{1}, A_{2}\dot{A}_{2}, A_{3}\dot{A}_{3}} = (-1)^{\#} \Co_{B_{1}\dot{A}_{1}}\Co_{B_{2}\dot{A}_{2}}\Co_{B_{3}\dot{A}_{3}}\mathcal{Z}_{A_{1}A_{2}A_{3}}^{B_{1}B_{2}B_{3}}\, ,
\eeq
with the overall sign%
\footnote{Its cyclic symmetry follows from the condition $F_A+F_B+F_C = 0$ mod 2. }
\beq
(-1)^{\#} = (-1)^{f_{A_{1}} + f_{A_{2}} + f_{A_{3}}}
(-1)^{f_{\dot{A}_{1}} f_{A_{2}}+f_{\dot{A}_{2}} f_{A_{3}}+ f_{\dot{A}_{3}} f_{A_{1}}} 
(-1)^{F_{1} + F_{2} F_{3}}  \, ,
\eeq
where $F_{i} = f_{A_{i}} + f_{\dot{A}_{i}}$. The core of the interaction is obtained by concatenating S matrices,
\beq\label{core}
\mathcal{Z}_{A_{1}A_{2}A_{3}}^{B_{1}B_{2}B_{3}} = 
(-1)^{f_{C_1}+f_{C_2}+f_{C_3}}  
\mathcal{S}(u,v)_{C_{1}A_{2}}^{B_{1}C_{2}}\mathcal{S}(v,w)_{C_{2}A_{3}}^{B_{2}C_{3}}\mathcal{S}(w,u)_{C_{3}A_{1}}^{B_{3}C_{1}}
 \, ,
\eeq
with a graded sum over the internal magnons' flavors $C_{1,2,3}$. For more magnons, one should dress with self-interactions the external legs, as shown in figure \ref{mp0}, scatter the three stacks together using the mutli-line uplift of the central vertex (\ref{core}) and finally contract left and right movers using the conjugation matrix. One could also remove magnons by sending lines to infinity. E.g., removing $w$ in (\ref{3M}), one gets
\beq\label{1to1}
\mathcal{M}(u, v)_{A_{1}\dot{A}_{1}, A_{2}\dot{A}_{2}} = (-1)^{f_{\dot{A}_{1}}f_{A_{2}}} \Co_{B_{1}\dot{A}_{1}}\Co_{B_{2}\dot{A}_{2}}\mathcal{S}(u,v)_{A_{1}A_{2}}^{B_{1}B_{2}}\, ,
\eeq
which appears to be the same matrix part as for the 2-body annihilation form factor, see Eq.~(\ref{simple}). (This well-known relation follows from the fact that the $\pm 4\gamma$ rotation acts trivially on the matrix part.)

\begin{figure}[t]
\begin{center}
\includegraphics[scale=0.45]{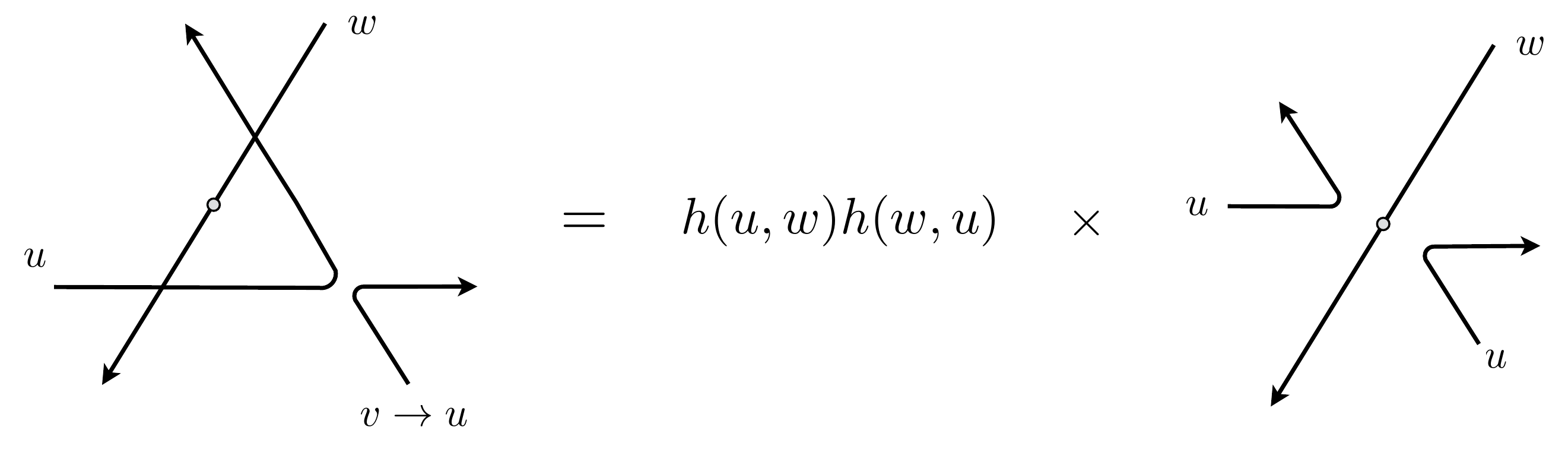}
\end{center}
\caption{Illustration of the decoupling of the matrix part for the three-magnon configuration. In the limit where $v\rightarrow u$ the $uv$ interaction reduces to a permutation, $\mathcal{S}\rightarrow -P$, and the $uw$ lines can be disentangled up to an overall abelian factor. The relation shown here is equivalent to the unitarity of $\mathcal{S}$ after crossing the magnon $w$. The abelian factor spit out by the matrix part completes the decoupling of the dynamical factor in (\ref{eq:hexagonsamesheet}) in the limit $v\rightarrow u$.}\label{decI} 
\end{figure}

The representation (\ref{eq:hexagonsamesheet}) also makes the kinematical singularities of the hexagon form factor manifest in the 3 channels. Namely, the form factor has a (simple) pole whenever two magnons, on different edges, take the same rapidity and have matching quantum numbers. This pole stems from the abelian factor in (\ref{eq:hexagonsamesheet}) and from the vanishing of $h(u, v)$ at $u = v$. Physically, it represents the situation where a magnon moves far away from the core of the hexagon and decouples. Its residue relates to the measure $\mu(u)$ normalising the magnon wave function. E.g., decoupling the leftmost particle, for simplicity, by taking $v_{n}\sim u_{1}$, one obtains
\beq\label{decleft}
h(\{u_{1}, \ldots \} ; \{\ldots,  v_{n}\}; \textbf{w}) \sim \frac{i \mathcal{I}}{\mu(u_{1})(v_{n}-u_{1})}\times h(\textbf{u}\backslash \{u_{1}\}; \textbf{v}\backslash\{v_{n}\}; \textbf{w})\, ,
\eeq
where $\mathcal{I}$ is a tensor contracting the indices of the decoupled pair of magnons. (The explicit expression for $\mathcal{I}$ will not be needed but could be read out from Eq.~(\ref{1to1}).) The factorisation of the matrix part underlying (\ref{decleft}) is depicted in figure \ref{decI} for the three-magnon configuration.

\subsection{Fishnet hexagon}\label{fishnet-hex}

The projection to the fishnet theory is done by selecting good scalar components and taking the weak coupling limit. More precisely, we shall select the SYM magnons carrying maximal charges under the $U(1)_{R}$ subgroup of $SU(2|2)_{D}$, distribute them along the edges of the hexagon as in figure \ref{fig:mp1}, and finally take the weak coupling limit. This choice of polarisation insures that the reservoir is transparent to the magnons and reduces to the domain-wall operator (\ref{third}), to leading order at weak coupling. These magnons are transverse, in the terminology of \cite{Basso:2015zoa}, and correspond to
\beq\label{fishnet-magnon}
\phi_{2}(u) = \varphi_{1}\otimes \varphi_{\dot{1}}(u)\, , \qquad \phi_{2}^{\dagger}(u) = \varphi_{2}\otimes \varphi_{\dot{2}}(u)\, ,
\eeq
for the elementary ones. Their relatives in the bound-state multiplets form higher representations of the Lorentz group, see e.g.~\cite{Ahn:2011xq,Fleury:2016ykk}, obtained by attaching derivatives to the scalar fields, e.g.,
\beq \label{DerivativesOneScalar} 
\partial_{\alpha_{1}\dot{\alpha}_{1}} \ldots \partial_{\alpha_{a-1}, \dot{\alpha}_{a-1}}\phi_{2}(u) = (-1)^{\tfrac{1}{2}(a-1)(a-2)}|\varphi_{1}\psi_{\alpha_{1}} \ldots \mathcal{i} |\varphi_{\dot{1}}\psi_{\dot{\alpha}_{1}} \ldots \mathcal{i} \, ,
\eeq
with $a$ the bound state label. They span, for given $a$, a symmetric traceless representation $V_{a}\otimes \dot{V}_{a}$ of $O(4)$, with spins $(\tfrac{1}{2}(a-1), \tfrac{1}{2}(a-1))$ and dimension $a^2$, and, altogether, they are enough to reconstruct the full 4d massless scalar fields of the fishnet theory. The energy $E_{a}(u)$ of a magnon, carrying momentum $p_{a}(u) = 2u$, is given by the SYM weak coupling formula~(\ref{Ea}).

The fishnet S matrix does not depend on our choice of polarisation and follows directly from the scalar component of the SYM S matrix (\ref{Ssym}), after taking the weak coupling limit in the mirror kinematics. As well known, the spin-chain interactions rationalise in the weak coupling limit, and, as a result, the fishnet S matrix factorises into two copies of the XXX $SU(2)$ R matrix, for the left and right Lorentz indices, respectively,
\beq\label{bbS}
\mathbb{S}_{ab}(u, v) = S_{ab}(u, v) \times R_{ab}(u-v) \otimes \dot{R}_{ab}(u-v)\, ,
\eeq
up to the scalar factor
\beq\label{Sab}
S_{ab}(u, v) = \frac{\tfrac{a+b}{2}+iu-iv}{\tfrac{a+b}{2}-iu+iv}\prod_{k=0, 1}\frac{\Gamma(k+\tfrac{a}{2}-iu)\Gamma(k+\frac{a-b}{2}+iu-iv)\Gamma(k+\tfrac{b}{2}+iv)}{\Gamma(k+\tfrac{a}{2}+iu)\Gamma(k+\frac{a-b}{2}-iu+iv)\Gamma(k+\tfrac{b}{2}-iv)}\, .
\eeq
Here, $R_{ab}$ is the standard R matrix  \cite{Kulish:1981gi,Sogo:1984bu,Faddeev:1996iy} acting on the tensor product of the $a$-th and $b$-th irrep of $SU(2)$, with dimension $a$ and $b$, respectively,
\beq
\begin{aligned}
&R_{ab} : V_{a}\otimes V_{b} \rightarrow V_{b}\otimes V_{a}\, , \\
&\qquad\,\,  |\alpha, \beta\mathcal{i} \rightarrow R_{ab}(u-v)_{\alpha\beta}^{\gamma\delta} |\delta, \gamma\mathcal{i}\, ,
\end{aligned}
\eeq
with $\alpha, ...$ the multi-spinor indices appropriate for totally symmetric tensors of rank $a-1$ and $b-1$. It can be obtained by fusing the fundamental (spin $1/2$) R matrix, with~$a=b=2$ in our notations,
\beq
R_{22}(u)_{\alpha\beta}^{\gamma\delta} = \frac{u}{u+i}\delta_{\alpha}^{\gamma}\delta_{\beta}^{\delta} + \frac{i}{u+i}\delta_{\alpha}^{\delta}\delta_{\beta}^{\gamma}\, .
\eeq
We spell it out in Appendix \ref{AppR} in the symmetric product basis (\ref{DerivativesOneScalar}). Alternatively, we can define it with no reference to a basis by collecting its eigenvalues,
\beq\label{Rab-eigen}
R_{ab}(u) = \sum_{j=0}^{\textrm{max}} (-1)^{j} \frac{\Gamma(\tfrac{a+b}{2}+iu)\Gamma(\tfrac{a+b}{2}-iu-j)}{\Gamma(\tfrac{a+b}{2}-iu)\Gamma(\tfrac{a+b}{2}+iu-j)} P_{a+b-1-2j}\, ,
\eeq
where $\textrm{max} = \frac{1}{2}(a+b-|a-b|)-1$ and with $P_{a+b-1-2j}$ the projector on the dim $(a+b-1-2j)$ irrep $\subset V_{a}\otimes V_{b}$. Its normalisation is such that $R_{ab} = 1$ in the symmetric channel, corresponding to $j = 0$, that it reduces to the identity matrix, $R_{ab}\rightarrow I$, when $u\rightarrow \infty$ and to the permutation operator at $u=0$ when $b=a$. 
Let us finally recall that it obeys the functional (crossing) relation
\beq\label{crossinge}
R_{ab}(u-i)_{\alpha\beta}^{\gamma\delta} = c_{ab}(u)\co_{b\,\beta \sigma}
R_{ba}(-u)_ {\rho \alpha}^{ \sigma\gamma}  \co_{b}^{\,\, \rho\delta} \, ,
\eeq
where $\co_{b}$ is the conjugation matrix defined by $\co_{2\,\alpha\beta} = \epsilon_{\alpha\beta}, \co_{2}^{\,\,\beta\alpha} = \epsilon^{\alpha\beta}$, with $\epsilon_{12} = \epsilon^{12} = 1$ for fundamental spins, and by suitable products thereof for higher $b$. The crossing factor is given by
\beq\label{crossingf}
c_{ab}(u^{+}) = \prod_{j= \frac{|a-b|}{2}}^{\frac{a+b}{2}-2} \frac{u^{-}-ij}{u^{+}+ij}\, ,
\eeq
where $u^{\pm} = u\pm i/2$.%
\footnote{It solves the fusion relations $c_{ab}(u^{+})c_{ab}(u^{-}) = c_{a+1, b}(u) c_{a-1, b}(u)$ with the initial conditions $c_{1b}(u) = 1, c_{2b}(u) = (u-\tfrac{ib}{2})/(u+\tfrac{i(b-2)}{2})$.}

\begin{figure}
\begin{center}
\includegraphics[scale=0.45]{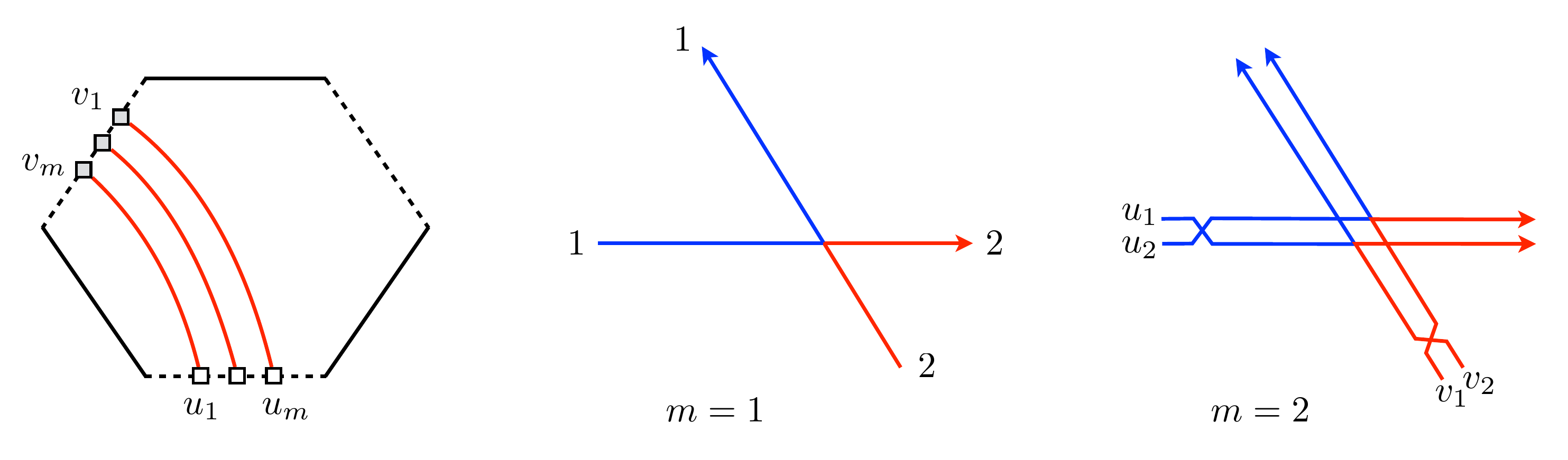}
\end{center}
\caption{Hexagon transition $H(\textbf{u}\rightarrow \textbf{v})$ with all magnons going to the left. The numbers of incoming and outgoing magnons must match 
for charge conservation. On the right panel, the SYM domain wall partition functions for the matrix parts when $m=1$ and $m=2$. At weak coupling the mirror S matrix is transmission-less and the partition function collapses to a single process where the magnons' flavors backscatter one each other.}\label{mp1} 
\end{figure}

Given the S matrix, the next step is to reduce the hexagon form factors. We shall proceed step-by-step starting with the simplest configurations where all the magnons are elementary and propagate on the left-hand side of the hexagon, as shown in figure \ref{mp1}. The computation of the corresponding form factor is an immediate application of the general formula given in the previous subsection. The most complicated component is the matrix part, which is represented by the partition function in figure \ref{mp1}. For a single magnon transition $u\rightarrow v$, we read out from (\ref{1to1}), using (\ref{fishnet-magnon}),
\beq
H(u\rightarrow v) = \mathcal{h}\mathfrak{h}|\phi_{2}(u)\mathcal{i}\otimes |0\mathcal{i}\otimes |\phi_{2}^{\dagger}(v)\mathcal{i} = \frac{1}{h(v, u)}\epsilon_{2\dot{1}}\epsilon_{1\dot{2}} \mathcal{S}(u, v)_{12}^{21}\, ,
\eeq
where, see e.g.~appendices in \cite{Caetano:2016keh},
\beq
\mathcal{S}(u, v)_{12}^{21} = \frac{1}{2}(A(u, v)+B(u, v))\, ,
\eeq
with $A$ and $B$ parameterising the symmetric and antisymmetric amplitudes of the scalar restriction of the S matrix. The $A$ amplitude is unitary and fulfills $A(u, v)A(v, u) = 1$ at any coupling. This is not a priori the case for the $B$ amplitude, since bosons and fermions can mix in the antisymmetric channel \cite{Beisert:2006qh}. However, as well known, this effect is absent to leading order at weak coupling. Moreover, in the mirror kinematics, the weak coupling scattering is transmission-less, and thus
\beq
B = A +O(g^2)\, .
\eeq
Hence, the hexagon form factor in the fishnet theory is simply given by
\beq
H(u\rightarrow v) = \frac{1}{H(v, u)}\, ,
\eeq
where $H(u, v) = -A(u, v)h(u, v)$ is the scalar hexagon amplitude \cite{Basso:2015zoa}. The analysis generalises straightforwardly to configurations involving more magnons, as shown in figure \ref{mp1}, thanks to the aforementioned properties of the scalar S matrix. The general formula is fully factorised and simply given by
\beq\label{left}
H(\textbf{u}\rightarrow \textbf{v}) = \frac{H_{<}(\textbf{u}, \textbf{u})H_{<}(\textbf{v}, \textbf{v})}{H(\textbf{v}, \textbf{u})}\, .
\eeq 
Similar simplifications are observed for bound states, although less transparently. In this case, the S matrix is more bulky and fermions must be included to represent the derivatives. Nonetheless, the scalar and Lorentz parts are seen to factorise and the final expression is a natural higher spin uplift of (\ref{left}). The abelian part is literally just (\ref{left}) up to $H\rightarrow H_{ab}$, with $H_{ab} = -A_{ab}h_{ab}$ and $A_{ab}$ the bound-state scalar amplitude, while the matrix part has a similar structure but in terms of R matrices. Putting all factors together, we get
\beq\label{Habuv}
H_{\textbf{a}\textbf{b}}(\textbf{u}\rightarrow \textbf{v}) = 
i^{f_\textbf{a}} i^{f_\textbf{b}} \frac{H^{<}_{\textbf{a}\textbf{a}}(\textbf{u}, \textbf{u})H^{<}_{\textbf{b}\textbf{b}}(\textbf{v}, \textbf{v})}{H_{\textbf{b}\textbf{a}}(\textbf{v}, \textbf{u})} \times R_{\textbf{a}\textbf{b}}(\textbf{u}, \textbf{v})R^{<}_{\textbf{b}\textbf{b}}(\textbf{v}, \textbf{v})R^{<}_{\textbf{a}\textbf{a}}(\textbf{u}, \textbf{u})\, ,
\eeq
where $f_{\textbf{a}} = \sum_{i}(a_{i}-1)$, and similarly for $f_{\textbf{b}}$. The indices enter as in the SYM formula, see, e.g., Eq.~(\ref{1to1}), with the dotted indices in the LHS obtained by lowering the outgoing indices of the R matrices using the conjugation matrix $\co$. E.g, for a single magnon transition, we have, using multi-spinor indices,
\beq\label{fullHab}
H_{ab}(u\rightarrow v)_{\alpha\dot{\alpha}, \beta\dot{\beta}} = i^{a+b-2}  \frac{C_{a\, \gamma \dot{\alpha}}C_{b\, \delta \dot{\beta}}}{H_{ba}(v, u)} R_{ab}(u-v)_{\alpha\beta}^{\gamma\delta}\, .
\eeq
The explicit expression for  $H_{ab}(u,v)$ will be given later on, see Eq.~(\ref{Hab}). The formula for the transition to the right-hand side of the hexagon follows from turning the picture around, i.e., by exchanging $u$ and $v$.

\begin{figure}
\begin{center}
\includegraphics[scale=0.45]{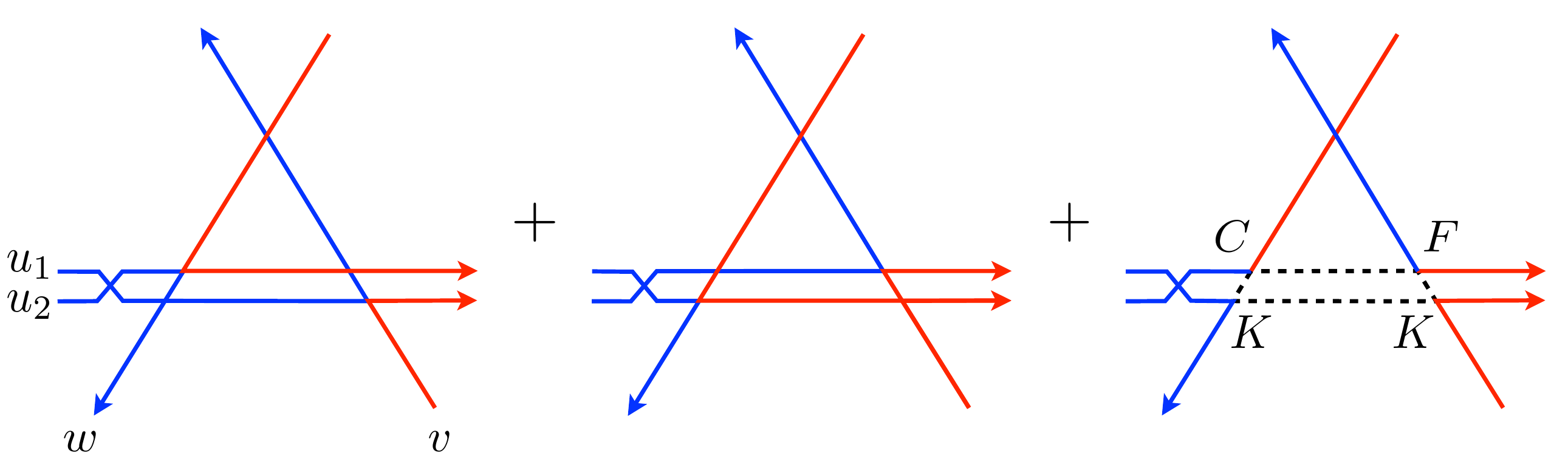}
\end{center}
\caption{Processes contributing to the matrix part for the hexagon splitting of a two-magnon wave function, with the magnons arranged as in left panel of figure \ref{mp0}. One of them displays a fermion loop represented by the dashed line. (The grading factors drop out since the number of fermions involved is even.)}\label{mp3} 
\end{figure}

We proceed with the more complicated situations where the beam of magnons is split in two, $\textbf{u}\rightarrow \textbf{v}|\textbf{w}$. The simplest such process is given by
\beq
\begin{aligned}
H(u_{1}, u_{2}\rightarrow v|w) &= \mathcal{h}\mathfrak{h}|\phi_{2}(u_{1})\phi_{2}(u_{2})\mathcal{i} \otimes|\phi_{2}^{\dagger}(w)\mathcal{i}\otimes |\phi_{2}^{\dagger}(v)\mathcal{i}\\
&= \frac{h(u_{1}, u_{2})\mathcal{M}(u_{1}, u_{2}, v, w)}{h(u_{1}, w)h(u_{2}, w)h(w, v)h(v, u_{1})h(v, u_{2})}\, ,
\end{aligned}
\eeq
with $\mathcal{M}$ the matrix part depicted in figure \ref{mp3}. Applying the general formula, we find that the matrix part $\mathcal{M}$ receives three contributions, one for each graph in figure \ref{mp3} and with the last one featuring a fermion loop. They yield
\beq
\begin{aligned}
&\mathcal{M} =  \frac{1}{2}A_{u_1u_2}(A_{vw}-B_{vw})\times \\
&\qquad \,\, \bigg[\frac{1}{8}(A_{u_1v}+B_{u_1v})(A_{u_2 v}-B_{u_2 v})(A_{wu_2}+B_{wu_2})A_{wu_1} \\
&\qquad \,\, +\frac{1}{8}A_{u_1v}(A_{u_2v}+B_{u_2v})(A_{wu_2}-B_{wu_2})(A_{wu_1}+B_{wu_1}) -\frac{1}{2}K_{u_1v}F_{u_2v}C_{wu_2}K_{wu_1}\bigg]\, ,
\end{aligned}
\eeq
where to save space we placed the arguments as subscripts, with $A,B$ the scalar amplitudes, $C, F \sim g$ the amplitude for creation and annihilation of a pair of fermions, and with $K\sim g^0$ the fermion-scalar reflection amplitude. All terms in brackets start at order $g^2$, including the one with fermions in the loop.%
\footnote{We should stress that the scaling with the coupling does not imply that the form factor is sub-leading. Indeed, a vanishing result would be in tension with the decoupling property of the fishnet hexagon form factors. The scaling with the coupling is merely reflecting the implicit normalisation of the external states in the SYM representation.} Straightforward algebra gives
\beq
H(u_{1}, u_{2}\rightarrow v|w) =\frac{H(u_{1}, u_{2})H(v, w)}{H(u_{1}, w)H(u_{2}, w)H(v, u_{1})H(v, u_{2})}\, .
\eeq
Remarkably, despite the several internal processes and the fermion loop, the result factorises and is expressed solely in terms of the basic scalar amplitude. Its structure is suggesting the general formula
\beq\label{Huvw}
H(\textbf{u} \rightarrow \textbf{v}|\textbf{w}) =\frac{H_{<}(\textbf{u},\textbf{u})H_{<}(\textbf{v},\textbf{v})H_{<}(\textbf{w},\textbf{w})H(\textbf{v}, \textbf{w})}{H(\textbf{u}, \textbf{w})H(\textbf{v}, \textbf{u})}\, ,
\eeq
for a generic distribution of elementary magnons, fulfilling charge conservation, $|\textbf{u}| = |\textbf{v}|+|\textbf{w}|$. We failed to find a proof of this ansatz, but we tested it extensively with Mathematica. As further evidence for its correctness, we notice that  it solves all the bootstrap axioms. It indeed transforms properly under permutation of the magnons in the states, as a result of the Watson relation,
\beq
H(u, v)/H(v, u) = S(u, v)\, ,
\eeq 
with $S(u, v) = S_{11}(u, v)$ the scalar S matrix, and it displays decoupling poles whenever rapidities in bottom and top sets become identical, again, thanks to the corresponding property of $H(u, v)$, see Eq.~(\ref{mu}) below. More precisely, one verifies that the decoupling condition (\ref{decleft}) is obeyed, with $\mathcal{I}\rightarrow 1$. Turning the logic around, the ansatz (\ref{Huvw}) appears as the simplest way of bringing together the left and right form factors, Eq.~(\ref{left}) and its right partner, while preserving the Watson relation and decoupling property. To enforce the latter requirement we simply added $H(\textbf{v}, \textbf{w})$ in the numerator.

\begin{figure}
\begin{center}
\includegraphics[scale=0.45]{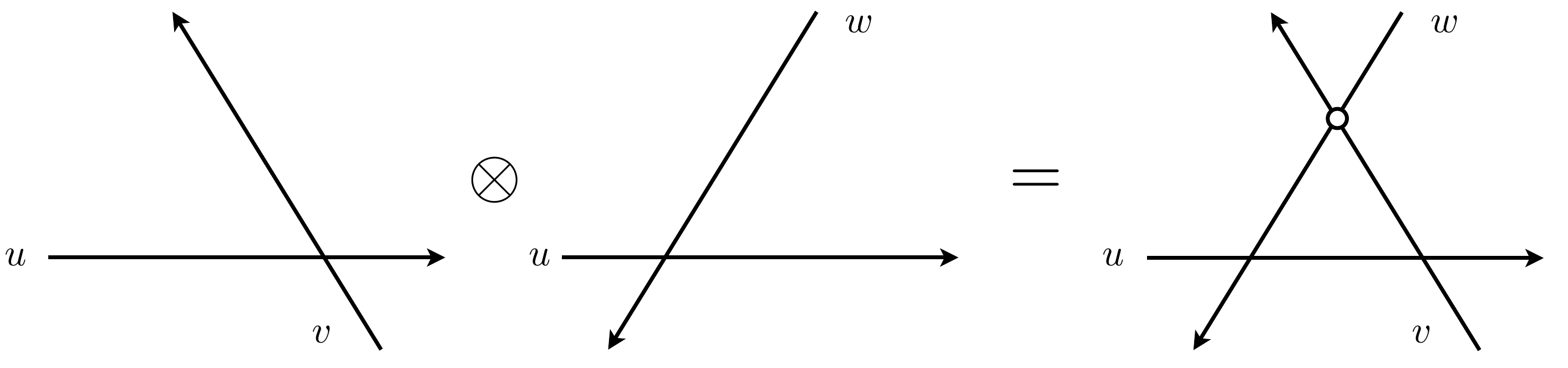}
\end{center}
\caption{Binding the left and right interactions in a decoupling friendly way fixes the third vertex, here shown as a blob, to be a shifted R matrix.}\label{binding} 
\end{figure}

\begin{figure}
\begin{center}
\includegraphics[scale=0.45]{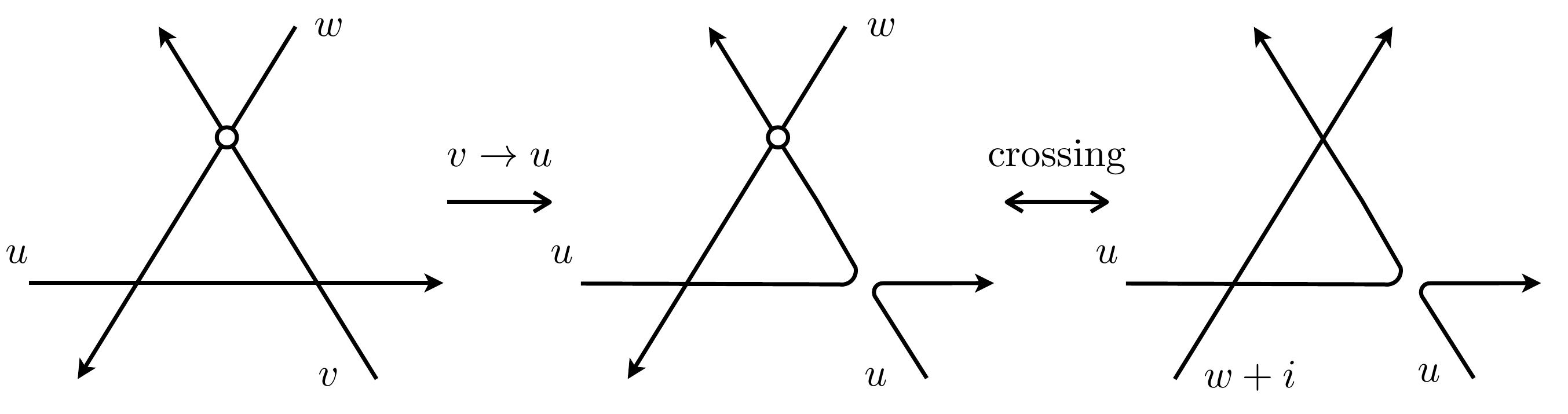}
\end{center}
\caption{Decoupling condition for the three-body matrix part. In the limit $v\rightarrow u$ the $uv$ interaction reduces to a permutation. After flipping the arrow on the $w$ line, using the crossing property of the R matrix, the interactions between $u$ and $w$ are seen to collapse thanks to the unitarity of the R matrix.}\label{mp2} 
\end{figure}

At last, we should include the bound states and their matrix degrees of freedom. Here also it proves easier to bootstrap the answer than to derive it from the SYM partition functions. Drawing inspiration from the structure of the result in the latter theory and assuming a factorised ansatz, one can uniquely determine the missing ingredient, that is, the vertex between the magnons $\textbf{v}$ and $\textbf{w}$, by imposing the decoupling axiom. More precisely, bringing together two R matrices, for the $uv$ and $wu$ scattering, as shown in figure \ref{binding}, we can then fix the $vw$ interaction point, denoted $R^{\circ}(v, w)$, by demanding that the latter vertex annihilates the left/right interaction in the right/left decoupling limit. This constraint is linear in $R^{\circ}(v, w)$ and it implies that $R^{\circ}$ is equal to the R matrix, up to a shift of its argument and a change of normalisation, 
\beq
R^{\circ}_{bc}(v, w) = \frac{c_{bc}(v-w)}{c_{bc}(v-w-i)}\, R_{bc}(v-w-2i) \, .
\eeq
To prove this relation, one simply needs to use the crossing property of the R matrix, see Eq.~(\ref{crossinge}), as shown in figure \ref{mp2}. Contrary to the SYM hexagon, here we find that the top vertex is inequivalent to the left and right ones; it goes along with the fact that the fishnet hexagon is not cyclic symmetric.

Crossing the lines permits to write the final result in the scattering form. E.g., after crossing the $\textbf{w}$'s, discarding the conjugation of their indices, we can write the core of the interaction as 
\beq\label{Mfishnet}
\mathcal{M}(\textbf{u}, \textbf{v}, \textbf{w})|_{\textrm{amputated}}= \frac{c(\textbf{v}, \textbf{w})}{c(\textbf{u}, \textbf{w})}\times R(\textbf{w}^{++}, \textbf{v})R(\textbf{u}, \textbf{v})R(\textbf{u}, \textbf{w}^{++})\, ,
\eeq
with $\textbf{w}^{++} = \textbf{w}+i$, with implicit bound state labels, and where
\beq
c(\textbf{u}, \textbf{v}) = \prod_{i, j}c_{a_{i}b_{j}}(u_{i}-v_{j})\, ,
\eeq
with $c_{ab}$ the crossing factor (\ref{crossingf}). For the sake of clarity, we removed the self-interactions on the external legs -- they can be inferred from (\ref{Habuv}) -- and the abelian prefactor is given by (\ref{Huvw}) with the $H$'s dressed with bound state indices. In the representation (\ref{Mfishnet}), the magnons $\textbf{v}$ and $\textbf{w}$ do not appear on an equal footing, but the left decoupling property of the matrix part is manifest, see figure \ref{mp2}.%
\footnote{A similar formula would be obtained by crossing the $\textbf{v}$'s, making the right decoupling obvious.} Finally, let us stress that we verified the bound state ansatz~(\ref{Mfishnet}) using Mathematica, for a few magnons and many different choices of bound state indices, starting from the SYM representation and using the mirror bound state $\mathcal{S}$ matrix obtained in \cite{Fleury:2017eph}.

This is it for the hexagon form factors to be used in this paper. To complete the picture, we quote the expression for the abelian factor $H_{ab}(u,v) = -A_{ab}(u, v)h_{ab}(u, v)$, which follows from the weak coupling limit of the fused SYM formula in the mirror kinematics,
\beq\label{Hab}
H_{ab}(u, v) = g^2 (-1)^{a-1} \frac{(\frac{a+b}{2}+iu-iv)\Gamma(1+\tfrac{a}{2}-iu)\Gamma(1+\frac{b-a}{2}+iu-iv)\Gamma(1+\tfrac{b}{2}+iv)}{(u^2+\tfrac{a^2}{4})^{\frac{3}{2}}\Gamma(\tfrac{a}{2}+iu)\Gamma(\frac{b-a}{2}-iu+iv)\Gamma(\tfrac{b}{2}-iv)(v^2+\tfrac{b^2}{4})^{\frac{3}{2}}}\, .
\eeq
Its zero at $v=u$ for $b=a$ equips the direct transition (\ref{fullHab}) with the decoupling pole
\beq\label{poleH}
\frac{1}{H_{ba}(v, u)} \sim \frac{(-1)^{a-1} \delta_{ab}}{i\mu_{a}(u)(u-v)}\, .
\eeq
The associated measure reads
\beq\label{mu}
\mu_{a}(u) = \frac{a g^2}{(u^2+a^2/4)^2}\, ,
\eeq
and it is identical to the SYM measure in the mirror kinematics at weak coupling. One also verifies the Watson relation, $H_{ab}(u, v)/H_{ba}(v, u) = S_{ab}(u, v)$, with the abelian S matrix (\ref{Sab}), as it should be.

\subsection{Charged hexagon}\label{charged}

There is an extra ingredient that we need for our investigation. It is associated to the insertion of magnons on the third operator. It appears natural indeed to enlarge the family of third operators by considering
\beq\label{Vnm}
\mathcal{O}_{3} \rightarrow V_{n, m, n_*} = \textrm{tr}\, \phi_{1}^{\dagger n_*}\phi_{2}^{\dagger m} \phi_{1}^{n}\phi_{2}^{m}\, .
\eeq
which includes, in particular, the dilaton,
\beq\label{Lag}
V_{1,1, 1} = \frac{1}{g^2} \mathcal{L}_{\textrm{int}} = \textrm{tr}\, \phi_{1}^{\dagger}\phi_{2}^{\dagger}\phi_{1}\phi_{2}\, .
\eeq
Owing to the specific ordering of the fields in the trace, the dynamics is frozen and the magnons cannot move in the background of the other fields. In sum, all these operators are protected.

\begin{figure}
\begin{center}
\includegraphics[scale=0.55]{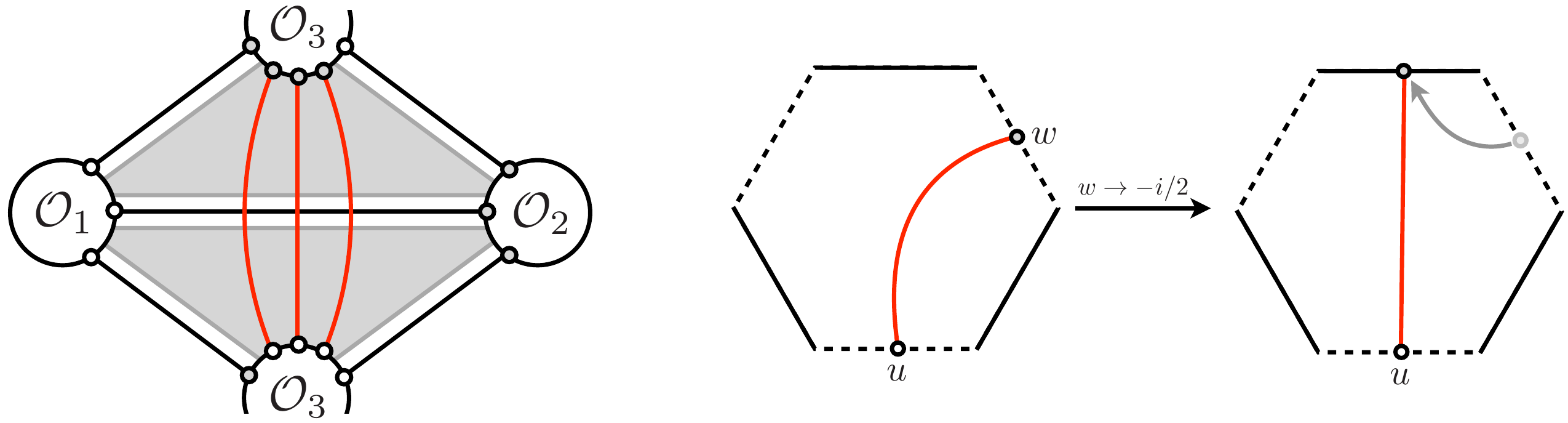}
\end{center}
\caption{Example of a fishnet structure constant with magnons ending on the third operator. We can bring a mirror magnon to this position by continuing its mirror momentum to $p(w) = 2w = -i$, as shown in the right panel.}\label{chargedhex} 
\end{figure}

From the integrability viewpoint, operator (\ref{Vnm}) acts as a sink or source for the mirror magnons. When placed inside a three-point function together with a pair of BMN operators, it leads to the diagram shown in the left panel of figure \ref{chargedhex}, to leading order at weak coupling. Importantly, the two sets of magnons, $\phi_2^{m}$ and $\phi_2^{\dagger m}$, split on two hexagons. Hence, to add the operator (\ref{Vnm}) to our story, we only need to charge the hexagon with a homogeneous reservoir of magnons on the edge associated to the third operator. The problem is reminiscent of the charging of the null pentagon Wilson loop \cite{Basso:2015rta}, used to embed the non-MHV amplitudes within the pentagon OPE framework in $\mathcal{N}= 4$ SYM. As we shall see, the outcome is essentially the same. 

For a unit of charge, we would like to place a single magnon on the edge associated to the third operator and set its spin-chain momentum $\bar{p}$ to zero. In this way, we are guaranteed that the magnon will not generate anomalous dimension. In $\mathcal{N}=4$ SYM, we could bring the magnon on the spin-chain edge starting from a neighbouring mirror edge, by using the mirror rotation. In the fishnet theory, because of the double scaling limit, the gates to the spin-chain kinematics pinch off at $\pm i/2$ on the mirror rapidity plane. Hence, the closest we have to a mirror move is to freeze a mirror magnon at either of these special points, as shown in the right panel of figure \ref{chargedhex}. The choice of the sign relates to which edge we charge.

The effect of this freezing operation on a spectator mirror magnon $u$, see figure \ref{chargedhex}, can be determined using equations (\ref{Huvw}) and (\ref{Hab}). We find
\beq\label{xi}
\lim_{w \rightarrow -i/2} \frac{\sqrt{\mu(w)}}{\sqrt{|\partial_w \bar{p}(w)|}}  \frac{1}{H_{a 1}(u,w)} = \frac{\sqrt{u^2+a^4/4}}{g} \equiv \xi_a(u) \, ,
\eeq
after switching to the spin-chain normalisation. The latter includes the measure $\mu$ and the Jacobian for the map between rapidity and spin-chain momentum, with $\bar{p} = iE$ and $E$ the mirror energy of the magnon. Note that one would obtain the same result starting from $\mathcal{N}=4$ SYM, placing a magnon on the relevant edge, and projecting to the fishnet theory.

More generally, each magnon present on the hexagon gets dressed by a factor that depends on its rapidity and representation. Labelling the magnons on the mirror edges as in figure \ref{fig:mp1}, with the third operator at the top, we obtain
\beq\label{xixi}
\xi_{\textbf{u}}/\xi_{\textbf{v}}\xi_{\textbf{w}}\, ,
\eeq
where $\xi(\textbf{u}) = \prod_{i} \xi_{a_{i}}(u_{i})$, etc. The generalization to the case where we insert $m$ magnons at the cusp follows from sending $m$ magnons to zero momentum, one after the other, and the dressing factor is obtained by raising (\ref{xixi}) to the power $m$.

\section{Tests and predictions}\label{Sect3}

In this section we carry out a battery of tests of our main formulae by comparing their predictions for structure constants and correlators with field theoretical calculations. We will also obtain a few predictions for a simple class of wheeled 3pt Feynman integrals. 

\subsection{The free propagator}\label{prop}

\begin{figure}
\begin{center}
\includegraphics[scale=0.45]{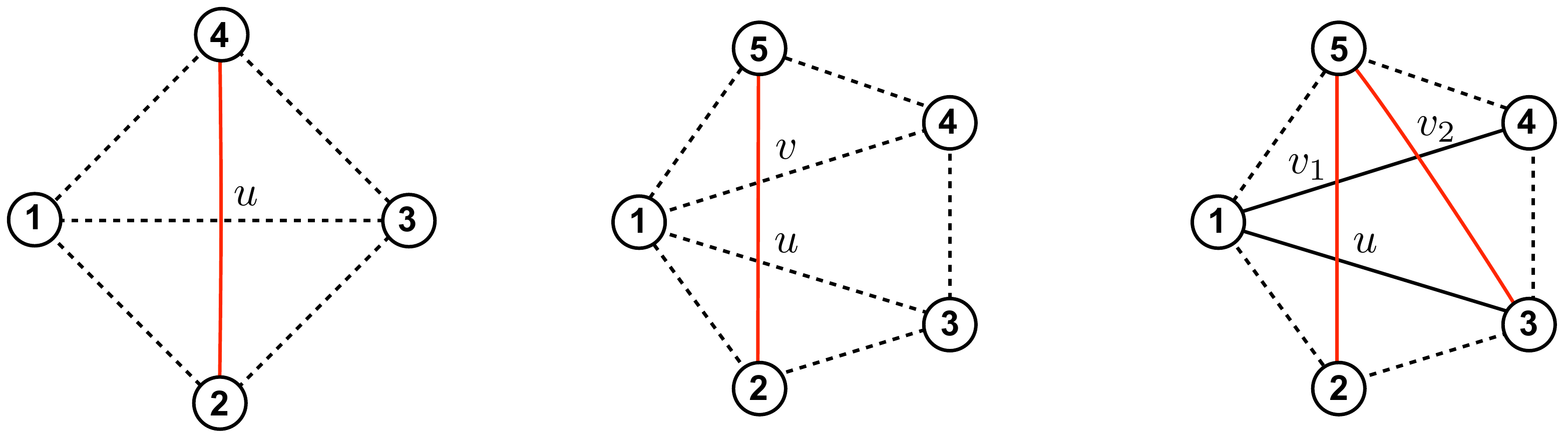}
\end{center}
\caption{Left and middle panels: Free propagator cut once and twice. We cut the interior of the polygon into two and three hexagons, respectively. 
The dashed middle lines denote bridges of length zero. Outer bridges / boundaries play no role here. For definiteness, one could give them arbitrarily large length to emphasize that nothing can leak out of the polygons. On the right panel, we give an example of a loop integral that could be hexagonalised by adding bridge lengths - for the horizontal propagators - and further magnons for the vertical ones.}\label{props} 
\end{figure}

We begin with the simplest fishnet correlator, the free propagator. Although elementary on the field theory side, its reconstruction using the hexagon factorisation is instrumental, as it gives a direct access to the hexagon building blocks. More precisely, by embedding the propagator inside a four- and five-point function and proceeding with its hexagonalisation~\cite{Fleury:2016ykk}, we shall be able to perform a direct test of the measure and 2-body form factor. The hexagon processes to be considered are displayed in figure \ref{props}, and, in all cases, the initial and final stages are the charged hexagons described in the previous section.

Let us start with the four-point function, which is an adaptation of the integrals considered in \cite{Fleury:2016ykk}, see also \cite{Basso:2017jwq}. It is obtained from the gluing of two hexagons, as shown in the leftmost panel in figure \ref{props}, and it involves a complete sum over the 1-magnon eigenstates along the middle cut 13. The spectral density to be integrated is
\beq
\xi_{a}(u)^2 \mu_{a}(u) \times {\rm{geometry}} \, ,
\eeq
where the first factor absorbs the amplitude for production and absorption of the mirror particle, on the bottom and top hexagon. The last factor is the geometrical weight for the dilatation and rotation of the magnon on the edge connecting the two hexagons. It reads~\cite{Fleury:2016ykk}
\beq
{\rm{geometry}} = \rho^{2iu} \times \chi_{a}(e^{i\phi}) \, , 
\eeq
where $\chi_{a}(e^{i\phi})$ is the $SU(2)$ character in the $a$-th irrep, i.e.,
\beq
\chi_{a} =  \textrm{tr}_{V_{a}}(e^{2i\phi J_{a}}) = \frac{\sin{(a\phi)}}{\sin{\phi}} \, , 
\eeq
with $J_a$ the spin operator on $V_{a}$. The dilation and rotation parameters, $\rho$ and $\phi$, are given by  
\beq
\rho = (z\bar{z})^{-\frac{1}{2}} \, , \qquad e^{-i\phi} = \sqrt{\frac{z}{\bar{z}}}\,, 
\eeq
where $z,\bar{z}$ are traditional 2d coordinates parameterizing the 4-point cross ratios,
\begin{equation}
z \bar{z} = \frac{x^2_{12} x^2_{34}}{x^2_{14} x^2_{23}} \, , \quad (1-z)(1- \bar{z}) = \frac{x^2_{13} x^2_{24}}{x^2_{14} x^2_{23}} \, .
\end{equation}
As described in \cite{Fleury:2016ykk,Eden:2016xvg}, we should also weight the scalar field insertions on the top and bottom cusps by including the factors
\beq
\label{weightscalars} 
\left(\frac{|x_{13}|}{|x_{43}||x_{41}|}\right)\times \left(\frac{|x_{13}|}{|x_{23}||x_{21}|}\right) =\frac{(1-z)(1- \bar{z})}{ \sqrt{z\bar{z}} \, x^2_{24}}\, .
\eeq
Alternatively, we can omit these extra weights and combine them with the propagator such as to define a conformally invariant propagator,
\beq \label{Propagator1}
\textrm{Propagator}_1 = \frac{\sqrt{x_{12}^2 x_{23}^2 x_{34}^2 x_{41}^2}}{x_{13}^2x^2_{24}} = \frac{\sqrt{z\bar{z}}}{(1-z)(1-\bar{z})}\, .
\eeq
Now, straightforwardly, after using the expression for the measure and $\xi$ factor, see Eqs.~(\ref{mu}) and~(\ref{xi}), and picking up the unique residue at $u = -ia/2$, we obtain
\beq
\begin{aligned}\label{Prop1}
&\sum_{a=1}^{\infty}\int \frac{du}{2\pi} \xi_{a}^2(u)\mu_{a}(u) \rho^{2iu} \textrm{tr}_{V_a}(e^{2i\phi J_{a}}) = \sum_{a=1}^{\infty}\int \frac{du}{2\pi} \frac{a}{u^2+a^2/4} \rho^{2iu} \textrm{tr}_{V_a}(e^{2i\phi J_{a}})\\
& = \sum_{a=1}^{\infty}\rho^{a} \chi_{a}(e^{i\phi}) = \textrm{Propagator}_1\, ,
\end{aligned}
\eeq
where the last equality is verified as a series expansion of (\ref{Propagator1}) around infinity.   
 
The ingredients for the five point function read the same but we have one more hexagon, the middle hexagon in the middle picture in figure \ref{props}. The magnon trajectory is now cut twice and we must sum over a complete basis of mirror states both along the zero length bridge $13$ and $14$. At each step the magnon wave function gets stretched and twisted by a dilation and a rotation, determined locally by the surrounding 4pt function. In order to perform the computation, we are going to consider the restriction to the 2d kinematics where all the points lie in the same plane, since the weight for moving away from the plane has not been determined yet.  Notice that distances in the plane can be written as $x^2_{ab} = x_{a,b} \, \bar{x}_{a,b}$ and we are going to use this notation below.  Only two pairs of cross ratios are needed and the weights are given by \cite{Fleury:2017eph}
\beq
\rho^{-2}_{i-1} = z_{(i-1)} \bar{z}_{(i-1)}= 
X_{1, i, i+1, i+2} \bar{X}_{1, i, i+1, i+2}\, , \qquad e^{-2 i\phi_{i-1}} = \sqrt{\frac{z_{(i-1)}}{\bar{z}_{(i-1)}}} = X_{1, i, i+1, i+2}/\bar{X}_{1, i, i+1, i+2}\,,
\eeq
where
\beq
X_{1, i, i+1, i+2} = \frac{x_{i,1}}{x_{i,i+1}}\frac{x_{i+1,i+2}}{x_{1,i+2}}\, , 
\eeq
and with $i=2,3$ for the bridge 13 and 14, respectively.  

Assembling all the ingredients together, we get the hexagon representation for the second propagator in figure \ref{props}. It reads
\begin{equation}
{\rm{Propagator}}_2 = 
\sum_{a, b = 1}^{ \infty }\int \frac{ d u}{ 2 \pi} \frac{ d v}{ 2 \pi}  \frac{\xi_{a}(u) \mu_a(u) \xi_{b}(v) \mu_b(v)}{H_{b a} (v+i0, u) }  
|\rho_{1}|^{2 i u} |\rho_{2}|^{2i v} 
\mathcal{F}_{ab} \, , 
\label{eq:ThePropagatorIntegral} 
\end{equation}
where $\mathcal{F}_{ab}$ originates from the R matrix in the middle transition, see Eq.~(\ref{fullHab}),
\beq
\mathcal{F}_{ab} = \textrm{tr}_{V_a\otimes V_b} \, (e^{2i\phi_{1} J_{a}}e^{2i\phi_{2} J_{b}} R_{ab}(u-v))\, ,
\eeq
with the trace taken over the tensor product of the $SU(2)$ modules, of total dimension $ab$. Using (\ref{Hab}) and (\ref{mu}), we obtain the dynamical part of the integrand
\beq\label{int5}
\frac{\xi_{a}(u)\mu_a(u)\xi_{b}(v)\mu_b(v)}{H_{b a} (v+i0, u) } = \frac{ab(-1)^{b-1} \Gamma(\tfrac{a}{2}-iu)\Gamma(\tfrac{a-b}{2}+iu-iv+0)\Gamma(\tfrac{b}{2}+iv)}{(\tfrac{a+b}{2}-iu+iv)\Gamma(1+\tfrac{a}{2}+i u)\Gamma(1+\tfrac{a-b}{2}-i u+iv)\Gamma(1+\tfrac{b}{2}-iv)}\, ,
\eeq
where the $i0$ prescription is needed to handle the decoupling pole at $u=v$ and $a=b$.%
\footnote{The contour is chosen in a such way that the 5pt integral reduces to the 4pt one in the limit $x_{3}\rightarrow x_{4}$.} We verify that the net integrand is of order $g^0$ as needed for a tree-level process. The scaling follows from, see Eqs.~(\ref{Hab}), (\ref{mu}) and (\ref{xi}), 
\begin{equation}
H_{ab}(u, v) = 
\mathcal{O}(g^2) \, , \qquad \mu_a(u) = \mathcal{O}(g^2) \, ,  \qquad \xi_{a} = \mathcal{O}(1/g)\, , 
\end{equation}
together with the fact that the matrix part is coupling independent. Note also that the $\xi$ factors for production and absorption of the magnon combine nicely with the square roots present in the middle transition $H_{ab}(u\rightarrow v)$, see Eqs.~(\ref{Hab}) and (\ref{xi}), such as to give a meromorphic function of the rapidities, as needed for any weak coupling expression.%
\footnote{It was observed in \cite{Fleury:2016ykk} by comparing hexagon  calculations with perturbation theory in the SYM theory that it is necessary to dress the mirror bound states with so called $Z$-markers to obtain an agreement. The general prescription for dressing the states, which passed all tests so far, was written down in the Appendix A of \cite{Fleury:2017eph}. In our case, since we deal with transverse scalar excitations, the $Z$-markers play no role and the dressing trivialises.}

We evaluate the integral (\ref{int5}) by closing the contours of integration in the lower half-planes and summing up the residues. (All the poles are simple; that would not be so if we had bigger bridge lengths.) We begin by picking up the residues in the lower half $u$ plane and then in the lower half $v$ plane. The former come from the single argument Gamma function in the numerator and are located at $u = -ia/2-ik$ with $k=0, 1,\ldots\,$. In principle, we should also worry about the simple poles coming from the matrix part, see Eq. (\ref{Rab-eigen}), at
\beq
u = v-i\frac{a+b-2j}{2}\, , \qquad j = 1, \ldots , \textrm{min}\{a-1, b-1\}\, ,
\eeq
to which we can add the pole at $u = v-i(a+b)/2$, which is visible in (\ref{int5}). However, the Gamma function of the difference of rapidities in the denominator removes them all, since
\beq
\frac{1}{\Gamma(1+\tfrac{a-b}{2}-iu+iv)} \rightarrow \frac{1}{\Gamma(1-b+j)}\, ,
\eeq
is zero at these points, whenever $j \leqslant b-1$. The next step is to pick up the residues in the lower half $v$ plane. Here, again, one verifies that they only come from the Gamma functions in the numerator, and, more specifically, from the Gamma function that depends on the difference of rapidities. Most of these poles are killed by the zeroes coming from the denominator, such that, in the end, the double integral can be taken at once by extracting the residues at
\beq
u = -ia/2 \qquad \textrm{and} \qquad  v = -ib/2\, .
\eeq
Moreover, $b\geqslant a$, as visible from the final expression for the double residue, which is given by a binomial coefficient. It yields
\beq
\rho_{1}\rho_{2}\sum_{b\geqslant 1} \rho_{2}^{b-1} \sum_{a=1}^{b} \frac{(-\rho_{1})^{a-1}\Gamma(b)}{\Gamma(a)\Gamma(1+b-a)} \, \textrm{tr}_{V_a\otimes V_b} ( e^{2i\phi_{1}J_{a}}e^{2i\phi_{2}J_{b}}R_{ab}(\tfrac{ib-ia}{2}))\, .
\eeq
The sum over $a$ can be viewed as generating the transfer matrices (at a specific point) for a twisted length-one spin chain with spin $(b-1)/2$ and it can be computed using the associated twisted Baxter equation. We refer the reader to Appendix \ref{AppT} for the detailed analysis and simply quote here the answer. Namely, after summing over $a$, we get that the 5pt integral reduces to the 4pt one, see Eq.~(\ref{Prop1}),
\beq \label{Propagator2}
\rho_{1}\rho_{2}\sum_{b\geqslant 1} (\rho'_{2})^{b-1} \chi_{b}(e^{2i\phi_{2}'}) = \frac{\rho_{1}\rho_{2}}{(1-1/z'_{2})(1-1/\bar{z}'_{2})} = \frac{\sqrt{z_1 \bar{z}_1} \sqrt{z_2 \bar{z}_2}}{(1-z_{1}+z_{2}z_{1})(1-\bar{z}_{1}+\bar{z}_{2}\bar{z}_{1})} \, , 
\eeq
up to a geometrical redefinition of the cross ratios,
\beq\label{zp}
(z'_{2})^{-1} = z_{2}^{-1}(1-z_{1}^{-1})\, , \qquad (\bar{z}'_{2})^{-1} = \bar{z}_{2}^{-1}(1-\bar{z}_{1}^{-1})\, .
\eeq
Expression (\ref{Propagator2}) is then immediately verified to match with the conformal propagator,
\begin{equation}
(\ref{Propagator2})  = \left( \frac{| x_{12} | | x_{23} |}{| x_{13} |} \right) \left( \frac{| x_{15} | | x_{45} |}{|x_{14}|} \right) \frac{1}{x_{25}^2} \, ,  
\end{equation}
after taking into account the aforementioned weights for the scalar insertions at the top and bottom.

One could keep going and insert the propagator in higher $n$ point function. The hexagon representation will then involve a sequence of transitions across the various mirror cuts. We expect the algebra to be similar to the one carried out here and to reduce to an iteration of the geometrical transformation (\ref{zp}). One could also consider products of free propagators stretching between different cusps of a polygon; the hexagon factorisation would give them in terms of convoluted integrals of products of multi-particle form factors. More ambitiously, one could add loops to the cocktail, of the type shown in figure \ref{props}, by dressing each magnon with the bridge factor $e^{-\ell E_{a}(u)}$, with $\ell$ measuring the number of horizontal propagators along the given cut. The resulting representations could be tested using the differential equations derived from the Yangian symmetry~\cite{Chicherin:2017cns,Chicherin:2017frs}, for specific bridge lengths.

\subsection{The bridge overlap}\label{spirals}

\begin{figure}[t]
\begin{center}
\includegraphics[height=5cm]{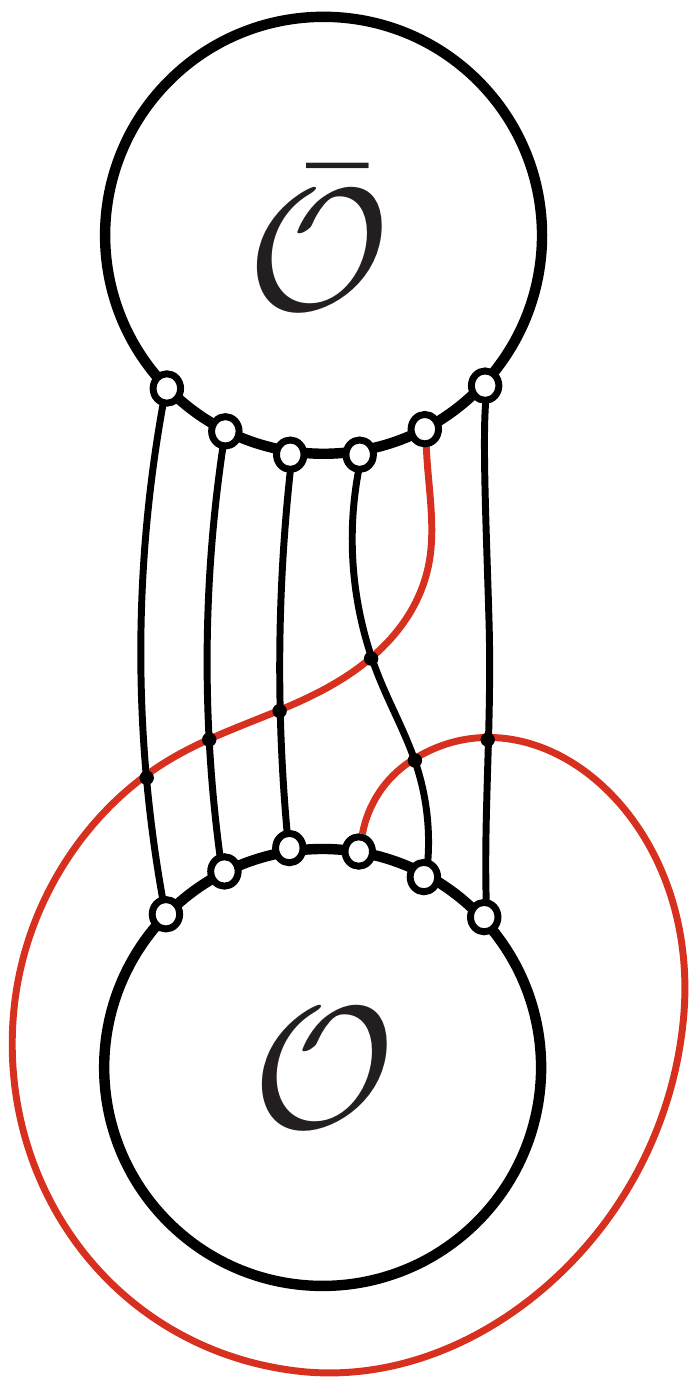}
\caption{The two point function of two spiraled states. The red lines correspond to the propagation of a $\phi_2$ field which is taken to be an excitation over the reference state made out of $\phi_1$, represented by the dark lines. The Feynman graph corrections wrapping the external operators have a configuration of a spiral.}
\label{fig:spiral}
\end{center}
\end{figure}  
As a simple and natural generalisation of our set-up, we shall consider spin-chain states with $\phi_2$ excitations propagating on top of the BMN vacuum,
\beq\label{spiraledstates}
\mathcal{O} \sim \textrm{tr}\, \phi_{1}^{L} \rightarrow \mathcal{O}_{\text{spiral}} \sim \sum_n \psi_n\, \text{tr} (\phi_1^{L-N} \phi_2^{N} )\,,
\eeq
where the RHS should be read as a linear superposition of $N$ insertions along the chain. These states are the fishnet counterparts of the states lying in the $SU(2)$ sector of $\mathcal{N}=4$ SYM \cite{Minahan:2002ve} (even though in the fishnet theory only a $U(1)$ subgroup remains). The Feynman graphs wrapping these operators look like spirals (see figure \ref{fig:spiral}) and for this reason we will refer to the operators in (\ref{spiraledstates}) as \textit{spiraled states}.

\begin{figure}[t]
\begin{center}
\includegraphics[height=4cm]{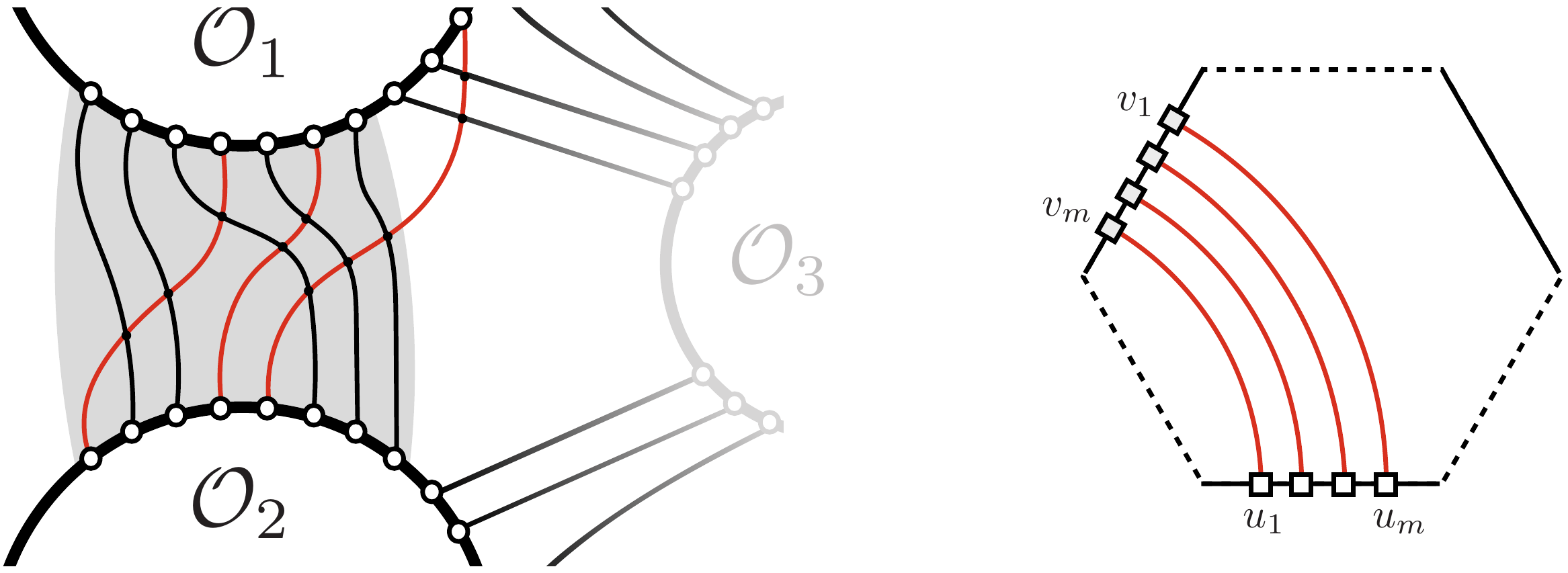}
\caption{In the left figure, the bridge overlap between the spiraled operators is represented in gray. Excitations cannot be contracted with the vacuum operator $\mathcal{O}_3$ so that the only nontrivial contractions occur in the gray region. On the right figure, we represent the excitation pattern of the hexagon used to compute the bridge overlap.} 
\label{bridgeoverlap}
\end{center}
\end{figure}  

Distributing magnons on the BMN vacua entering the structure constant (\ref{prototype}) leads to graphs of the type shown in figure \ref{bridgeoverlap}. Hence, contrary to the previous setup, where all quantum corrections came from virtual particles moving across the bridges, structure constants for spiraled states receive nontrivial corrections in the form of a perturbative tail in $g^2$, before the wrapping corrections $\sim g^{2L_{1,2}}$ kick in. We will limit ourselves to the asymptotic regime in the following, obtained by neglecting the wrapping corrections. This scenario is realised when the bridges connecting the BMN operators to the third operator are asymptotically thick, i.e.~$\ell_{13}, \ell_{23} \rightarrow \infty$. In these circumstances, the nontrivial part only comes from the bridge overlap between the two excited states, as illustrated in figure \ref{bridgeoverlap}. 

The spectrum of spiraled states was thoroughly studied in \cite{Caetano:2016ydc}. It is described, asymptotically, by the double scaling limit of the twisted Beisert-Staudacher equations \cite{Beisert:2005if}. The main outcome of this analysis is that the fishnet limit amounts to performing an infinite (imaginary) boost on the magnons, which pushes them all the way to the mirror kinematics. Therefore, in the end, the magnons sourcing the spirals are just mirror magnons, like the ones discussed throughout this paper. The sole difference is that the Bethe ansatz equations subject them to have imaginary energies and momenta. More precisely, all the Bethe roots originate at the same canonical point $p = -i$ at weak coupling, see Subsection \ref{charged}, and then spread out along the mirror plane as the coupling increases. (Turning the flow around, one could say that the Bethe roots are pushed to the spin chain edge, represented by a single point on the mirror sheet, when the coupling is sent to zero.) They admit the expansion
\beq \label{fluctuations}
u_{k}=- \frac{i}{2}+\sum_{j=1}^{\infty} \delta u_{k}^{(j)} g^{2j}\,,
\eeq
with an infinite tail of perturbative corrections $\delta u_{k}$. The latter are determined iteratively by solving the Bethe ansatz equations,
\beq\label{BAE}
e^{i \phi_j}= \left( \frac{g^2}{u_{j}^2+1/4} \right)^{L} \prod_{k\neq j}^{N} S^{s}(u_j, u_k)\,,
\eeq
where $S^{s}(u, v) = \xi(v)^{2}\xi(u)^{-2}S(u, v)$ is the scalar mirror S matrix (\ref{Sab}), with $a=b=1$, in the spin-chain normalisation and where we used that each spiral carries an imaginary spin-chain momentum $\bar{p}$ equals to its mirror energy $E$,
\beq\label{spin}
e^{i \bar{p}}  = e^{-E} = \frac{g^2}{u^2+1/4} \, .
\eeq
Similarly, although the magnons populate different edges of the hexagons, as shown in the right panel of figure \ref{bridgeoverlap}, the hexagon amplitude takes exactly the same form as before, if not for the conversion to the spin-chain normalisation.%
\footnote{This conversion is by no means necessary, but is conventional for spin-chain states.}
The translation between the string and spin-chain frames boils down to inserting $\xi$ factors, as described in \cite{Basso:2015zoa}, and the hexagon amplitude showed in figure \ref{bridgeoverlap} is given by (\ref{left}) up to the replacement
\beq \label{spiraledh}
H(u, v) \rightarrow H^s(u, v) = \frac{\xi(v)}{\xi{(u)}}H(u, v)\, .
\eeq
It obeys Watson relation for the spin-chain framed S matrix, $H^{s}(u, v)/H^{s}(v, u) = S^{s}(u, v)$.

Asymptotically, the hexagon prescription to compute the structure constant consists in attaching two hexagons together along the bridge $12$ and summing over all the ways of distributing magnons on both sides of the cut \cite{Basso:2015zoa}. It yields
\beq
\begin{aligned} \label{excitedhaxegon}
\frac{C^{\bullet\circ \bullet}_{132}}{\sqrt{L_{1}L_{2}}} &= \mathcal{N}({\bf u})\, \mathcal{N}({\bf v})\sum_{\alpha \cup \overline\alpha ={\bf u}, \beta \cup \overline{\beta} ={\bf v}} e^{i (\bar{p}(\overline{\alpha})- \, \bar{p}(\overline{\beta}) )\ell_{12}}  
S^{s}_{<}(\bar\alpha,\alpha) S^{s}_{<}(\beta,\bar\beta)\, H^{s^{\prime}}(\alpha, \beta) H^{s^{\prime}}(\bar\beta, \bar\alpha)\,,
\end{aligned}
\eeq
where $|\alpha| = |\beta|$ for charge conservation. Here, $H^{s^{\prime}}(\alpha ,\beta) = H^{s}_{<}(\alpha, \alpha)H^{s}_{<}(\beta, \beta)/H^{s}(\beta, \alpha)$, $\bar{p}$ is the spin chain momentum defined in (\ref{spin}), and the splitting factor is given by
\beq\label{splitting}
S^{s}_{<}(\bar{\alpha}, \alpha) = \prod_{i\in  \bar{\alpha}, j\in \alpha, i<j} S^{s}(u_{i}, u_{j})\, ,
\eeq
and similarly for $\textbf{v}$. The normalisation factor $\mathcal{N}(\textbf{u})$ is given by the Gaudin norm of the spin-chain state, up to the hexagon measures, see Eq.~(\ref{mu}),
\beq
\mathcal{N}({\bf u})^2 =  \frac{\prod_{i}\mu(u_{i})}{\det {\partial_{u_{i}} \phi_{j}}}\,,
\eeq 
with $\phi_{j}$ the quasi-momentum of the $j$-th magnon, Eq.~(\ref{BAE}) with $L$ replaced by the length of the operator supporting the magnon.

Plugging the Bethe roots for the two Bethe states, $\textbf{u}$ and $\textbf{v}$, inside (\ref{excitedhaxegon}) should produce all the perturbative corrections to the structure constants below wapping order. The hexagons themselves depend trivially on the coupling constant, which enters as an overall factor. Hence, the nontrivial dependence on the coupling $g^2$ come entirely from the Bethe roots (\ref{fluctuations}), as in the case of the anomalous dimension. Let us also add that the bridge length appearing above is measured in the spin chain frame, and thus counts the total numbers of lines in the bridges, for the two types of fields, $\phi_{1}$ and $\phi_{2}$. (In comparison, in the string frame, only the vacuum lines would be counted.)

To perform a field theoretic check of the hexagon formula we need the precise definition of the conformal operators, that is, we must determine their wave-functions $\psi_{n}$ in (\ref{spiraledstates}). The relevant spin chain Hamiltonian was computed through four loops in \cite{Caetano:2016ydc}. We will only need to known the first two terms here, to carry out a test at one loop. They read
\beq
H= -2g^2 \sum_{j} \sigma_{j}^{+}\sigma_{j+1}^{-}-2g^4 \sum_{j} \sigma_{j}^{+}\sigma_{j+1}^{-}\sigma_{j+1}^{+}\sigma_{j+2}^{-} + O(g^6)\, , \label{fishnetH}
\eeq
with $\sigma^{\pm}_{j}$ the operator creating or annihilating a magnon at the $j$-th site and where we assume periodic boundary conditions. This system can be solved by means of the Bethe ansatz, perturbatively in $g^2$, with the S matrix and anomalous dimension,%
\footnote{These expressions look different from the ones given earlier, but are nonetheless identical. The difference comes from the fact that we are expanding around $g^2 = 0$ at finite spin chain momentum $\bar{p}$.}
\beq \label{BAdata}
S^{s}(\bar{p}_1,\bar{p}_2)=-1+2g^2 \left(e^{-i \bar{p}_1}- e^{-i \bar{p}_2}\right)\,,\,\,\,\, \gamma(\bar{p})=-2g^2 e^{-i\bar{p}}-2g^4 e^{-2 i \bar{p}}\,,
\eeq
Contrary to what happens in $\mathcal{N}=4$ SYM \cite{Staudacher:2004tk}, we find no need of introducing contact terms in the higher-loop wave functions. In other words, the eigenstates of (\ref{fishnetH}) are just plain Bethe wave functions built out of the S matrix (\ref{BAdata})

\begin{figure}[t]
\begin{center}
\includegraphics[height=3.4cm]{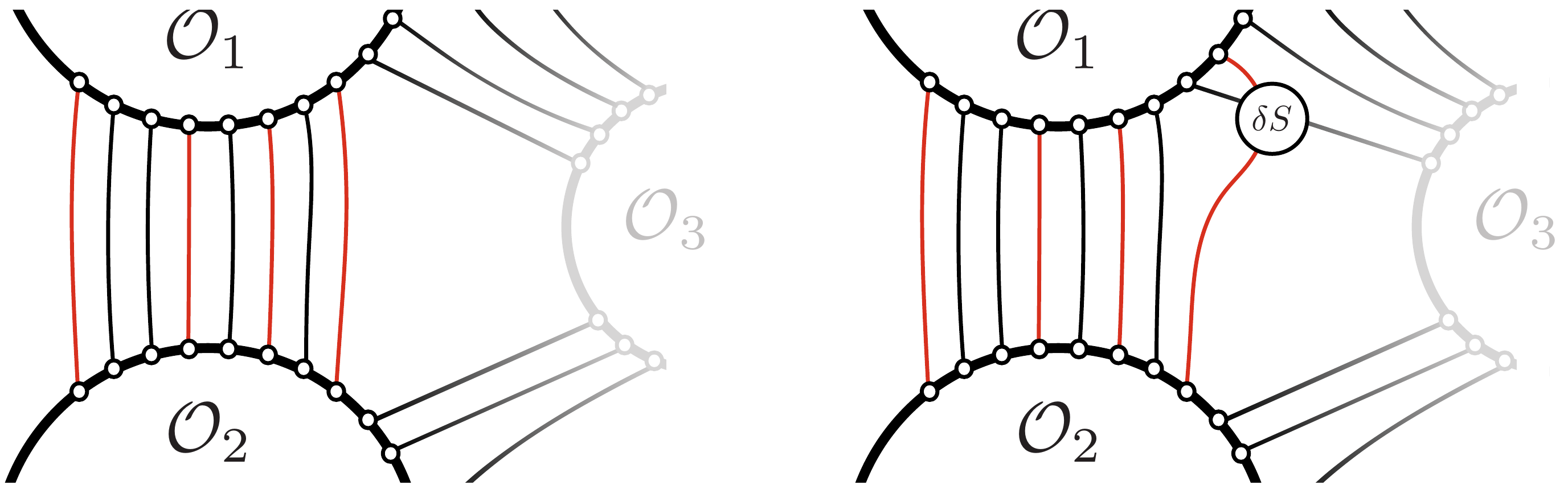}
\caption{The perturbative structure constant can be obtained by tree level Wick contractions of the three operators as represented in the left figure. In the right figure, the one-loop result can be concisely accommodated by inserting a lagrangian density at the splitting points where all the three operator are involved. This insertion simply accounts for the result of the Feynman diagram computation.}
\label{fig:insertion}
\end{center}
\end{figure}  

In the field theory, the tree level structure constant is readily obtained by overlapping the wave-functions of the two spiraled states. At one loop, one should in addition dress the tree-level Wick contraction with Feynman diagrams which stem from the insertions of the single $\phi^4$ vertex of the fishnet theory. These corrections move the magnons away to the neighbouring sites. Normalising by the two-point functions, the effect of the one-loop diagrams can be cast as the insertions of a local operator $\delta S_{ij}$ at the locations of the splitting points in the tree-level diagrams (see figure \ref{fig:insertion}), with
\beq \label{splitop}
\delta S_{ij}= g^2 \sigma^{+}_{i}\sigma^{-}_{j}\,.
\eeq
We refer the reader to \cite{Caetano:2014gwa} for a detailed one-loop  computation in a similar set up. 

In order to confront the field theory computations with the hexagon predictions, we expand (\ref{excitedhaxegon}) to one loop taking into account the perturbative corrections to the rapidities given in (\ref{fluctuations}). Imposing the Bethe equations (\ref{BAE}) is not instrumental for these checks so that we can keep the fluctuations $\delta u_k^{(i)}$ arbitrary. In practice, the comparison boils down to match the expression (\ref{excitedhaxegon}) with the  overlap  along the bridge of two Bethe wave-functions,
for which we use the coordinate frame,  with the additional contribution of the one-loop splitting insertions (\ref{splitop}). To ensure the same normalization on both sides, we use the fact that the Gaudin norm in the coordinate normalization contains the Jacobian for the exchange of momentum and rapidity space, namely
\beq
\langle {\bf{u}} | {\bf{u}} \rangle_{\text{coord}}= \frac{ \det {\partial_{u_{i}} \phi_{j}}}{\prod_i \partial_{u_i} \bar{p}_i}\,,
\eeq 
where $\langle {\bf{u}} | {\bf{u}} \rangle_{\text{coord}}$ denotes the norm of a coordinate Bethe state.
Up to a factor of total momentum which trivializes for physical states, we obtain a perfect match.

\subsection{Half structure constants}\label{Half}

\begin{figure}
\begin{center}
\includegraphics[scale=0.45]{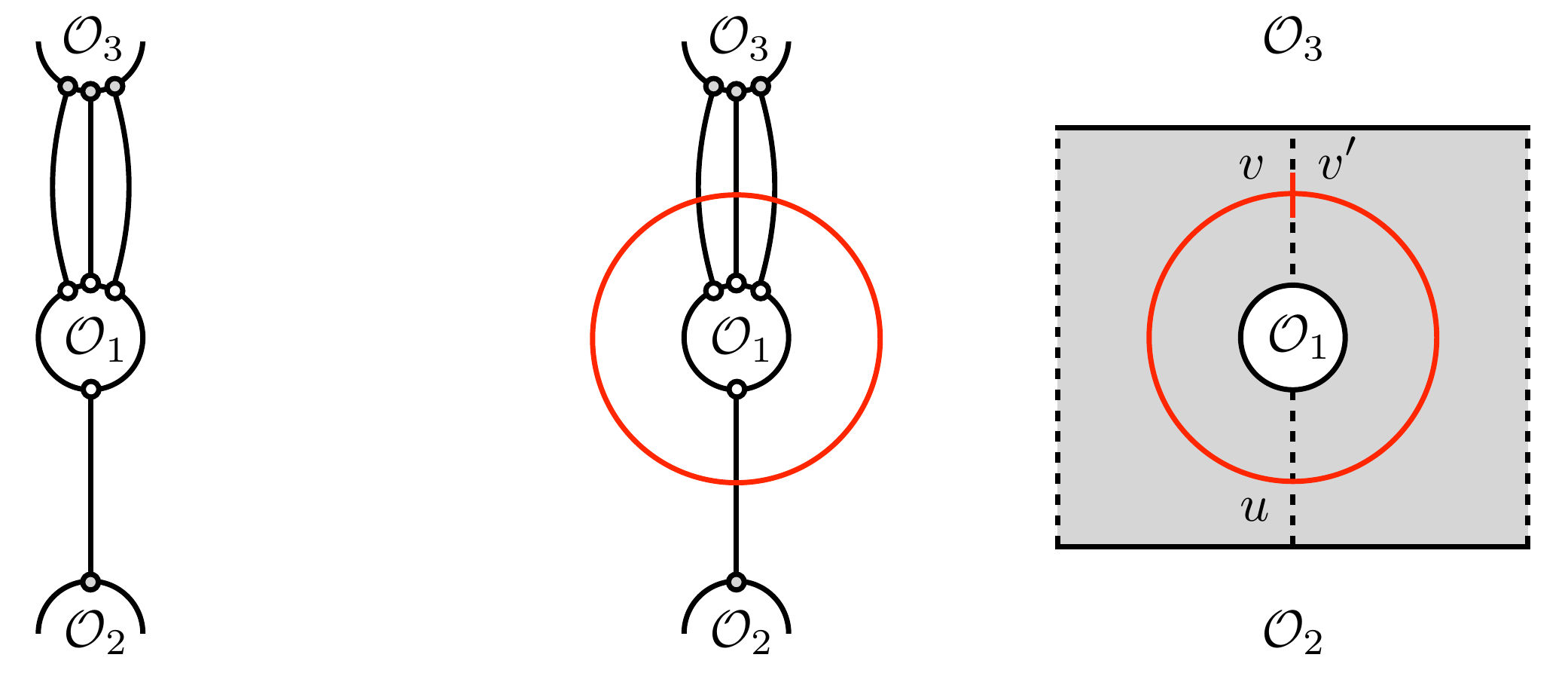}
\end{center}
\caption{Tree and one-wheel graph contributing to the half structure constant. The one wheel graph has logarithmic divergence when the wheel shrinks on the central operator. The counterpart of this singularity in the hexagon framework is a double pole in the rapidity difference $u-v$. One can regularise the divergences by opening up the wheel along a mirror cut and remove the polar part which accompanies the coinciding limit $\epsilon = v'-v\rightarrow 0$.}\label{sw} 
\end{figure}

In this subsection, we consider the structure constant that splits the \textit{single-trace} BMN operator $\mathcal{O}_{1} = \textrm{tr}\, \phi_{1}^{L}$ in two conjugate \textit{untraced} BMN operators, $\mathcal{O}_{2} = \phi_{1}^{\dagger\ell_{2}}$ and $\mathcal{O}_{3} = \phi_{1}^{\dagger\ell_{3}}$, with $\ell_{2}+\ell_{3} = L$, for charge conservation. Two hexagons are needed to cover this closed-$(\textrm{open})^2$ correlator in the planar limit, but only two edges are stitched together, as shown in figure~\ref{sw}. This correlator can be understood as a limit of the three-point function introduced in Section \ref{Sect2}, describing the situation where the bridge 23 is arbitrarily thick and thus impenetrable to the magnons. Feynman diagrammatically, this is equivalent to removing the latter bridge and only including the graphs that stay within the perimeter of interest.

Obviously, the perturbative expansion of the structure constant takes the form of a sum over the number of wheels surrounding the closed string operator. The hexagon form factor expansion follows the same pattern,
\beq\label{TheTwistedC123}
C^{\bullet\circ\circ}(\ell_{2}, \ell_{3}) = \sqrt{L}\times (1 + \mathcal{A}_{\textrm{1-wheel}} + ...)\, ,
\eeq
where the first term is the tree result, etc. In the hexagon picture, the 1-wheel amplitude is given by a double integral over the rapidities $u$ and $v$ that the mirror magnon takes on the mirror cuts; its integrand can be read out from~Eq.~(\ref{intB}). However, this amplitude, which is of the wrapping type, is not immediately meaningful. Its integrand has a double at $u=v$, when $a=b$, as a result of the kinematical singularities of the hexagon form factors, and the naive integration is divergent. This divergence has a simple interpretation and resolution on the field theory side: it maps to the short-distance singularity of the one-wheel diagram and is removed by renormalising the BMN operator at its center.    

Since the one-wheel graph has no subdivergent graphs, any procedure that opens up the wheel should remove the problem. In particular, the divergence goes away if we open up a mirror cut, point split the rapidity of the magnon sitting there, and integrate properly the magnon in the other bridge, see figure~\ref{sw}. So defined, the sub-amplitude is regular but has a pole $\sim 1/\epsilon$ when $\epsilon = v'-v\sim 0$. The full amplitude is renormalised by subtracting the polar part and integrating the finite part over $v$. We refer the reader to Section \ref{Sect4.2} for a detailed implementation of this procedure in a more general set-up. Here, we simply need to note that this renormalisation procedure was performed under similar conditions in $\mathcal{N}=4$ SYM \cite{Basso:2017muf} and the formula derived in this context immediately applies to our amplitude, after specialising it to the fishnet theory.

This formula yields the renormalised amplitude as the sum of two contributions,%
\footnote{Note that $C_{1}$ is defined differently than in \cite{Basso:2017muf}, as we stripped out the factor $1/2$ for aesthetic reasons.}
\beq\label{renorA}
\mathcal{A}_{\textrm{1-wheel}}  = B_{1} +\frac{1}{2}C_{1}\, .
\eeq
The bulk of the answer has the exact same integrand as the bare amplitude,
\beq\label{intB}
\begin{aligned}
B_{1} =& \sum_{a, b =1}^{\infty} \dashint \frac{du dv}{(2\pi)^2} \frac{ab\,\mu_{a}(u)\mu_{b}(v)}{H_{ab}(u, v)H_{ba}(v, u)} e^{-E_{a}(u)\ell_{2}-E_{b}(v)\ell_{3}} \\
=& \sum_{a, b =1}^{\infty} \dashint \frac{du dv}{(2\pi)^2} \frac{g^{2\ell_{2}}}{(u^2+\frac{a^2}{4})^{\ell_{2}}} \frac{g^{2\ell_{3}}}{(v^2+\frac{b^2}{4})^{\ell_{3}}} \frac{a^2 b^2}{((u-v)^2+\tfrac{1}{4}(a+b)^2)((u-v)^2+\tfrac{1}{4}(a-b)^2)}\, ,
\end{aligned}
\eeq
but is equipped with a principal value for integrating the singularity at $v=u$, when $b=a$.%
\footnote{One could also avoid the double pole using a $\pm i0$ prescription; the two options are equivalent here.} The second term $C_{1}$ is a contact term, which results from the subtraction of the short-distance singularity. It only depends on the total length, $L = \ell_{1}+\ell_{2}$, and is given as a single integral,
\beq\label{intC}
C_{1} = \sum_{a=1}^{\infty} \int \frac{du}{2\pi} \frac{a^2 g^{2L}}{(u^2+\frac{a^2}{4})^{L}} K_{aa}(u, u)\, .
\eeq
It is controlled by the scattering kernel
\beq\label{Kab}
K_{ab}(u, v) = \frac{1}{a^2 b^2} \textrm{tr}\, \mathbb{S}_{ba}(v, u)\frac{\partial}{i\partial u} \mathbb{S}_{ab}(v, u)\, ,
\eeq
with the trace running over the $a^2\times b^2$ states in the module $(V_{a}\otimes \dot{V}_{a})\otimes (V_{b}\otimes \dot{V}_{b})$. The kernel is easily evaluated using the factorisation of the S matrix (\ref{bbS}) and the explicit expressions for its diagonal and matrix parts, see Eqs.~(\ref{Sab}) and (\ref{Rab-eigen}). It yields, for coinciding arguments,
\beq
K_{aa}(u, u) = 2+k_{a}(u)\, ,
\eeq
where
\beq\label{ka}
k_{a}(u) = K_{aa}(u, u)\big|_{\textrm{diag}} =  -H(\tfrac{a}{2}+iu) - H(\tfrac{a}{2}-iu) -  H(\tfrac{a}{2}-1+iu) - H(\tfrac{a}{2}-1-iu)\, ,
\eeq
and with $H(z)$ the analytically continued harmonic sum.

The integrals in (\ref{renorA}) can be evaluated by the method of residues and the accompanying sums can be expressed in terms of multiple zeta values. A general algorithm for carrying out these steps is given in Appendix \ref{AppZeta}, and the expressions so-obtained are presented in table \ref{intprediction}, for several values of the bridge lengths. Interestingly, they only involve odd zeta values and products thereof.

\begin{table}
\makegapedcells
\begin{center}
\begin{tabular}{c|c|c}
\toprule
$\ell_{2}$ & $\ell_{3}$ & $\mathcal{A}_{\textrm{1-wheel}} $  \\
\midrule
1 & 2 & $6 \zeta_3$ \\ 
\hline
1 & 3 & $20 \zeta_5 $  \\ 
\hline
1 & 4 & $ 70 \zeta_7$ \\ 
\hline
1 & 5 & $252 \zeta_9$  \\ 
\hline
2 & 2 & $- 6 \zeta_3^2 + 20 \zeta_5$  \\ 
\hline
2 & 3 & $- 30 \zeta_3 \zeta_5 + 70 \zeta_7$  \\ 
\hline
2 & 4 & $-10 \zeta_5^2 - 112 \zeta_3 \zeta_7 + 252 \zeta_9$  \\ 
\hline
3 & 3 & $- 290 \zeta_5^2 + 112 \zeta_3 \zeta_7 + 252 \zeta_9 $  \\ 
\hline
3 & 4 & $- 1176 \zeta_5 \zeta_7 + 420 \zeta_3 \zeta_9 + 924 \zeta_{11}$ \\ 
\hline
4 & 4 & $-3178 \zeta_7^2 - 1680 \zeta_5 \zeta_9 + 1584 \zeta_3 \zeta_{11} + 3432 \zeta_{13}$ \\ 
\bottomrule
\end{tabular}
\end{center} 
\caption{$\mathcal{A}_{\textrm{1-wheel}}$ 
for various bridge lengths, $\ell_2$ and $\ell_3$, up to the overall factor of the coupling, $g^{2(\ell_2 + \ell_3)}$. The method used for generating these expressions is described in Appendix~\ref{AppZeta}. 
}  \label{intprediction}
\end{table}

Another interesting pattern of table \ref{intprediction} concerns the transcendentality, which appears almost uniform, at a given loop order $L = \ell_{2} +\ell_{3}$. In fact, the $L$-loop expressions are seen to have uniform weight $2L-2$, after subtracting the term linear in $\zeta$. The latter is proportional to $\zeta(2L-3)$ and is identical to the one-wheel anomalous dimension \cite{Broadhurst:1985vq,Gurdogan:2015csr}, up to a factor $-2$. Moreover, this linear piece is the only contribution that remains when one bridge length is set to $1$, regardless of the length of the other bridge. This feature can actually be proven for any $L$ by integrating out the excitation on the small bridge in (\ref{intB}),
\beq
\frac{a^2}{(u^2+\tfrac{a^2}{4})^{L-1}}\times \sum_{b\geqslant 1} \dashint \frac{dv}{2\pi}\frac{b^2}{(v^2+\tfrac{b^2}{4})((u-v)^2+\tfrac{(a-b)^2}{4})((u-v)^2+\tfrac{(a+b)^2}{4})} = -\frac{a^2 k_{a}(u)}{2(u^2+\tfrac{a^2}{4})^{L}}\, .
\eeq
The bulk integral is then seen to neutralise most of the contact term, if not for a tiny remainder,
\beq\label{B1C1}
B_{1} + \frac{1}{2}C_{1} =  \sum_{a=1}^{\infty} \int \frac{du}{2\pi} \frac{a^2 g^{2L}}{(u^2+\frac{a^2}{4})^{L}}\, , 
\eeq
which reproduces the anomalous dimension of the length $L$ operator, see~\cite{Broadhurst:1985vq,Gurdogan:2015csr}. As an additional comment, let us point out that our formula breaks down for the shortest operator, with $L=2$ (or whenever a bridge length vanishes). In this circumstance, the summation-integration is divergent and the divergence is indicative of the length-two mixing between single- and double-trace operators, as discussed in detail in \cite{Grabner:2017pgm,Gromov:2018hut,Korchemsky:2015cyx}.

\begin{table}
\makegapedcells
\centering
\begin{tabular}{c | c | c |c |c}
\toprule
 \strut $\ell_{2}$ & $\ell_{3}$ & \text{{\footnotesize{Three-point integral constant}}} & \text{{\footnotesize{Two-point integral constant}}} & $\mathcal{A}_{\textrm{1-wheel}}$  \\
\midrule
  1 & 2 & $ 12 \zeta_3-\frac{\pi ^4}{30}$ &$-12 \zeta_3-\frac{\pi ^4}{15}$ & $6 \zeta_3$   \\ 
\hline
  1 & 3 & $\zeta_3^2+45 \zeta_5-\frac{5 \pi ^6}{378} $ & \multirow{2}{*}{$2 \zeta_3^2+50 \zeta_5-\frac{5 \pi ^6}{189}$} & $20 \zeta_5$  \\ 
\cline{1-3}\cline{5-5}  
   2 & 2 & $-5 \zeta_3^2+45 \zeta_5-\frac{5 \pi ^6}{378}$ & &$- 6 \zeta_3^2 + 20 \zeta_5$ \\ 
 \bottomrule
\end{tabular}
\caption{Terms $\sim \epsilon^0$ for the dimensionally regularised three- and two-point integrals in Eq.~(\ref{2and3ptintegrals}), with the spacetime dependence stripped off.  The integrals were computed using the $G$-scheme normalization \cite{Baikov:2010hf}. The last column gives the normalised structure constants, in perfect agreement with the integrability predictions in table~\ref{intprediction}.} \label{3ptcheck}
\end{table}

We were able to reproduce the results in table \ref{intprediction} through four loops by a direct field theory calculation.\footnote{We thank Vasco Gon\c{c}alves for help with the Feynman integrals.} In the field theory, the normalised structure constant is computed by combining two- and three-point Feynman integrals, see e.g.~\cite{Caetano:2014gwa}. Namely, one adds up all the Feynman integrals contributing to the 3-point function, keeping only the constant terms in the regulator expansion and subtracting half of the constants for the diagrams obtained by merging two of the three external points. (The outcome does not depend on the regularisation used.) In the present case, the fishnet theory trims the diagrammatics down to a single wheel integral and the structure constant of interest is given by
\beq
C^{\bullet\circ\circ}(\ell_{2}, \ell_{3}) =\left[\vcenter{\hbox{\includegraphics[width=0.14\linewidth]{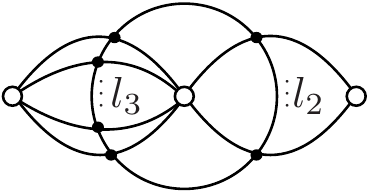}}}\right]_{\text{constant} } -\frac{1}{2} \,\left[\vcenter{\hbox{\includegraphics[width=0.14\linewidth]{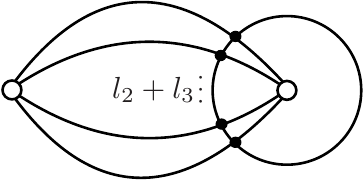}}}\right]_{\text{constant} }\,, \label{2and3ptintegrals}
\eeq
where ``constant" refers to the constant term in the regulator expansion. (We are dropping the space-time dependence of the integral, which are fixed by conformal symmetry.) We computed the Feynman integrals in (\ref{2and3ptintegrals}) up to four loops, using dimensional regularisation and the so-called $G$-scheme normalisation \cite{Baikov:2010hf}. The results for the $\epsilon^0$ terms of the corresponding two- and three-point integrals are listed in table \ref{3ptcheck}. When put together, as in~(\ref{2and3ptintegrals}), we obtain a perfect match with the integrability output listed in table \ref{intprediction}. The higher-loop expressions on the integrability side readily map to predictions for the corresponding three-point Feynman integrals, after carrying out one subtraction (e.g., one could conveniently remove the linear $\zeta$-piece $= \mathcal{A}_{\textrm{1-wheel}}(1, L-1)$ on both sides).

\section{Wrapped structure constants and dilaton insertion}\label{Sect4}

In this section we push the analysis further by considering wheel corrections to the structure constant
\beq\label{C1234}
C^{\bullet\circ\bullet}_{132} \sim \mathcal{h} \textrm{tr}\, \phi_{1}^{L_{1}}(x_{1})\, V_{n, m, n^*}(x_{3})\, \textrm{tr}\, \phi_{1}^{\dagger L_{2}}(x_{2}) \mathcal{i}
\eeq
where $V_{n, m, n^*}$ is the protected operator defined in (\ref{Vnm}) with dimension $\Delta_{V} = 2m+n+n_*$. Conservation of $\phi_{1}$ charge requires that $n - n_* = L_{2}-L_{1}$ and the structure constant is characterized by three quantum numbers: the lengths $L_{1,2}$ of the BMN operators and the number $m$ of zero-momentum magnons inserted on each side of the puncture $V$. The diagonal structure constants, to be discussed at length later on, are obtained by setting $L_{1} = L_{2}$ or, equivalently, $n = n_*$, and the dilaton insertion is the special case~$m=n=1$.

\begin{figure}
\begin{center}
\includegraphics[scale=0.45]{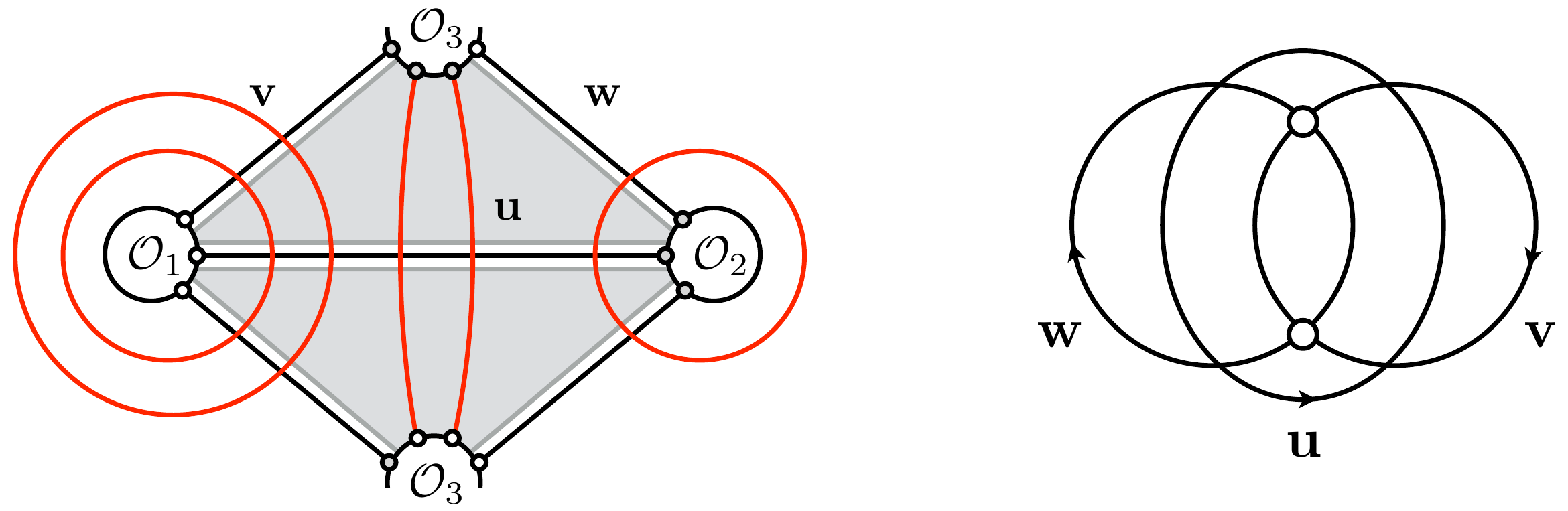}
\end{center}
\caption{On the left, a Feynman diagram contributing to the structure constant between a pair of BMN operators (left- and right-hand sides) and the protected puncture $\mathcal{O}_{3} \sim V_{n, m, n_*}$. We cut it down into two hexagons as shown here. The magnons circulate along the wheels surrounding operators 1 and 2, if not for $m$ of them, which terminate on $\mathcal{O}_{3}$. We denote by $\textbf{u}$ the set of rapidities in the ``bottom" channel (12) and by $\textbf{v}$ and $\textbf{w}$ those corresponding to the ``adjacent" channels $(13)$ and $(23)$; by charge conservation, $|\textbf{u}| = |\textbf{v}|+|\textbf{w}|+m$. In the right panel, we represent the hexagon matrix part for the process. Each circle stands for a stack of lines with corresponding rapidities. Crossings represent R matrices and blobs their shifted versions.}\label{Vins} 
\end{figure}

The diagrams contributing to (\ref{C1234}) are shown in figure \ref{Vins}. At leading order, $m$ magnons are produced at the bottom and sent to the top where they are absorbed. The perturbation theory amounts to dressing this process with wheels encircling the first or the second operator. The associated hexagon series is given by
\beq\label{hseries}
C^{\bullet\circ \bullet}_{132} = \sqrt{L_{1} L_{2}}\times (\mathcal{A}_{(0, m, 0)} + \mathcal{A}_{(1, m+1, 0)}+ \mathcal{A}_{(0, m+1, 1)} + ...)\, ,
\eeq
where the first term contains no wheels, the following ones 1 wheel around the left or the right operator, etc. (Note that the leading term $\mathcal{A}_{(0, m, 0)}$ is insensitive to the left and right bridges, $13$ and $32$, and only probes the bottom bridge $12$.) In this section we will explain how to make sense of the first few terms in the series (\ref{hseries}), and of all of them in a particular regime.

For the Lagrangian insertion~(\ref{Lag}) an exact field theory formula is known. This formula expresses the structure constant as the derivative w.r.t.~coupling constant of the scaling dimension $\Delta_{L}(g)$ of the BMN operator $\textrm{tr}\, \phi_{1}^L$. More precisely, after stripping out an inessential factor,
\beq\label{csmall}
c^{\bullet\circ\bullet} = \frac{g^2}{L} C^{\bullet\circ \bullet} = -\frac{1}{2} \frac{\partial \Delta_{L}(g)}{\partial \log{g^{2L}}}\, ,
\eeq
where $L = L_{1} = L_{2}$. This formula was discussed at length in \cite{Costa:2010rz} and more recently in \cite{Cavaglia:2018lxi}. We shall use it as a testing ground for our formulae, in the following.

\subsection{Bare hexagon series}\label{Sect4.1}

To begin with, let us spell out the hexagon prediction for the generic term in (\ref{hseries}). It follows from taking the general expressions for the hexagon form factors, attaching legs together, summing over indices and integrating over the rapidities. Taking all the steps at a time, we get
\beq\label{generic}
\mathcal{A}_{(i, k, j)} = \int \frac{d\textbf{v}d\textbf{u}d\textbf{w}}{i!k!j!(2\pi)^{i+k+j}} \tilde{\mu}_{L}(\textbf{v})\tilde{\mu}_{B}(\textbf{u})\tilde{\mu}_{R}(\textbf{w}) \frac{\Delta_{<}(\textbf{u}, \textbf{u}) \Delta_{<}(\textbf{v}\cup \textbf{w}, \textbf{v}\cup \textbf{w})}{\Delta(\textbf{u}, \textbf{v}\cup \textbf{w})} \mathcal{R}(\textbf{u}, \textbf{v}, \textbf{w})\, ,
\eeq
where $i, k, j$ counts the number of magnons per channel, with $i+j = k-m$ for charge conservation. Integration is taken over each rapidity $u_{i}, ...$ and an implicit sum is made on the associated bound state label $a_i, ...$ Owing to the specific form of the abelian parts of the hexagon form factors, see~(\ref{Huvw}), we could combine together the magnons $\textbf{v}$ and $\textbf{w}$ in the left and right channels. The property does not extend to the matrix part $\mathcal{R}(\textbf{u}, \textbf{v}, \textbf{w})$, which is nonetheless left-right symmetrical, $\mathcal{R}(\textbf{u}, \textbf{v}, \textbf{w}) = \mathcal{R}(\textbf{u}, \textbf{w}, \textbf{v})$. It is depicted in the right panel of figure~\ref{Vins} and can be written concisely by squaring the matrix in (\ref{Mfishnet})
\beq\label{calR}
\begin{aligned}
&\mathcal{R}(\textbf{u}, \textbf{v}, \textbf{w}) = \frac{1}{D}\,  \textrm{tr}\,\, \{\mathcal{M}(\textbf{u}, \textbf{v}, \textbf{w})^{\dagger}\mathcal{M}(\textbf{u}, \textbf{v}, \textbf{w})\}\, \\
& = \frac{1}{D} \frac{r(\textbf{v}, \textbf{w})}{r(\textbf{u}, \textbf{w})}\textrm{tr}\,\{ R(\textbf{w}^{--}, \textbf{u})R(\textbf{v}, \textbf{u})R(\textbf{v}, \textbf{w}^{--})R(\textbf{w}^{++}, \textbf{v}) R(\textbf{u},\textbf{v}) R(\textbf{u}, \textbf{w}^{++})\}\, ,
\end{aligned}
\eeq
where $\textbf{w}^{\pm\pm} = \textbf{w}\pm i$, with the trace taken over the tensor product of the $SU(2)$ modules, with dimension $D = \prod_{i, j, k}a_{i}b_{j}c_{k}$, and with $r_{ab}(u, v) = r_{ab}(u-v)$,
\beq\label{rab}
\begin{aligned}
r_{ab}(u) = c_{ab}(u)c_{ba}(-u) = \frac{u^2+\tfrac{1}{4}(a+b-2)^2}{u^2+\tfrac{1}{4}(a-b)^2}\, .
\end{aligned}
\eeq
Note that the matrix part (\ref{calR}) collapses if $\textbf{w}$ is empty,
\beq\label{calR2}
\mathcal{R}(\textbf{u}, \textbf{v}, \emptyset) = 1 \, .
\eeq
and similarly for $\textbf{v} = \emptyset$, thanks to the left-right symmetry. (The symmetry is not manifest in the representation (\ref{calR}) but is visible in figure \ref{Vins}.) The bulk of the interaction in (\ref{generic}) comes from the dynamical part of the hexagon form factors, which we normalized such as to be independent of the coupling and function of differences of rapidities,
\beq
\begin{aligned}
\Delta_{ab}(u, v) &= \frac{1}{g^4}(u^2+\tfrac{a^2}{4})^2(v^2+\tfrac{b^2}{4})^2 H_{ab}(u, v)H_{ba}(v, u) \\
&= ((u-v)^2+\tfrac{1}{4}(a-b)^2)((u-v)^2+\tfrac{1}{4}(a+b)^2)\,.
\end{aligned}
\eeq
The effective measure $\tilde{\mu}$ collects the remaining factors. It depends on the channel, through the bridge length $\ell$ and $\xi$ factors, and reads
\beq\label{tildemu}
\tilde{\mu}_{a}(u) = \frac{a^2 g^{2\ell}}{(u^2+a^2/4)^{\ell \pm m}}\, ,
\eeq
with $+/-$ applying to bottom and adjacent channels, respectively. The overall power of the coupling constant readily counts the total number of intersection points on all the bridges, $\# = k\ell_{12}+i\ell_{13}+j\ell_{32} = k\ell_{B}+i\ell_{L}+j\ell_{R}$, as it should be.

As already mentioned, due to the decoupling singularities at $u = v$ or $w$, the integral (\ref{generic}) is not properly defined, in general. The sole exception is the leading term, with no wheels, i.e., $i = j = 0$ and $k = m$. For this choice there is no denominator in (\ref{generic}) and the integral is unambiguous. The integration can be done explicitly by taking the pinching limit $z, \bar{z}\rightarrow 1$ of the fishnet four-point function studied in \cite{Basso:2017jwq}, which gives the answer in the form of a determinant,
\beq\label{LOf}
\begin{aligned}
\mathcal{A}_{(0, m, 0)}  = \frac{\textrm{det}\, M}{\prod_{k=1}^{m}(\ell-m+2k-1)!(\ell-m+2k-2)!}\, ,
\end{aligned}
\eeq
where $\ell = \ell_{12} =\ell_{B}$ is the bottom bridge length. Here, $M$ is a $m\times m$ Hankel matrix of Riemann $\zeta$-values,
\beq
M_{ij} = p!(p-1)! \times \mathcal{A}_{(0, 1, 0)}(p)\, ,
\eeq
with $p = \ell-m+i+j-1$, and $\mathcal{A}_{(0, 1, 0)}(p)$ relates to the period of the one-wheel graph with $p+1$ spokes \cite{Broadhurst:1985vq,Gurdogan:2015csr},
\beq\label{period}
\mathcal{A}_{(0, 1, 0)}(p) = \sum_{a\geqslant 1} \int \frac{du}{2\pi} \frac{a^2 g^{2p}}{(u^2+a^2/4)^{p+1}} = \frac{(2g)^{2p}\Gamma(\tfrac{1}{2}+p)}{\Gamma(\tfrac{1}{2})\Gamma(1+p)} \zeta(2p-1)\, .
\eeq
We should add that formula (\ref{LOf}) breaks down with the divergence of the top-left corner of $M$, when $\ell \rightarrow m$,
\beq
\mathcal{A}_{(0, m, 0)}  \propto \zeta (2(\ell-m)+1)\sim 1/(\ell-m)\, .
\eeq
A similar phenomenon was encountered in Subsection \ref{Half}, see comment after (\ref{B1C1}), and the pole is indicative of a mixing with double-trace operators. The extremality condition is indeed reached as soon as the dimension of the puncture exceeds the total dimension of the pair of BMN operators. At weak coupling, the condition translates into
\beq
0 = \Delta_{1}+\Delta_{2}-\Delta_{3} = 2(\ell-m) \, ,
\eeq
and, to stay on the safe side, one should impose that $\ell > m$.%
\footnote{The singularity is shifted away by the anomalous dimensions of the BMN operators at finite coupling. However, controlling this effect requires re-summing the wheel graphs inducing the anomalous dimensions.}

For the dilaton, we set $m = 1, p = \ell = L-1$ in (\ref{period}) and verify, in agreement with (\ref{csmall}), that the structure constant measures the 1-wheel anomalous dimension of the length $L$ operator \cite{Broadhurst:1985vq}, up to the overall factor $-2g^2$. The comparison can also be done at the integrand level using the L\"uscher formula for the scaling dimension \cite{Gurdogan:2015csr,Ahn:2011xq}
\beq\label{Delt1}
\Delta = L -2\sum_{a\geqslant 1} \int \frac{du}{2\pi} \textbf{Y}_{a}(u) + O(g^{4L})\, .
\eeq
Here $\textbf{Y}_{a}(u)$ is the asymptotic value of the vacuum Y function,
\beq
\textbf{Y}_{a}(u) = e^{-LE_{a}(u)}\textrm{tr}_{V_{a}\otimes \dot{V}_{a}}(1) = \frac{a^2 g^{2L}}{(u^2+a^2/4)^{L}}\, .
\eeq
It fixes the initial condition for the low temperature, $1/L \ll 0$, iteration of the TBA equations, determining $\Delta$ to all orders in the wheel expansion, see Eq.~(\ref{TBAall}) below. Combining the TBA formula (\ref{Delt1}) with the field theory one (\ref{csmall}), and using $\partial \textbf{Y}_{a}(u)/\partial \log{g^{2L}} = \textbf{Y}_{a}(u)$, we obtain
\beq
c^{\bullet\circ\bullet} = \sum_{a\geqslant 1} \int \frac{du}{2\pi} \textbf{Y}_{a}(u) + O(g^{4L})\, ,
\eeq
in agreement with the bottom channel hexagon measure, $g^2 \tilde{\mu}_{B}(u) = \textbf{Y}_{a}(u)$, when $m = 1$ and $\ell = L-1$.

\subsection{Renormalizing the leading wheels}\label{Sect4.2}

\begin{figure}
\begin{center}
\includegraphics[scale=0.45]{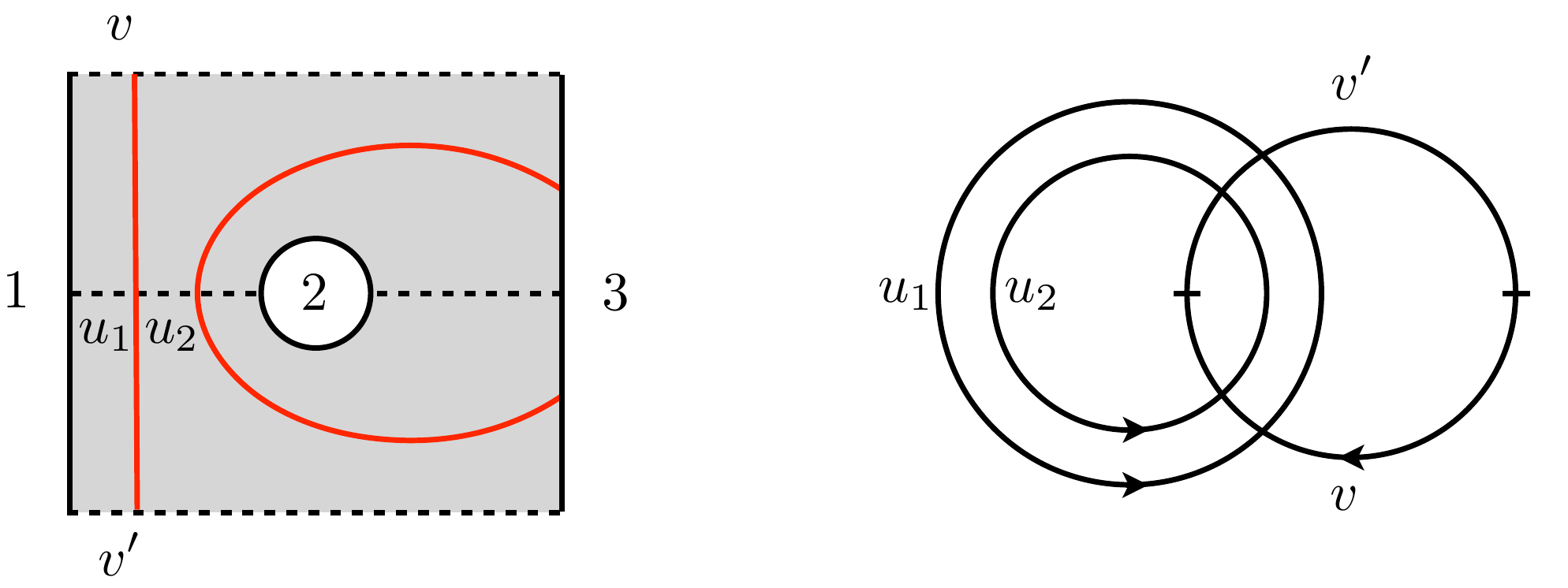}
\end{center}
\caption{The leading wrapping contribution comes from a single wheel surrounding either operator 1 or 2. The short-distance singularity can be handled by point splitting the rapidity along a mirror cut, as shown here for a wheel around operator $1$. The divergence appears then as a simple pole $\sim 1/\epsilon$ in the regulator $\epsilon = v'-v\sim 0$. The finite part $\sim \epsilon^0$ can be understood as dressing with finite-size corrections the spectator magnons in the channel 12. On the right panel, we show the contraction of R matrices yielding the matrix part for the point-split process. The flavours circulate freely along the loops, but the rapidity jumps from its incoming to outgoing values, $v'$ and $v$, along one of them.}\label{LW} 
\end{figure}

We move to the leading wheels. To handle them properly we must subtract their divergences. The procedure was briefly recalled in Subsection \ref{Half}. Here, we will generalise it to the case of the $m$-charged puncture.

The regularisation is performed by considering the point-split process shown in figure~\ref{LW}. There we focus on the channel $12$ where two hexagons are attached together. The puncture produces a beam of $m$ magnons crossing the channel. On top of that, there is a magnon that is propagating from bottom to top, from a rapidity $v'$ to a rapidity $v$. Compactifying the picture along the channel $13$, the end-points of the latter magnon get identified, $\epsilon = v'-v \rightarrow 0$, and a wheel forms around the operator 1, as desired. There is nothing wrong with the point-split process, as long as $\epsilon\neq 0$; the problem shows up in the diagonal limit $\epsilon \rightarrow 0$, in the form of a pole $\sim 1/\epsilon$. The renormalised amplitude is obtained by removing this pole and integrating the remainder $\sim \epsilon^0$ over $v$. Of course, a similar picture applies for a wheel around the operator $2$.

We focus on the case where we have only two magnons $u_{1,2}$ in the (bottom) channel~$12$; the generalisation to more magnons is straightforward and will be given later on. The amplitude for the regularised process is
\beq\label{int121}
\frac{H_{a_{1}a_{2}}(u_{1}, u_{2})H_{a_{2}a_{1}}(u_{2}, u_{1})}{\prod_{i=1,2}H_{ba_{i}}(v+i0, u_{i})H_{a_{i}b}(u_{i}+i0, v')} \times M_{a_{1}a_{2}b}(u_{1,2}, v)\, .
\eeq
It should be weighted with appropriate measures and energy factors, integrated over $u_{1,2}$ and summed over~$a_{1,2}$. The matrix part $M_{a_{1}a_{2}b}$ is depicted in the right panel of figure \ref{LW}, and reads
\beq\label{m121}
M_{a_{1}a_{2}b}(u_{1,2}, v) = (a_{1}a_{2}b)^{-1}\, \textrm{tr}\, \{R_{a_{1}b}(u_{1}, v)R_{a_{2}b}(u_{2}, v)R_{ba_{2}}(v', u_{2})R_{ba_{1}}(v', u_{1})\}\, ,
\eeq
with the trace taken over $V_{a_{1}}\otimes V_{a_{2}}\otimes V_b$. It trivialises in the limit $\epsilon \rightarrow 0$, in agreement with~(\ref{calR2}). The integration over the $u$'s is well-defined thanks to the $i0$ prescription. (Note that this is the same $i0$'s as used for the computation of the propagator in Subsection \ref{prop}.) The amplitude is divergent when $\epsilon\rightarrow 0$, since then the upper and lower half-plane singularities, coming from the denominator in (\ref{int121}), pinch the contours of integration. The pole it produces can be isolated from the rest by deforming the contours, in e.g.~the upper half-planes; the pole will then reside in the residues at $u_{1,2} = v+i0$. Owing to the permutation symmetry of the integrand, we can concentrate on the residue at $u_{1} = v+i0$, with $b = a_{1}$. It yields 
\beq
\begin{aligned}
\frac{(-1)^{b-1}}{H_{bb}(v+i0, v')}\times \frac{H_{a_{2}b}(u_{2}, v)}{a_{2} bH_{a_{2}b}(u_{2}+i0, v')} \times \textrm{tr}\, \{R_{a_{2}b}(u_{2}, v)R_{ba_{2}}(v', u_{2})R_{bb}(v', v)\}\, ,
\end{aligned}
\eeq
with the pole $\sim 1/\epsilon$ sitting in the first factor, see (\ref{poleH}). The Laurent expansion gives then
\beq\label{interm}
\frac{1}{i\epsilon} + \frac{1}{2}K_{bb}(v, v) +i\partial_{v} \log{H_{a_{2}b}(u_{2}+i0, v)}+\frac{1}{ia_{2}b} \partial_{v}\textrm{tr}_{V_{a_{2}}\otimes V_{b}}\, \log{R_{ba_{2}}(v-u_{2})} + O(\epsilon)\, ,
\eeq
up to overall measures, and with $K$ as defined in (\ref{Kab}). Dropping the first term, we read out the remainder produced by the renormalisation. To find their effects on the structure constant, we must weight them properly and integrate. The weight of the wheel is easy to remember since it has to match with the asymptotic Y function $\textbf{Y}^{L}_{b}(v)$ for the left BMN operator. Integrating the first term $\sim \epsilon^0$ in (\ref{interm}) against $\textbf{Y}^{L}_{b}(v)$ reproduces the contact term $\tfrac{1}{2}C^{L}_{1}$ met earlier, see (\ref{intC}). The other terms encode the interaction between the wheel $v' = u_{1} = v$ and the leftover magnon $u_{2}$ in the bottom channel. We can interpret them as shifting the measure of the latter magnon,
\beq
\tilde{\mu}_{a_{2}}(u_{2}) \rightarrow \tilde{\mu}_{a_{2}}(u_{2})y^{L}_{a_{2}}(u_{2})\, ,
\eeq
with the left finite-size corrections
\beq\label{yL}
y^{L}_{a}(u) = 1+\sum_{b\geqslant 1} \frac{1}{iab}\int\limits_{\mathbb{R}-i0} \frac{dv}{2\pi}\, \textbf{Y}^{L}_{b}(v)\, \textrm{tr}_{V_{b}\otimes V_{a}}\, \partial_{v}\log{\big[\frac{R_{ba}(v-u)}{H_{ab}(u, v)}\big]} + O(\textbf{Y}^2)\, .
\eeq
Note that we cannot ignore the leftover $i0$ shift in the contour of integration. It is needed to avoid the pole triggered by the zero of $H_{ab}(u, v)$, see (\ref{poleH}). A similar analysis applies to the right wheeled amplitude $\mathcal{A}_{(0, 2, 1)}$; one replaces $v, b\rightarrow w, c$, complex conjugate and pick up the residue in the lower half-plane, at $u_{1} = w-i0$. It yields
\beq\label{yR}
y^{R}_{a}(u) = 1+\sum_{c\geqslant 1} \frac{1}{iac}\int\limits_{\mathbb{R}+i0} \frac{dw}{2\pi}\, \textbf{Y}^{R}_{c}(w)\,\textrm{tr}_{V_{c}\otimes V_{a}}\, \partial_{w}\log{\big[H_{ca}(w, u)R_{ca}(w-u)\big]}  + O(\textbf{Y}^2)\, .
\eeq
Finally, owing to the decoupling property of the hexagon form factors, the general formula for a generic state $\textbf{u}$ in the bottom channel is simply obtained by adding up the individual left and right shifts, that is,
\beq\label{muyy}
\tilde{\mu}_{B}(\textbf{u}) \rightarrow \tilde{\mu}_{B}^{LR}(\textbf{u}) = \tilde{\mu}_{B}(\textbf{u}) \prod_{i}y^{L}_{a_{i}}(u_{i})y^{R}_{a_{i}}(u_{i})\, .
\eeq

Summarising, besides the need to evaluate integrals with $\mp i0$ prescriptions, for left and right channels, respectively, we must also dress each measure in the bottom channel by the finite size corrections sourced by the left and right BMN operators, using (\ref{muyy}), (\ref{yR}), (\ref{yL}). At last, adding the left and right contact terms, $\tfrac{1}{2}C_{1}^{L,R}$, we obtain the hexagon series
\beq
\begin{aligned}\label{wrappedC}
&\frac{C_{132}^{\bullet\circ\bullet}}{\sqrt{L_{1}L_{2}}}= e^{\frac{1}{2}(C^{L}_{1}+C^{R}_{1})} \\
&\,\,\,\, \times \bigg[\int \frac{du_{1}\ldots du_{m}}{m!(2\pi)^m} \tilde{\mu}^{LR}_{B}(\textbf{u})\Delta_{<}(\textbf{u}, \textbf{u}) \\
&\,\,\,\,\,\,\,\,\,\, +\int \frac{du_{1}\ldots du_{m+1}}{(m+1)!(2\pi)^{m+1}}\tilde{\mu}^{LR}_{B}(\textbf{u})\Delta_{<}(\textbf{u}, \textbf{u})\bigg\{\int\limits_{\mathbb{R}-i0}\frac{dv}{2\pi} \frac{\tilde{\mu}_{L}(v)}{\Delta(\textbf{u}, v)}
+\int\limits_{\mathbb{R}+i0}\frac{dw}{2\pi} \frac{\tilde{\mu}_{R}(w)}{\Delta(\textbf{u}, w)}\bigg\} \\
&\,\,\,\,\,\,\,\,\,\,+ O(\textbf{Y}^2_{L}, \textbf{Y}_{L}\textbf{Y}_{R}, \textbf{Y}^2_{R})\bigg]\, ,
\end{aligned}
\eeq
with an implicit summation over the bound state labels and with the higher order corrections standing for amplitudes with two or more wheels, $\mathcal{A}_{(..., m+2, ...)},$ etc.

The terms displayed in the form factor expansion (\ref{wrappedC}) are now perfectly well defined. One verifies, in particular, that the formula reduces to the one for the half structure constant analyzed in Subsection \ref{Half}, when $m=0$. More precisely, setting $m=0$, the closed string structure constant is seen to factorize into two half structure constants, for the left and right wheel, respectively,
\beq
C^{\bullet\circ\bullet}_{132} = C^{\bullet\circ\circ}_{132}(\ell_{L}, \ell_{B})\times C^{\circ\circ\bullet}_{132}(\ell_{B}, \ell_{R})\, ,
\eeq
and only one factor remains if one sends an adjacent bridge length, either $\ell_{L}$ or $\ell_{R}$, to infinity. The algebraic problem of evaluating the integrals in (\ref{wrappedC}) for higher values of $m$ is beyond the scope of this paper. Here we will bypass the difficult problem of integrating over the $\textbf{u}$ rapidities and carry a test at the integrand level by specializing to the dilaton and comparing the outcome with the TBA prediction.

One first notices that in the diagonal case, $\textbf{Y}^{L} = \textbf{Y}^{R}:=\textbf{Y}$, the two shifts can be combined together and given in terms of the TBA data,
\beq\label{ydiag}
y_{a}^{L}(u)y_{a}^{R}(u) = 1 -\textbf{Y}_{a}(u) + \sum_{b\geqslant 1} \int \frac{dv}{2\pi} \textbf{Y}_{b}(v)K_{ba}(v, u) + O(\textbf{Y}^2)\, ,
\eeq
with $K$ the flavour averaged scattering kernel (\ref{Kab}). This relation follows from
\beq
i\partial_{v} \log{\bigg[\frac{H_{ab}(u, v-i0)}{H_{ba}(v+ i0, u)}\bigg]}= -2\pi \delta_{ab} \delta(u-v) -i\partial_{v}\log{S_{ba}(v, u)}\, ,
\eeq
paying attention to the $i0$'s in the arguments. More precisely, the smooth part in the RHS originates from the permutation property of the hexagon form factor, $H_{ab}/H_{ba} = S_{ab}$, with $S_{ab}$ the abelian component of the S matrix (\ref{Sab}), while the singular part is coming from the zero of $H_{ab}(u, v)$ at $u = v$ and $a=b$,
\beq
i\delta_{ab}\partial_{v} \log{\bigg[\frac{v-u-i0}{v-u+i0}\bigg]} = -2\pi \delta_{ab} \delta(u-v)\, .
\eeq
Equation (\ref{ydiag}) can also be written in terms of the thermodynamic filling fractions
\beq
y_{a}^{L}(u)y_{a}^{R}(u) \simeq \frac{Y_{a}(u)}{\textbf{Y}_{a}(u)(1+Y_{a}(u))}\, ,
\eeq
by using the two universal terms in the IR expansion of the Y functions, see~\cite{Balog:2001sr,Ahn:2011xq} and references therein,
\beq\label{NLO}
Y_{a}(u)/\textbf{Y}_{a}(u) =  1 + \sum_{b\geqslant 1}\int \frac{dv}{2\pi} \textbf{Y}_{b}(v)K_{ba}(v, u) +O(\textbf{Y}^2)\, .
\eeq
The appearance of TBA filling fractions in the dressing of the asymptotic measure is in line with the expectations for finite volume diagonal form factors. The phenomenon is further discussed in the following subsection.

We are now equipped to verify our formula for the dilaton. Setting $m = \ell_{L} = \ell_{R} =1$, the effective weight for an adjacent magnon reduces to $\tilde{\mu}^{L, R}_{a} = a^2 g^2$ and, after transferring all the coupling dependence to the bottom magnons, we obtain
\beq
g^{2k-2}\prod_{i=1}^{k}\tilde{\mu}_{a}(u_{i})y_{a_{i}}^{L}(u_{i})y_{a_{i}}^{R}(u_{i}) =\frac{1}{g^2} \prod_{i=1}^{k}\frac{Y_{a_{i}}(u_{i})}{(1+Y_{a_{i}}(u_{i}))}\, .
\eeq
It yields
\beq\label{got}
\begin{aligned}
c^{\bullet \circ \bullet} =&\, e^{C_{1}}\bigg[\sum_{a\geqslant 1}\int \frac{du}{2\pi} \frac{Y_{a}(u)}{1+Y_{a}(u)} \\
&+ \sum_{a_{1,2}\geqslant 1}\int \frac{du_{1}du_{2}}{2(2\pi)^2}\frac{Y_{a_{1}}(u_{1})Y_{a_{2}}(u_{2})}{(1+Y_{a_{1}}(u_{1}))(1+Y_{a_{2}}(u_{2}))} \times \mathcal{B}_{a_{1}a_{2}}(u_{1}, u_{2}) + O(Y^3)\bigg]\, ,
\end{aligned}
\eeq
where the two-body integrand $\mathcal{B}$ combines the integrals for the left and right channels. It reads
\beq\label{int12}
\mathcal{B}_{a_{1}a_{2}}(u_{1}, u_{2}) = 2\sum_{b\geqslant 1}\dashint\frac{dv}{2\pi} \frac{b^2 \Delta_{a_{1}a_{2}}(u_{1}-u_{2})}{\Delta_{a_{1}b}(u_{1}-v)\Delta_{a_{2}b}(u_{2}-v)}\, ,
\eeq
where the principal value refers to the (double) pole at $v = u_{1}$ or $v= u_{2}$ and is only needed for $b = a_{1}$ or $b = a_{2}$.%
\footnote{The integral is defined for $u_{1}\neq u_{2}$ when $b = a_{1} = a_{2}$ and elsewhere by analytical continuation.}
Note that $\mathcal{B}$ is a function of the difference of the rapidities. 

The field theory formula (\ref{csmall}) predicts, on the other hand, that
\beq\label{TBA2}
c^{\bullet \circ \bullet} = \sum_{a\geqslant 1}\int \frac{du}{2\pi} \frac{Y_{a}(u)}{1+Y_{a}(u)} + \sum_{a_{1}, a_{2}\geqslant 1}\int \frac{du_{1}du_{2}}{(2\pi)^2} \frac{Y_{a_{1}}(u_{1})K_{a_{1}a_{2}}(u_{1}, u_{2})Y_{a_{2}}(u_{2})}{(1+Y_{a_{1}}(u_{1}))(1+Y_{a_{2}}(u_{2}))} + O(Y^{3})\, ,
\eeq
after invoking the all-order TBA equation for the scaling dimension,
\beq\label{TBAall}
\Delta = L -2\sum_{a\geqslant 1}\int \frac{du}{2\pi} \log{(1+Y_{a}(u))}\, ,
\eeq
and expanding the logarithm using (\ref{NLO}).

The field theory formula (\ref{TBA2}) and the hexagon prediction (\ref{got}) are strikingly similar. To conclude the test, we should evaluate the adjacent channel hexagon integral (\ref{int12}) and show that it can be expressed in terms of the scattering kernel. Straightforward integration, see Appendix \ref{CVP}, yields
\beq
\mathcal{B}_{a_{1}a_{2}}(u_{1}, u_{2}) = 2K'_{a_{1}a_{2}}(u_{1}-u_{2}) := 2K_{a_{1}a_{2}}(u_{1}, u_{2}) - 2K_{a_{1}a_{1}}(u_{1}, u_{1})\, .
\label{subtractedkernel}
\eeq
The disconnected term in (\ref{got}) removes the undesired second term in the RHS, see  (\ref{intC}), proving the agreement with the 2-body TBA integrand in (\ref{TBA2}).

\subsection{Diagonal form factors and Leclair-Mussardo series}

In this subsection we push the analysis to higher orders for the diagonal structure constants by using the Leclair-Mussardo (LM) formula \cite{Leclair:1999ys}. The formula allows one to obtain the complete form factor series for diagonal matrix elements of local operators in finite volume, or, equivalently, their expectation values at finite temperature. It is best understood for factorised scattering theories with abelian S matrices, although generalisations to higher rank models also exist \cite{Hegedus:2017zkz}. To meet this requirement, we shall limit ourselves to the singlet sector, by setting all the magnons in the bottom channel to scalar fields with $a_{i} = 1$. The magnons in the adjacent channels will remain unconstrained, since they will be integrated and summed over. Let us also mention that the (abelian) LM formula was put on firm ground in \cite{Pozsgay:2007gx,Pozsgay:2010cr} and proved in \cite{Pozsgay:2010xd} using thermodynamic arguments; see also~\cite{Bajnok:2017bfg} for a recent discussion and~\cite{Pozsgay:2009pv} for a nice review. Our following considerations also relate to studies performed in the context of the string-SYM theory and notably to \cite{Bajnok:2014sza} and \cite{Hollo:2015cda}.

\begin{figure}
\begin{center}
\includegraphics[scale=0.5]{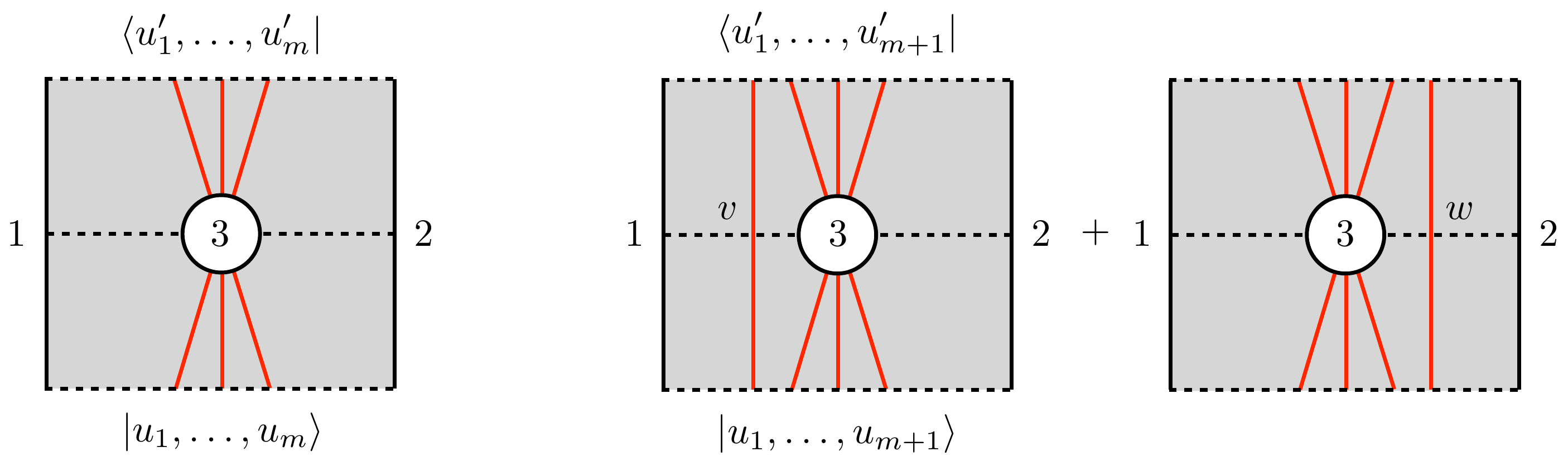}
\end{center}
\caption{Examples of form factors for the charge $m$ dilaton-like operator. Form factors with $k>m$ magnons have kinematical singularities stemming from magnons decoupling on the far left or far right. In the diagonal set up the left and right boundary are identical and one obtains, after averaging over all the decoupling paths, the same residue as for the matrix elements of a local operator in a local QFT. The left panel shows the first non zero form factor with $k=m$; all the magnons hit the operator and there is no kinematical singularity. The next panels show the form factor with $k = m+1$; a magnon can then travel on the left- or right-hand side of the operator.}\label{DFF} 
\end{figure}

The abstract operator $V$ that we will consider is obtained by attaching two hexagons together around the symmetric dilaton-like operator $V_{n, m, n}(0)$. As shown in figure \ref{DFF}, and as part of the definition of $V$, a resolution of the identity is inserted on each mirror cut ending on $V_{n, m, n}$. The bottom channel, connecting the BMN operators on the far left and far right, stays open and is used to prepare asymptotic states in the past and future of $V$. The operator is then defined through its form factors, themselves given as integrals over the magnons in the adjacent channels. Schematically, dropping bound state indices, measures, etc., we have
\beq\label{abstract}
\begin{aligned}
\mathcal{h}u'_{1}, ..., u'_{k}|V|u_{1}, ..., u_{k}\mathcal{i} &= \sum_{i+j = k-m} \int \frac{d\textbf{v}d\textbf{w}}{i!j!(2\pi)^{i+j}}\, e^{-n(E(\textbf{v})+E(\textbf{w}))}\\
&\qquad\qquad \qquad \times H(\textbf{u}-i0\rightarrow \overleftarrow{\textbf{v}}|\overleftarrow{\textbf{w}})H(\overleftarrow{\textbf{u}}'+i0\rightarrow \textbf{w}|\textbf{v})\, ,
\end{aligned}
\eeq
where the arrow is used to indicate outgoing ordering of rapidities, e.g., $\overleftarrow{\textbf{u}} = \{u_{k},\ldots, u_{1}\}$, and where $|\textbf{u}| = |\textbf{u}'|$ for the magnon number conservation. Note also that the integrals here are perfectly well defined, as long as $\textbf{u}\neq \textbf{u}'$, thanks to the $i0$ shifts. (Note also that the form factor is zero if $k<m$, for charge conservation again, see figure \ref{DFF}.)

The LM formula allows one to make sense of the finite volume vacuum expectation value of $V$ as an infinite series over the diagonal form factors, with $\epsilon = \textbf{u}'-\textbf{u} = 0$. Although originally designed for local operators in a local 2d integrable QFT, the formula also applies to our set up. The sole requirement is that the form factors exhibit the same kinematical singularities in the decoupling limit as the matrix elements of a local operator. More precisely, in the limit $\epsilon_{1} = u'_{1}-u_{1}\rightarrow 0$, taking the first particle for simplicity, the form factors should obey the recurrence relation%
\beq\label{kin}
\mathcal{h}u'_{1}, u'_{2}, ...|V|u_{1}, u_{2}, ...\mathcal{i} \sim \mu_{V}(u_{1})^{-1}\left[\frac{i}{\epsilon_{1}+i0}-\frac{i\prod_{j\neq 1}S(u_{1}, u_{j})S(u'_{j}, u_{1}))}{\epsilon_{1}-i0}\right]\times \mathcal{h}u'_{2}, ...|V|u_{2}, ...\mathcal{i}\, ,
\eeq
where $\mu_{V}$ relates to the normalisation of the free particle, with $S$ the diagonal S matrix, and where the $\pm i0$'s are needed to accommodate the disconnected delta-function supported on $\epsilon_{1} = 0$, see e.g.~\cite{Smirnov:1992vz,Bajnok:2018fzx}.

Relation (\ref{kin}) is easily seen to be respected by our abstract operator $V$. The reason is simply that there are two paths contributing to the kinematical residue of its matrix elements, corresponding to a particle moving freely on the far left or far right of the operator, respectively. In a diagonal configuration, the paths are weighted equally, if not for the universal phase in (\ref{kin}) which reflects the ordering of the particles in the states. Namely, if the left path is set to have unit residue then the right path must come with the opposite residue, by parity, up to the scattering phase for bringing the particle back and forth across the remaining magnons.

Then, given an operator obeying (\ref{kin}), the LM formula separates connected and disconnected contributions and expresses the operator expectation value at temperature $1/L$ as
\beq\label{LM}
\mathcal{h}V\mathcal{i}_{L} =  \sum_{k=0}^{\infty} \int \frac{du_{1}\ldots du_{k}}{k!(2\pi)^k} \prod_{i=1}^{k}\frac{Y(u_{i})}{1+Y(u_{i})} \times \mathcal{V}_{k}(u_{1}, \ldots , u_{k})\, ,
\eeq
where $Y$ is the solution to the vacuum TBA equation and where the integrand $\mathcal{V}_{k}(\textbf{u})$ is the so-called connected evaluation of the diagonal form factor. The latter is defined by the contour integral
\beq\label{conn}
\mathcal{V}_{k}(\textbf{u}) = \mu_{V}(\textbf{u})\oint \frac{d\epsilon_{1}\ldots d\epsilon_{k}}{(2\pi i)^k\epsilon_{1}\ldots \epsilon_{k}}\mathcal{h}u_{1}+\epsilon_{1}, \ldots , u_{k}+\epsilon_{k}|V|u_{1}, \ldots , u_{k}\mathcal{i}_{\textrm{conn}}\, ,
\eeq
where each $\epsilon$ is integrated anti-clockwise along a small contour around $0$ and where the subscript indicates that the distributional part should be discarded. Note that the integration is transparent to contributions that are smooth in the diagonal limit $\epsilon_{i} = 0, i = 1, \ldots\,$, as naively expected. The prescription is nonetheless required to address situations where the diagonal limit is ambiguous, see e.g. Eq.~(\ref{sing}) below.

Let us, for illustration, revisit the computation of the leading terms in the wheel expansion using the LM formula. The simplest form factor has $m$ magnons, which are absorbed-produced by the operator on the bottom-top hexagon, and it is factorised, 
\beq
\mathcal{h}u'_{1}, u'_{2}, ...|V|u_{1}, u_{2}, ...\mathcal{i} = \prod_{i=1}^{m}\xi(u_{i})^{2m}\prod_{1\leqslant i<j\leqslant m}H(u_{i}, u_{j})H(u'_{j}, u'_{i})\, .
\eeq
It is smooth in the diagonal limit $\epsilon = \textbf{u}'-\textbf{u}\rightarrow 0$ and evaluates to
\beq\label{less}
\mathcal{V}_{m} (\textbf{u}) = \mu_{V}(\textbf{u}) \xi(\textbf{u})^{2m} H_{\neq}(\textbf{u}, \textbf{u})\, .
\eeq
The next form factor has $m+1$ particles and features a pole whenever a particle decouples. In the hexagon picture, we have $m+1$ magnons in the split bottom channel and one intermediate magnon, $v$ or $w$, on the adjacent cut on the left- or right-hand side of the operator. The pole stems from processes where $v \sim u_{i} \sim u'_{i}$ and similarly for $w$. We isolate these non-analytic contributions to the form factor by splitting the integration contours in (\ref{abstract}) into a contour integral around $u_{i}$ and an integral avoiding the singularities. Namely, we write
\beq\label{split}
\int = \sum_{i}\oint_{u_{i}\pm i0} + \int_{\mathbb{R}\pm i0}\, ,
\eeq
with the up and down choice corresponding to the $w$- and $v$-integral, respectively, and with the circulation chosen accordingly. The amplitude over $\mathbb{R}\pm i0$ is smooth around $\epsilon = 0$ and thus goes through the connected evaluation. For the non-analytic piece, one can use that the residue only exists if the intermediate magnon has the same quantum numbers as the external ones, allowing us to set the bound state label to 1 in the contour integrals. Taking it into account, the amplitude for the left transition is given by
\beq\label{residues}
\xi(\textbf{u})^m \xi(\textbf{u}')^{m}H_{<}(\textbf{u}, \textbf{u})H_{>}(\textbf{u}', \textbf{u}')\times \sum_{i}\oint_{u_{i}-i0} \frac{dv}{2\pi} \frac{\mu_{A}(v)}{H(\textbf{u}'+i0, v)H(v, \textbf{u}-i0)}\, ,
\eeq
with $\mu_{A}(u) := \mu(u) e^{-\ell_{A}E(u)}/\xi(u)^{2m}$ the effective weight of a magnon in the adjacent channel, where $\ell_{A} = n$. The right channel amplitude follows from exchanging the roles of the primed and un-primed rapidities in the denominator and replacing $v\rightarrow w$ to comply with our general notations. Now, fixing $i=1$, for simplicity, and collecting the residues using (\ref{poleH}), we obtain the non-analytic part of the process
\beq\label{sing}
\begin{aligned}
&\frac{\mu_{A}(u_{1})(\xi\xi')^{m}}{\mu(u_{1})}\prod_{1<i<j}H(u_{i}, u_{j})H(u'_{j}, u'_{i})\prod_{i>1}\frac{H(u'_{i}, u'_{1})}{H(u'_{i}, u_{1})}\\
&\qquad \qquad \qquad  \times (\frac{1}{H(u'_{1}+i0, u_{1})}+\frac{\prod_{i=2}^{m+1}S(u_{1}, u_{i})S(u'_{i}, u_{1})}{H(u_{1}, u'_{1}-i0)})\, ,
\end{aligned}
\eeq
where the first and second terms in brackets come from the left and right amplitudes, respectively. The result manifestly obeys the kinematical axiom (\ref{kin}) when $\epsilon_{1} = u'_{1}-u_{1} \rightarrow 0$, using (\ref{poleH}), with the measure
\beq
\mu_{V}(u) = \mu(u)^2/\mu_{A}(u)\xi(u)^{2m} = \mu(u)e^{\ell_{A}E(u)}\, .
\eeq
To read out the diagonal form factor, we drop the $\pm i0$ shifts, factor out $1/H(u'_{1}, u_{1})$, and expand (\ref{sing}) around $(\epsilon_{1}, \ldots , \epsilon_{m+1}) = \vec{0}$. We find
\beq\label{eps}
\textrm{sing} \sim \frac{1}{\epsilon_{1}}(\sum_{i=1}^{m+1}\epsilon_{i}K_{i1} )\times \mathcal{V}_{m} (\textbf{u}\backslash\{u_{1}\})/\mu_{V}(\textbf{u})\, ,
\eeq
using $S(u'_{1}, u_{1}) = H(u'_{1}, u_{1})/H(u_{1}, u'_{1}) = -(1+i\epsilon_{1}K_{11}+O(\epsilon^2_{1}))$, and
\beq
K_{ij} = K(u_{i}, u_{j}) = -i\frac{\partial}{ \partial u_{i}}\log{S(u_{i}, u_{j})}\, .
\eeq
Notice that although the pole in the RHS of (\ref{eps}) is formally cancelled by the zeros in the numerator, the diagonal limit depends on the undetermined ratios $\epsilon_{i}/\epsilon_{1}$ and, as such, the limit does not exist. This behaviour is typical for form factors of local operators and follows from the left-right interference in the kinematical residue; see \cite{Leclair:1999ys,Pozsgay:2009pv}. In the end, only the term $\propto \epsilon_{1}$ in the numerator survives the connected evaluation (\ref{conn}), which returns
\beq
\mu_{V}(\textbf{u}) \oint \frac{d\epsilon_{1}}{2\pi i \epsilon_{1}} \, \textrm{sing}|_{\epsilon_{i\neq 1} = 0} = K_{11}\times \mathcal{V}_{m} (\textbf{u}\backslash\{u_{1}\})\, .
\eeq
Similar expressions are found for the other residues in (\ref{residues}), with $u_{i}$ replacing $u_{1}$. Adding them up, we finally obtain
\beq\label{calV1}
\mathcal{V}_{m+1}(\textbf{u}) = \sum_{l=1}^{m+1} K_{ll} \times \mathcal{V}_{m}(\textbf{u}\backslash \{u_{l}\})+ \textrm{smooth part}\, .
\eeq
The first term is proportional to the diagonal form factors with less magnons, see Eq.~(\ref{less}), and can be taken out of the LM sum (\ref{LM}),
\beq\label{LOLM}
\mathcal{h}V\mathcal{i}_{L} = e^{C_{1}}\int \frac{du_{1}\ldots du_{m}}{(2\pi)^m m!} \prod_{i}\frac{\mu_{V}(u_{i})\xi(u_{i})^{2m}Y(u_{i})}{1+Y(u_{i})} \prod_{i\neq j}H(u_{i}, u_{j}) + \ldots 
\eeq
with
\beq
e^{C_{1}} = 1 +\int \frac{du}{2\pi} K(u, u) \frac{Y(u)}{1+Y(u)} + \ldots\, .
\eeq
This factor matches with the contact term obtained earlier, up to the filling fraction and singlet restriction. Moreover, one verifies that the magnons are properly weighted in (\ref{LOLM}), since $\mu_{V}(u) = \mu(u) e^{-\ell_{B}E(u)}/\textbf{Y}(u)$, with $\textbf{Y}(u) = e^{-LE(u)}$ the asymptotic value of the Y function. The remaining contribution with $m+1$ magnons comes from the regular term in (\ref{calV1}). It accounts both for the ab initio regular contributions from intermediate bound states with $a>1$ and for the leftover contributions from intermediate singlets, integrated along $\mathbb{R}\pm i0$, see (\ref{split}). Adding everything together, we recover the formula obtained in the previous subsection by a slightly different method.

\begin{figure}
\begin{center}
\includegraphics[scale=0.5]{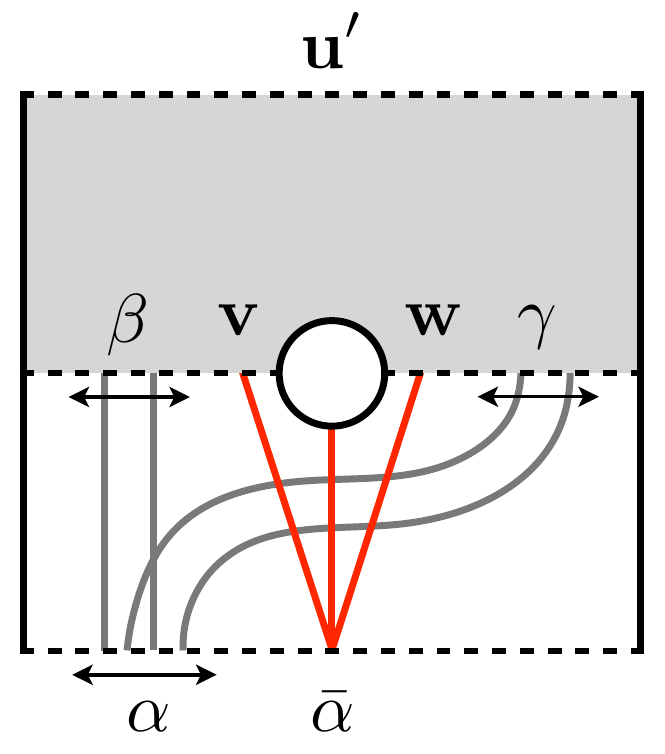}
\end{center}
\caption{Process contributing to the singular part of the form factor. A subset $\alpha = \beta\cup \gamma$ of the incoming magnons $\textbf{u} = \alpha\cup \bar{\alpha}$ moves straight to the next hexagon with a fraction $\beta$ of it landing on the far left of the operator and its complement $\gamma$ on the far right. The remaining magnons $\bar{\alpha}$ in the incoming state are absorbed by the operator or smoothly diffused into the adjacents magnons $\textbf{v}\cup \textbf{w}$ in its surroundings. For the top process, we turn the picture around. Paying attention to our convention for ordering the magnons around the hexagon, we read out the form factor $H(\{u'_{k}, \ldots , u'_{1}\} \rightarrow \textbf{w}\cup \gamma|\beta\cup \textbf{v})$.}\label{Diag} 
\end{figure}

We shall now generalise the analysis to form factors with arbitrarily many magnons. A generic form factor with $k$ magnons has multiple kinematical singularities since it can support the simultaneous decoupling of up to $k-m$ magnons. However, the strategy for taking care of the non-analytic terms triggered by these processes applies to any~$k$. Using (\ref{split}) we decompose the full process into a sum of amplitudes labelled by the subset of magnons $\alpha \subseteq \textbf{u}$ that we want to decouple. With no loss of generality, we choose $\alpha$ to come first in the state. We then split $\alpha$ in two subsets, $\alpha = \beta\cup \gamma$, for the magnons decoupling on the left- and right-hand sides of the operator, respectively, see figure \ref{Diag}. In response to this splitting, we integrate $|\beta|$ rapidities around $\beta-i0$ in the left intermediate channel and $|\gamma|$ rapidities around $\gamma+i0$ on the other side. The leftover intermediate rapidities, denoted $\textbf{v}$ and $\textbf{w}$, are integrated along $\mathbb{R} \mp i0$ and respond to the magnons $\in \bar{\alpha} = \textbf{u}\backslash \alpha$, which are absorbed, or smoothly diffused, by the operator.

Picking up the residues at $\beta-i0\cup \gamma+i0$ has the effect of decoupling the magnons on the bottom hexagon. It yields
\beq\label{weight}
S(\gamma, \bar{\alpha})S_{<}(\gamma, \beta) \times \frac{H(\overleftarrow{\textbf{u}}'\rightarrow \textbf{w}\cup \gamma|\beta\cup \textbf{v})}{H(\overleftarrow{\textbf{u}}' \rightarrow \textbf{w}|\alpha\cup \textbf{v})}\, ,
\eeq
where the amplitude has been normalised to 1 when $\gamma = \emptyset$, i.e., when all the magnons are decoupling on the left. The S matrices come from the bottom hexagon and accounts for the scattering shown at the bottom of figure \ref{Diag}, with the splitting factor $S_{<}$ as given in (\ref{splitting}) with $S^s\rightarrow S$. The numerator in the last factor is the amplitude on the top hexagon, with $\overleftarrow{\textbf{u}}' = \{u'_{k}, \ldots , u'_{1}\}$. Plugging the factorised ansatz (\ref{Huvw}) into (\ref{weight}) and using Watson relation, the dependence on the intermediate rapidities $\textbf{v}\cup \textbf{w}$ drops out and the weight (\ref{weight}) takes the simple form
\beq
(-1)^{|\gamma|}S(\gamma, \textbf{u})S(\textbf{u}', \gamma)\, .
\eeq
The sum over the paths, i.e., partitions of $\alpha$, follows straightforwardly,
\beq
\sum _{\beta\cup \gamma = \alpha}(-1)^{|\gamma|}S(\gamma, \textbf{u})S(\textbf{u}', \gamma) = \prod_{i\in \alpha} (1-S(u_{i}, \textbf{u})S(\textbf{u}', u_{i}))\, .
\eeq
We can then approach the diagonal limit by expanding around $\epsilon_{i}\sim 0$ for all $i\in \textbf{u}$. Restoring the normalisation, we obtain%
\footnote{The poles come from the diagonal limit of the reference amplitude, with $\gamma = \emptyset$, which enforces the decoupling of the $\alpha$-magnons on the top hexagon, $H(\overleftarrow{\alpha'\cup \bar{\alpha}} \rightarrow \textbf{w}|\alpha\cup \textbf{v}) \sim H(\overleftarrow{\bar{\alpha}} \rightarrow \textbf{w}|\textbf{v})/\prod_{i\in \alpha}(-i\mu_{V}(u_{i})\epsilon_{i})$.}
\beq
\begin{aligned}
\textrm{sing}^{(\alpha, \bar{\alpha})}\sim \frac{1}{\epsilon_{1}\ldots \epsilon_{|\alpha|}} (\prod_{i\in \alpha}\sum_{j\in \textbf{u}}\epsilon_{j}K_{ji})\times \mathcal{B}_{|\bar{\alpha}|}(\bar{\alpha})/\mu_{V}(\textbf{u}) \, ,
\end{aligned}
\eeq
where $\mathcal{B}_{|\bar{\alpha}|}(\bar{\alpha})$ is the bulk part of the amplitude, for the leftover rapidities \{$\bar{\alpha}, \textbf{v}, \textbf{w}$\}. We observe, again, that the numerator formally neutralises the zeros in the denominator. The connected evaluation sets $\epsilon_{i} = 0$ for all $i\notin \alpha$ and returns the term $\propto \epsilon_{1}\ldots \epsilon_{|\alpha|}$ in the numerator, i.e.,
\beq\label{gen}
\begin{aligned}
\mathcal{V}_{k}^{(\alpha, \bar{\alpha})}(\textbf{u}) &= \mu_{V}(\textbf{u})\oint \frac{d\epsilon_{1}\ldots d\epsilon_{|\alpha|}}{(2\pi i)^{|\alpha|}\epsilon_{1}\ldots \epsilon_{|\alpha|}}\textrm{sing}^{(\alpha, \bar{\alpha})}|_{\epsilon_{i \notin \alpha} =  0}\\
&= \sum_{\sigma\in S_{|\alpha|}}K_{1\sigma(1)}K_{2\sigma(2)}\ldots K_{|\alpha|\sigma{(|\alpha|)}}\times \mathcal{B}_{|\bar{\alpha}|}(\bar{\alpha})\, ,
\end{aligned}
\eeq
with $S_{|\alpha|}$ the permutation group of the $\alpha$-indices.

The other partitions of $\textbf{u}$ can be obtained by permuting the indices in (\ref{gen}). Hence, below the integral signs in (\ref{LM}), we can write
\beq
\mathcal{V}_{k}(\textbf{u}) = \sum_{|\alpha|=0}^{k} \frac{k!}{|\alpha|!|\bar{\alpha}|!} \, \mathcal{V}^{(\alpha, \bar{\alpha})}_{k}(\textbf{u})\, ,
\eeq
generalising (\ref{calV1}) to $k>m+1$. An immediate consequence of these formulae is that the LM series factorises and takes the pleasant form
\beq\label{Cfinal}
C^{\bullet\circ\bullet}/L|_{\textrm{singlet}} = \frac{1}{\mathcal{N}} \sum_{k=m}^{\infty} \int \frac{du_{1}\ldots du_{k}}{k!(2\pi)^k}\prod_{i=1}^{k}\frac{Y(u_{i})}{1+Y(u_{i})}\mathcal{B}_{k}(u_{1}, \ldots , u_{k})\, ,
\eeq
where $\mathcal{N}$ is the Fredholm-like determinant generating the $K$ factors,
\beq\label{fredholm}
\log{\mathcal{N}} = -\sum_{k=1}^{\infty}\frac{1}{k} \int \,  \frac{du_{1}\ldots du_{k}}{(2\pi)^{k}}\prod_{i=1}^{k}\frac{Y(u_{i})}{1+Y(u_{i})} K_{12}K_{23}\ldots K_{k1}\, .
\eeq
Finally, the bulk integrand $\mathcal{B}_{k}$ can be read out from the bare hexagon formula in (\ref{generic}), after restricting the $u$'s to the singlet sector, stripping out $\prod_{i}\textbf{Y}(u_{i})$, summing over the $i+j=k-m$ ways of distributing the intermediate magnons in the two adjacent cuts and integrating them along $\mathbb{R}\mp i0$. The singlet part of the dilaton formula follows from setting $m= n =1$ everywhere and rescaling the series by $g^2$.

Formula (\ref{Cfinal}) generalizes (\ref{got}) to all orders in the singlet sector. Before testing it, let us point out that we could extend it to the infinite tower of bound states if we limit ourselves to the abelian components of the hexagon form factors. The generalisation boils down to dressing with bound state indices $\textbf{a}$ all the functions of $\textbf{u}$ and adjoining to every integral sign over $d\textbf{u}$ a corresponding summation over $\textbf{a}$. The incorporation of the matrix degrees of freedom, present for $a>1$, is more delicate. If not for the single wheel, which proceeds from a single trace upgrading, a full-fledged nested Bethe ansatz procedure might be needed for a comprehensive treatment; see e.g.~\cite{Kostov:2018dmi} for a recent study. As it stands, formula (\ref{Cfinal}) might also be applied to diagonal structure constants with spirals, still in the singlet sector, by invoking the analytical continuation trick, as done recently in~\cite{Pozsgay:2014gza} for excited-state matrix elements of local operator. The contours of integration in (\ref{Cfinal}) should then be deformed such as to enclose the roots of $(1+Y(u))$, which we expect to map to the spirals ending on the BMN operators, see Subsection \ref{spirals}.

\subsection{Comparison with the field theory formula}

As a conclusion for this section, we shall carry out a test of our general expression through a comparison with the field-theory-TBA formula. The latter formula, once reduced to the singlet sector, expresses the structure constant (\ref{csmall}) as a sum over linear trees, with the nodes representing the filling fractions and the links the scattering kernels,%
\footnote{This follows from the TBA equation $\log{Y/\textrm{\textbf{Y}}} = \int dv\log{(1+Y(v))}K(v, u)/2\pi$ and singlet restriction of the free energy (\ref{TBAall}).}
\beq
c^{\bullet\circ\bullet}|_{\textrm{singlet}} = \int \frac{du}{2\pi}\frac{Y(u)}{1+Y(u)} +\sum_{k=2}^{\infty}\int \frac{du_{1}\ldots du_{k}}{(2\pi)^{k}}\prod_{i=1}^{k}\frac{Y(u_{i})}{1+Y(u_{i})} K_{12}K_{23}\ldots K_{k-1, k}\, .
\eeq
Phrased in terms of the bulk integrand $\mathcal{B}_{k}(\textbf{u})$, this is saying that
\beq
\begin{aligned}\label{Bseries}
&\mathcal{B}_{1}(\textbf{u}) = 1\, ,\\
&\mathcal{B}_{2}(\textbf{u}) = 2K'_{12}\, , \\
&\mathcal{B}_{3}(\textbf{u}) = 2K'_{12}K'_{23} + 2K'_{21}K'_{13}+2K'_{13}K'_{32} -K'_{12}K'_{21} -K'_{23}K'_{32} -K'_{31}K'_{13}\, , \\
&\ldots \, ,
\end{aligned}
\eeq
where $K'_{ij} = K'_{ji} = K_{ij}-K_{ii}$ is the subtracted scattering kernel.

Equation (\ref{Bseries}) is predicting that the integrals over the two adjacent bridges in $\mathcal{B}_{k}(\textbf{u})$ assemble to give a linear combination of products of scattering kernels. This structure is in line with the fact that the abelian component of the integrand can be cast in the form of a Cauchy-Vandermonde determinant, as discussed in Appendix \ref{CVP}. However, this observation alone is not enough for a precise match; mysterious cancellations, related to the structure of the matrix part, are also at work. Below we illustrate the computation for $k=3$, leaving the study of the generic term in (\ref{Bseries}) to a future investigation. The lower cases, with $k=1,2$, were already explained in Subsections \ref{Sect4.1} and \ref{Sect4.2}.

There are three integrals contributing to $\mathcal{B}_{3}(\textbf{u}) = \mathcal{B}_{3}(u_{1}, u_{2}, u_{3})$, for the three different ways of distributing two magnons in the left and right channels,
\beq
\mathcal{B}_{3}(\textbf{u})  = \frac{1}{2}I_{2|0} + I_{1|1} + \frac{1}{2}I_{0|2}\,,
\eeq
where the combinatorial factors have been stripped out for convenience. The first and third integrals are identical and their integrand does not involve a matrix part, see (\ref{calR}). This is not the case for the middle integral, with one magnon on each cut, 
\beq
I_{1|1}(u_{1,2,3}) = \sum_{b, c\, \geqslant 1}\, \int\limits \frac{dvdw}{(2\pi)^2}\, \frac{\prod_{i<j}\Delta_{11}(u_{i}-u_{j})\Delta_{bc}(v-w)}{\prod_{i}\Delta_{1b}(u_{i}-v)\Delta_{1c}(u_{i}-w)} (bc)^2\mathcal{R}_{bc}(v-w)\, ,
\eeq
with the contour $\mathbb{R}\mp i0$ for $v, w$, respectively. Eqs.~(\ref{calR}) and (\ref{rab}) give,
\beq\label{Rbc}
\mathcal{R}_{bc} =r_{bc}(v-w)\,  (bc)^{-1}\, \textrm{tr}_{V_{b}\otimes V_{c}}\{ R_{bc}(v-w^{--})R_{cb}(w^{++}-v)\} = 1+\frac{(b^2-1)(c^2-1)}{3\Delta_{bc}(v-w)}\, ,
\eeq
where in the last equality the trace was evaluated using the eigenspace decomposition of the R matrix (\ref{Rab-eigen}). We can split the integral into two for each term in the matrix part. The trivial term returns the same integral as for two intermediate magnons on either the left or right channel. Combining them together, it yields
\beq\label{double}
\frac{1}{2}(I_{2|0}+I_{0|2}) +I_{1|1}^{(1)} =2 \sum_{b_{1,2}\geqslant 1}b_{1}^2b_{2}^2\, \dashint \frac{dv_{1}dv_{2}}{(2\pi)^2} \frac{\prod_{i<j}\Delta_{11}(u_{i}-u_{j})\Delta_{b_{1}b_{2}}(v_{1}-v_{2})}{\prod_{i, j}\Delta_{1b_{j}}(u_{i}-v_{j})}\, ,
\eeq
where the sum over the $\pm i0$'s was replaced by the principal values. The second term in (\ref{Rbc}) vanishes whenever $b$ or $c$ is equal to $1$. Therefore, in the leftover integral, the sums over the bound states can be restricted to $b, c\geqslant 2$. There are no decoupling poles to worry about, and the $i0$'s are not needed. Furthermore, the denominator of the matrix part cancels the abelian $vw$ interaction and the integral factorises,
\beq
I^{(2)}_{1|1} = \frac{1}{3}\, \Delta_{<}(\textbf{u})\times J^2\, ,
\eeq
where
\beq
J(u_{1,2,3}) = \sum_{b\geqslant 2}\int \frac{dv}{2\pi}\, \frac{b^2(b^2-1)}{\prod_{i}\Delta_{1b}(u_{i}-v)}\, .
\eeq
The integral can be taken directly by picking up the residues and, remarkably, the sum over $b$ telescopes, yielding a simple rational function,
\beq\label{rational}
I^{(2)}_{1|1}  = 12\prod_{i<j}\frac{(u_{i}-u_{j})^2}{(1+(u_{i}-u_{j})^2)}\, .
\eeq
The double integral (\ref{double}) is computed in Appendix \ref{CVP}, using the Cauchy determinant representation for its integrand, see equation (\ref{n3}). It produces the sought-after expression, if not for a tiny rational piece, which is precisely minus the one in (\ref{rational}). Thanks to this mysterious property, we finally get
\beq
\frac{1}{2}(I_{2|0}+I_{0|2}) +I_{1|1}^{(1)} + I^{(2)}_{1|1} = -((K'_{12})^2+(K'_{13})^2+(K'_{23})^2) +2(K'_{13}K'_{12}+K'_{12}K'_{23}+K'_{13}K'_{23})\, ,
\eeq  
in complete agreement with the field-theory-TBA prediction (\ref{Bseries}).

\section{Conclusion}\label{Sect5}

In this paper, we presented conjectures for hexagon form-factors in the 4d fishnet theory. 
The formulae were deduced from the ones proposed in $\mathcal{N}=4$ SYM  by selecting the field components carefully and taking the weak coupling limit. 
Interestingly, the simplicity of the SYM ansatz was not altered by the truncation to the fishnet theory. The answer remains, for its most complicated part, entirely written in terms of the S matrix, which in the fishnet theory is just the standard rational R matrix. This type of ansatz is certainly the simplest solution to all the bootstrap axioms. However, its validity is harder to assess in the fishnet theory, than it was in the mother theory, since e.g.~there is no crossing move in the former theory. Moreover, the simplicity of the general fishnet formula is merely emerging, from the underlying microscopic SYM description, after eliminating the contributions from the fermions running in the loops, and is not visible from the onset.

We made several tests of our conjectures by applying standard recipes for building correlators and comparing the outcomes with direct Feynman diagrammatic computations in the fishnet theory. We also extracted higher-loop predictions for 1-wheel 3pt functions in the fishnet theory. This calculation entailed subtracting the divergences (double poles) which plague the hexagon amplitudes at wrapping order, allowing us to explore the prescription proposed to address this issue in the SYM context.

We also extended the renormalisation procedure such as to obtain the leading wrapping corrections for a large class of structure constants involving higher-charge generalisations of the dilaton. We could complete the hexagon series for diagonal structure constants, in the scalar sector, using the Leclair-Mussardo formula and check its validity in the case of the dilaton through comparison with the field theory prediction. It would be interesting to examine this all order formula in the continuum limit, where the truncation to the scalar sector is fully justified, and explore its connection with the sigma model description. This dual viewpoint could shed light on the method to be used to re-sum the magnon series, as it is orthogonal to the form factor expansion and involves gapless modes. It is however not immediately clear what the dilaton and its higher-charge siblings correspond to in the sigma model.

\begin{figure}[t]
\begin{center}
\includegraphics[scale=0.45]{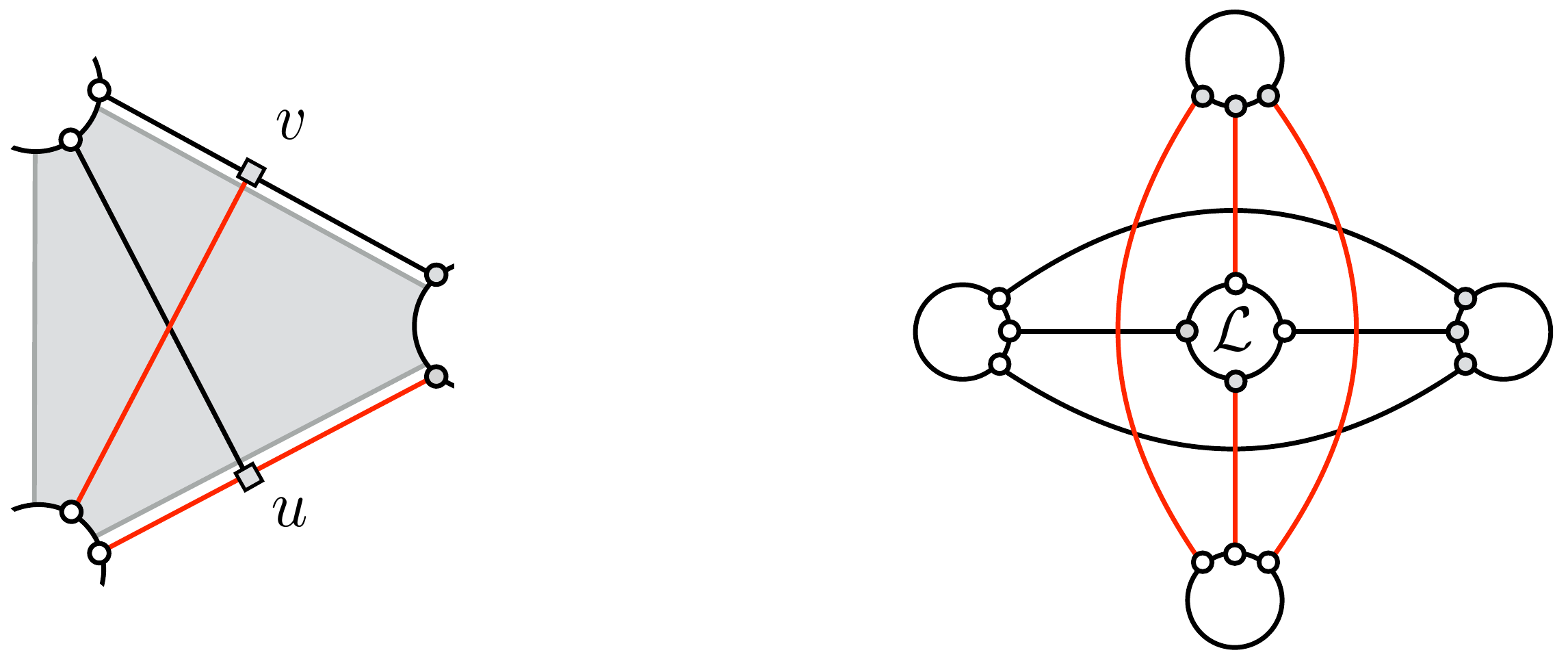}
\end{center}
\caption{In the left panel, we show an exotic form factor with two scalars of different flavours along the edges of a hexagon. The right panel displays a 5pt function obtained by inserting the dilaton in the fishnet 4pt function. We could factorise it using the hexagon form factors constructed in this paper by telling the magnonic story all the way from the bottom to the top. The exotic form factors would provide an alternative representation, where the hexagons are glued all around the dilaton, making the crossing symmetry manifest.}\label{dec} 
\end{figure}

We focused in this paper on a particular class of hexagon form factors where all the magnons where charged w.r.t.~to the symmetries preserved by the hexagon. However, it is not excluded that longitudinal magnons - which are naturally associated to the vacuum lines in the picture used in this paper - can be added to the excitations propagating along the mirror edges. An example of an ``exotic" hexagon carrying both types of fields along its edges is shown in figure \ref{dec}. This hexagon would provide alternative, and perhaps more tractable, representations for certain correlators of the theory, like the one shown in the right panel of figure \ref{dec}. Furthermore, having the vacuum and magnonic lines entering on an equal footing could make some underlying symmetries of the formalism manifest and pave the way to a more covariant formulation.

Recently, it was shown in \cite{Grabner:2017pgm} 
by an explicit multi-loop calculation that the   
planar fishnet theory has a nontrivial fixed point 
(depending on the couplings of the double traces).
The theory is integrable, conformal 
and non-unitary at
the fixed point.  Much less is known about the conformal symmetry, and a fortiori the integrability, of the fishnet theory at the non-planar level. The direct computation involves more 
types of double traces and the existence or not of a fixed point has yet to be shown. 
The SYM hexagons were used to compute non-planar
quantities (four-point functions) in \cite{Bargheer:2017nne, Bargheer:2018jvq}. 
The strategy was to cut the torus with four operators into eight hexagons and promote
each of them to hexagon form factors. It is conceivable that non-planar fishnet graphs can be cut down similarly and it would be interesting to examine the consistency of this procedure through a comparison with the direct evaluation of the corresponding Feynman integrals. 

Finally, let us mention that several observables are known exactly in the fishnet theory~\cite{Grabner:2017pgm, Gromov:2018hut, Korchemsky:2018hnb}. In particular, exact representations for four-point functions of short operators were derived using purely field theory techniques and, proceeding with the OPE, infinitely many structure constants for arbitrarily excited operators could be generated.  
Reproducing these results, at the three- or four-point level, using our fishnet hexagons may help developing general methods for re-summing the infinite tail of mirror corrections. (Note that for spinning operators one would have to generalise the analysis performed in this paper and include derivatives along the spin-chain edges of the hexagon.).
This in turn could unveil the relation between the hexagons and the more abstract ``non-magnonic" formalisms, like the Quantum  Spectral Curve \cite{Gromov:2013pga,Kazakov:2015efa,Kazakov:2018hrh} or the method of Separation of Variables \cite{Sklyanin:1991ss,Sklyanin:1995bm,Derkachov:2001yn,Belitsky:2014rba}; see e.g.~\cite{Derkachov:2018rot,Cavaglia:2018lxi,Giombi:2018qox,Giombi:2018hsx} for recent applications of these methods to correlation functions.

\section*{Acknowledgements}

We thank Zoli Bajnok, Sergey Derkachov, Vasco Gon\c calves, Arpad Hegedus, Volodya Kazakov, Grisha Korchemsky, Ivan Kostov, Enrico Olivucci, Didina Serban and Deliang Zhong for interesting discussions. T.F. is thankful to the LPTENS and Nordita for the warm hospitality during this work's completion. The research of B.B. was supported by the French National Agency for Research grant ANR-17-CE31-0001-02. The research of J.C. and T.F. was supported by the People Programme (Marie Curie Actions) of the European Union's Seventh Framework Programme FP7/2007-2013/ under REA Grant Agreement No 317089 (GATIS), by the European Research Council (Programme ``Ideas" ERC-2012-AdG 320769 AdS-CFT-solvable), and by the ANR grant Strong Int (BLANC-SIMI-4-2011). T.F. also thanks the CAPES (Coordenação de Aperfeiçoamento de Pessoal de Nível Superior) grant INCTMAT 88887.143256/2017-00 for financial support. J.C. is supported by a Simons Collaboration grant.

\appendix

\section{R matrix in matrix form}\label{AppR}

The mirror bound-state $\mathcal{S}$ matrix was  computed using the hybrid convention relevant for the hexagon formalism in \cite{Fleury:2017eph}. This computation was an adaptation of the one done in \cite{Arutyunov:2009mi} for the bound states that are physical from the spin-chain kinematical viewpoint.  The $\mathcal{S}$ matrix has a block diagonal form and the blocks are divided into three classes: I, II and III, following the terminology used in the Appendix B of \cite{Fleury:2017eph}. The R matrix of interest appears already in case I. The latter involves the scattering  of states of the form
\begin{equation}
| \{u,k \} ,\{v,l\} \rangle_{a,b}^{I}  = | \phi_i \psi_1^{a-k-1} \psi_2^k (u)\rangle
\otimes | 
\phi_i \psi_{1}^{b-l-1} \psi_2^l (v) \rangle \, , 
\end{equation}       
with $i=1$ being the case Ia and $i=2$ being the case Ib. The fields in the kets are implicitly symmetrized and the states can be obtained by acting with the supercharges on symmetrized states entirely made out of $\psi's$. The non-vanishing matrix elements at leading order in the mirror sheet are of the form 
\begin{equation}
\mathcal{S} \cdot | \{u,k \} ,\{v,l\} \rangle_{a,b}^{I} 
= \sum_{n=0}^{N=k+l} H^{k,l}_n(u,v)_{a,b}  
| \{v,N-n \} ,\{u,n\} \rangle_{a,b}^{I} \, ,
\end{equation}
with
\begin{equation}
H^{k,l}_n(u,v)_{a,b} = D_{ab}(u,v) R_{ab}[k,l,n](u,v) \, ,
\label{eq:TheCaseIH} 
\end{equation}
and
\begin{equation}
D_{ab}(u, v) = - (-1)^{(a-1)(b-1)} \frac{\sqrt{\frac{b^2}{4}+v^2}} {\sqrt{\frac{a^2}{4}+u^2}} \, \frac{u - i \frac{a}{2}}{v - i \frac{b}{2}} \, .
\end{equation}
The R matrix is given by
\begin{equation}
\begin{aligned}
R_{ab} \, [k,l,n](u,v) =  \frac{{\rm{N}}^1_a( \{u, n\} ) {\rm{N}}^1_b( \{v, N-n\} ) }{{\rm{N}}^1_a( \{u, k\} ) \, {\rm{N}}^1_b( \{v, l\} )} \times \frac{\prod_{p_1=1}^{n} p_1 \prod_{p_2=1}^{k+l-n} p_2}{
\prod_{p_3=1}^{k+l} ( i \delta u - \frac{a+b}{2} + p_3) \prod_{p_4=1}^k p_4 \prod_{p_5=1}^{l} p_5} \times \hspace{3mm} \\
\sum_{m=0}^{k} \binom {k}{k-m} 
\binom {l}{n-m} 
\prod_{p=1}^{m} c^{+}(p) \prod_{p = 1-m}^{l-n} c^{-} (p) \prod_{p=1}^{k-m} d \left( \frac{k-p+2}{2} \right) \prod_{p=1}^{n-m} \tilde{d} \left( \frac{k+l-m-p+2}{2} \right) \, , 
\end{aligned}
\label{eq:DefinitionR} 
\end{equation}
with $\delta u = u -v $, and we defined
\begin{equation}
\begin{aligned}
c^{+}(t) = i \delta u - \frac{(a-b)}{2} + t -1 \; , \quad 
d(t) = - (a+ 1 - 2t) \, , \\
c^{-}(t) = i \delta u + \frac{(a-b)}{2} + t -1 \; , \quad 
\tilde{d}(t) = - (b+1 - 2 t) \, , 
\end{aligned}
\end{equation}
and
\begin{equation}
{\rm{N}}^i_a( \{u, k\} )^2 = \langle \phi_i \psi_{1}^{a-k-1} \psi_2^k (u) 
| \phi_i \psi_{1}^{a-k-1} \psi_2^k (u) \rangle
= M^i(u) \frac{(a-1)!}{(a-k-1)! k!} \, .   
\end{equation}
The function $M^i(u)$ drops out in $R_{ab}$ and its explicit expression is not needed. The factors of $N^i_a$ are absent in the formulae given in \cite{Fleury:2017eph}. They appear here because we are normalising the states to one. Note finally that the sum appearing in (\ref{eq:DefinitionR}) can be evaluated explicitly and written as a hypergeometric function with unit argument. 

\section{Computing half structure constants}\label{AppZeta}
In this appendix, we present a routine for evaluating the 1-wheel amplitude,
\beq
\mathcal{A}_{\text{1-wheel}}(\ell_{2}, \ell_{3}) =  B_1 + \frac{1}{2} C_1 \, , 
\eeq
where $B_1$ and $C_1$ are given in (\ref{intB}) and (\ref{intC}), respectively. We will illustrate it on the particular case $\ell_2=\ell_3=3$, which is generic enough for our purposes.

We begin with $C_1$, which involves a single sum and a single integral. Its integrand contains harmonic sums, see Eq.~(\ref{ka}), which we can split according to their arguments, depending on whether they produce poles in the upper or in the lower half-plane. The two halves give the same result, by parity, and each of them can be integrated by closing the contour in such a way that only the pole in the energy factor in (\ref{intC}), at $u=\pm i a/2$, is enclosed. Its residue is a combination of polygamma functions of order $\ell_2+\ell_3-1$ and lower, which we immediately translate into generalized harmonic numbers $H_{n}^{(m)}$, using
\beq
\psi ^{(k)}(n)= (-1)^{k+1} k! \zeta (k+1)+(-1)^{k+2}\Gamma (k+1) H_{n-1}^{(k+1)} \, , 
\eeq
where $\psi ^{(k)}(n)$ is the polygamma function of order $k$. The resulting sum over bound states is of Euler type and can be expressed in terms of multiple zeta functions,
\begin{equation}
\zeta(s) = \sum_{n=1}^{\infty} \frac{1}{n^s} \, , \quad \quad  \zeta(s, t) = \sum_{n=1}^{\infty} \frac{H_{n}^{(t)}}{(n+1)^s} \,.
\end{equation}
E.g., taking all the steps at a time, for our specific example, yields
\beq
\begin{aligned}
\frac{1}{2} C_1=&-2 \zeta(4,6)-12 \zeta(5,5)-42 \zeta(6,4)-112 \zeta(7,3)-252 \zeta(8,2)+264 \zeta (5)^2\\
&+616 \zeta (3) \zeta (7)+252 \zeta (9)-\frac{4399 \pi ^{10}}{467775}\,.
\end{aligned}
\eeq
Note that this expression could be simplified using identities among multiple zeta values and given entirely in terms of Riemann zeta values, as done at the end of this appendix for the full amplitude.

The double integral $B_1$ can be split in two using
\beq
\begin{aligned} \label{B1}
B_1=
 \sum_{a, b =1}^{\infty} \dashint \frac{du dv}{(2\pi)^2}   \frac{ ab}{ \left(\frac{a^2}{4}+u^2\right)^{\ell_2} \left(\frac{b^2}{4}+v^2\right)^{\ell_3}} \Biggl(&\frac{1}{ \frac{1}{4} (a-b)^2+(u-v)^2}-\frac{1}{ \frac{1}{4} (a+b)^2+(u-v)^2} \Biggr) \, . 
\end{aligned}
\eeq
The second term, denoted $B_{12}$, is the simplest one, and no principal value is needed. The integral can be taken by first picking up the residue at $u= ia/2$ and $u=v+i (a+b)/2$, and then at $v=ib/2$ and $v=i(2 a+b)/2$. The next steps are the same as before; the sum over $b$ is straightforward and produces generalized harmonic numbers, etc. It yields, for $\ell_{2}=\ell_{3} = 3$,
\beq
B_{12}=-12 \zeta(4,6)-24 \zeta(5,5)-6 \zeta(6,4)+42 \zeta(7,3)+150 \zeta (5)^2-\frac{19 \pi ^{10}}{14175} \, . 
\eeq
For the first term in (\ref{B1}), which we denote by $B_{11}$, it is convenient to consider separately the cases $a>b$, $a<b$ and $a=b$. The first two cases, $a>b$ and $a<b$, are in all respects similar to $B_{12}$ and, in the case at hand, produce identical results,
\beq
\begin{aligned}
B_{11}^{a>b} = B_{11}^{a<b}  &= 42 \zeta(3,7)+132 \zeta(4,6)+252 \zeta(5,5)+252 \zeta(6,4)+252 \zeta(7,3)+252 \zeta(8,2)\\
&-378 \zeta (5)^2-546 \zeta (3) \zeta (7)+\frac{64 \pi ^{10}}{5775}\,.
\end{aligned}
\eeq
Finally, there is the case $a=b$, which contains the singularity at $u=v$ regularized by principal part integration, i.e.,
\beq
B_{11}^{a=b}=\frac{1}{2}\sum_{a=1}^{\infty} \int \frac{du dv}{(2\pi)^2} \frac{a^2}{ \left(\frac{a^2}{4}+u^2\right)^{\ell_{2}}\left(\frac{a^2}{4}+v^2\right)^{\ell_{3}}}  \Biggl(\frac{1}{(u-v+i0)^2}+\frac{1}{(u-v-i0)^2}\Biggr)\,.
\eeq
The integral over $u$ is taken by picking up the residues at $u=\pm ia/2$ for the first and second terms, respectively. The remaining integral and sum are as straightforwardly performed and produce,
\beq
B_{12}^{a=b}=-\frac{20 \pi ^{10}}{6237}\,,
\eeq
for our specific example.

At last, we combine all the terms together and use well-known identities,
\begin{equation}
\zeta(s) \zeta(m) = \zeta(s,m) + \zeta(m,s) + \zeta(m+s) \, , \,  \,
\end{equation}   
together with similar ones derived from the shuffle algebra, to simplify the expression.  E.g., for $\ell_{2} = \ell_{3} = 3$, using
\beq
\begin{aligned}
\zeta(7,3)&=\frac{11}{10}\zeta (10)-\zeta (5)^2-\zeta (6,4)\,,\\
\zeta(8,2)&=-\frac{70}{277} \zeta(10)+\frac{2}{7} \zeta(6,4)+\frac{10 \zeta (5)^2}{7}+2 \zeta (3) \zeta (7)\,,
\end{aligned}
\eeq
we immediately obtain
\beq
\mathcal{A}_{\text{1-wheel}} = B_{11}^{a>b}+B_{11}^{a<b}+B_{11}^{a=b}- B_{12} + 
\frac{1}{2} C_1 =-290 \zeta (5)^2+112 \zeta (3) \zeta (7)+252 \zeta (9)\,.
\eeq
The other results in table \ref{intprediction} are obtained similarly.

\section{Twisted transfer matrix}\label{AppT}

\begin{figure}
\begin{center}
\includegraphics[scale=0.5]{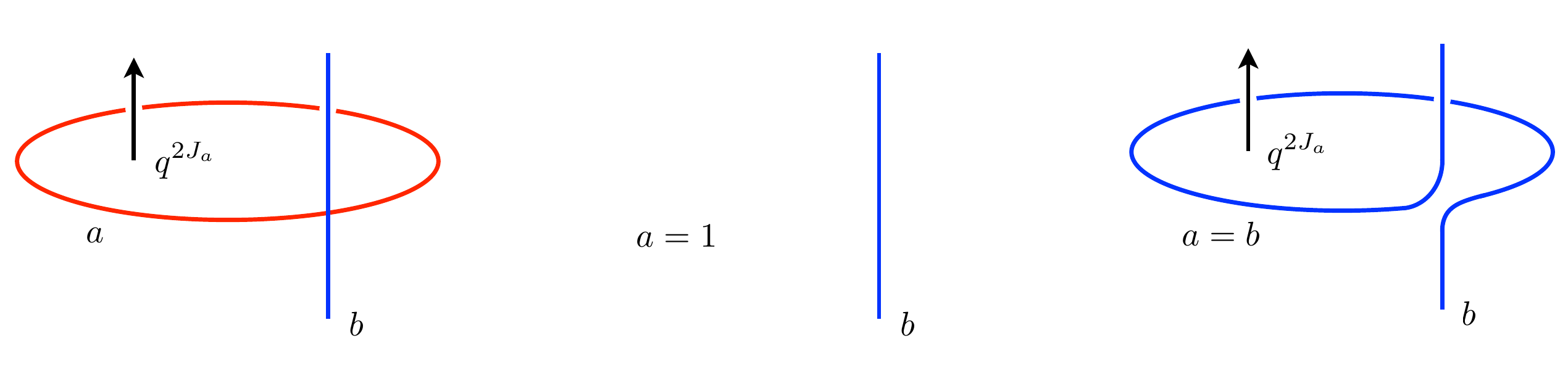}
\end{center}
\caption{Twisted transfer matrix in the $a$-th irrep for a length one spin chain with spin in the $b$-th irrep of $SU(2)$. When $a=1$ the trace is empty while when $a=b$ and for a specific choice of the rapidity it opens up.}\label{twisted} 
\end{figure}

In this appendix we derive the expression for the generating function of twisted transfer matrices, used in Subsection \ref{prop} to reproduce the free propagator,
\beq
P(\rho) = \sum_{a=1}^{b} \frac{\Gamma(b)}{\Gamma(a)\Gamma(b-a+1)}(-\rho)^{a-1}\textrm{tr}_{V_a}\, q^{2J_{a}}R_{ab}(\tfrac{i(b-a)}{2}) = (1-\rho q)^{\frac{b-1}{2}+J_{b}} (1-\rho/q)^{\frac{b-1}{2}-J_{b}}\, .
\eeq
Here the trace is taken over the $a$-th irrep of $SU(2)$, with spin $(a-1)/2$, and the identity holds as an operator identity on the Hilbert space $V_{b}$ of the $b$-th irrep. Note that $P(\rho)$ is by definition a polynomial in $\rho$ of degree $b-1$ and that it transforms as
\beq\label{parity}
P|_{J_{b}\rightarrow -J_{b}} =  P|_{q\rightarrow 1/q}\, , 
\eeq
under Weyl reflection. Also, obviously, see figure \ref{twisted},
\beq
P(0) = \textrm{tr}_{V_1}\, q^{2J_{1}}R_{1b}(\tfrac{i(b-1)}{2}) = 1\, ,
\eeq
since $a=1$ is the trivial representation, and
\beq
\lim_{\rho\rightarrow \infty} P(\rho)/(-\rho)^{b-1}  = q^{2J_{b}}\, ,
\eeq
since $R_{bb}(0)$ is the permutation operator on $b\otimes b$. Our goal is to fill the gap between these two extreme behaviours.

Let us denote by
\beq
T_{ab}(u) = \textrm{tr}_{V_a}\, q^{2J_{a}}R_{ab}(u)
\eeq
the twisted transfer matrix with twist parameter $q$, auxiliary space $V_{a}$, and quantum space~$V_{b}$. The eigenvalues of $T_{ab}$ are in one-to-one correspondence with the polynomial solutions of the twisted Baxter equation, for the associated chain with a single spin $\frac{1}{2}(b-1)$,
\beq
q (u+i\tfrac{b-1}{2}) Q(u-i) + \frac{1}{q} (u-i\tfrac{b-1}{2})Q(u+i) = t(u)Q(u)\, ,
\eeq
with the degree $M$ of the Baxter polynomial $Q(u) = \prod_{i=1}^{M}(u-u_{i})$ corresponding to the eigenvalue of the spin operator,%
\footnote{The solution is unique at given $M$ in the case at hand.}
\beq
J_{b} = \frac{b-1}{2} -M\, .
\eeq
Here $t(u)$ is a polynomial of degree $1$, which is fixed by the large $u$ behaviour of the LHS of the Baxter equation,
\beq
t(u) = (q+\frac{1}{q}) u +i(q-\frac{1}{q}) J_{b}\, ,
\eeq
and which coincides \cite{Faddeev:1996iy} with the eigenvalue of the fundamental transfer matrix (Lax matrix), up to a shift of the rapidity and an overall factor,
\beq
T_{2b}(u) = \textrm{tr}_{V_2}\, q^{\sigma_{3}}\frac{(u+\tfrac{i}{2}+i\vec{\sigma}\cdot \vec{J}_{b})}{u+\tfrac{ib}{2}} = \frac{t(u+\tfrac{i}{2})}{u+\frac{ib}{2}}\,.
\eeq
Another well-known relation, used typically to compute the spin chain energy, is
\beq
T_{2b}(u) = q\frac{Q(u-\frac{i}{2})}{Q(u+\frac{i}{2})} + O(u-\tfrac{i(b-2)}{2})\, .
\eeq
It follows from the structure of the LHS of the Baxter equation, and the neglected terms are linear in $u-i(b-2)/2$, since the chain has length $1$. One can access to the higher transfer matrices through fusion and obtain the more general formula
\beq
T_{ab}(u) = q^{a-1}\frac{Q(u-\frac{i(a-1)}{2})}{Q(u+\frac{i(a-1)}{2})} + O(u-\tfrac{i(b-a)}{2})\, .
\eeq
Nicely, the point $u = i(b-a)/2$ is precisely where we need to evaluate the transfer matrices, and the above identity allows us to write
\beq
P(\rho) = \sum_{a=1}^{b}(-\rho q)^{a-1}\frac{\Gamma(b)}{\Gamma(a)\Gamma(b-a+1)}\frac{Q(\tfrac{i(b-2a+1)}{2})}{Q(\tfrac{i(b-1)}{2})}\, .
\eeq
For $Q$ a polynomial of degree $M$, $P(\rho)$ must have a zero of degree $b-1-M$ at $\rho  =1/q$,
\beq
P(\rho) \propto (1-\rho q)^{b-1-M}\, .
\eeq
(This is obvious for the vacuum solution,
\beq
P(\rho)|_{Q\rightarrow 1} = \sum_{a=1}^{b}(-\rho q)^{a-1}\frac{\Gamma(b)}{\Gamma(a)\Gamma(b-a+1)} = (1-\rho q)^{b-1}\, ,
\eeq
while, for $M$ magnons, we should act on this function with a differential operator in $\rho$ of maximal degree $M$.) The remaining factor of degree $M$ is determined using (\ref{parity}), and, fixing the overall normalization at $\rho = 0$, we get
\beq
P(\rho) = (1-\rho q)^{b-1-M} (1-\rho/q)^{M} = (1-\rho q)^{\frac{b-1}{2}+J_{b}}(1-\rho/q)^{\frac{b-1}{2}-J_{b}}\, ,
\eeq
as desired.

\section{Cauchy et al.}\label{CVP}

In Section \ref{Sect4}, it was necessary to perform two integrations involving the factorized interaction among magnons given below, see (\ref{int12}) and (\ref{double}). 
In this appendix, we explicitly carry out these integrals. The typical interaction is given by     
\beq \label{interactionBulk}
\Delta^{(n,m)}(\textbf{u},\textbf{v}) := \frac{\prod_{i<j}^{n}\Delta_{a_{i}a_{j}}(u_{i}-u_{j})\prod_{i<j}^{m}\Delta_{b_{i}b_{j}}(v_{i}-v_{j})}{\prod_{i, j} \Delta_{a_{i}b_{j}}(u_{i}-v_{j})}\, ,
\eeq
with
\beq
\Delta_{ab}(u-v) = (u^{[+a]}-v^{[+b]})(u^{[+a]}-v^{[-b]})(u^{[-a]}-v^{[+b]})(u^{[-a]}-v^{[-b]})\, ,
\eeq
where $u^{[\pm a]} = u\pm ia/2$ and $v^{[\pm b]} = v\pm ib/2$.  Although concise, this representation is not convenient for integration. 
The algebra can be simplified by proceeding as follows. Assume firstly 
that $m = n$ and define the $2n+2n$ variables $x$'s and $y$'s by 
\beq
x_{2i-1} = u^{[+a_{i}]}_{i}\,, \qquad x_{2i} = u^{[-a_{i}]}_{i}\, , \qquad y_{2i-1} = v^{[+b_{i}]}_{i}\,, \qquad y_{2i} = v^{[-b_{i}]}_{i}\, .
\eeq
Then the above interaction can be written as
\beq
\Delta^{(n,n)}(\textbf{u},\textbf{v}):= \frac{1}{\prod_{i}^n a_i b_i} \, C_{2n|2n}(\textbf{x}|\textbf{y}) \, ,
\eeq
where $C$ is the Cauchy determinant
\beq
C_{2n|2n}(\textbf{x}|\textbf{y}) = \textrm{det}\, \left( \frac{1}{y_{j}-x_{i}} \right)_{i, j} = \frac{\prod_{i<j}^{2n}(x_{i}-x_{j})\prod_{i>j}^{2n}(y_{i}-y_{j})}{\prod^{2n}_{i, j}(y_{i}-x_{j})}\, .
\eeq
Integrands containing the interaction (\ref{interactionBulk}) for $m<n$ are readily obtained as a limit. In the following we concentrate on the case $m = n-1$, which is the situation encountered in Section \ref{Sect4}. Eliminating two $y$'s, say $y_{2n}$ and $y_{2n-1}$, by sending them to $\infty$, one after the other, we get
\beq \label{firstLimiit}
C_{2n|2n-2}(\textbf{x}|\textbf{y}) = \lim\limits_{y_{2n}\, \gg\, y_{2n-1}\rightarrow \infty} y_{2n}y_{2n-1}^2 C_{2n|2n}(\textbf{x}|\textbf{y}) = \frac{\prod_{i<j}^{2n}(x_{i}-x_{j})\prod_{i>j}^{2n-2}(y_{i}-y_{j})}{\prod_{i=1}^{2n-2} \prod_{j=1}^{2n}(y_{i}-x_{j})}\, .
\eeq
This expression can be equivalently defined as the determinant of a $2n\times 2n$ matrix obtained by replacing the two bottom rows of the Cauchy matrix by the corresponding ones in a Vandermonde matrix,
\beq
\left\{\begin{array}{ccc} (y_{2n-1}-x_{1})^{-1} & (y_{2n-1}-x_{2})^{-1} & \ldots \\ (y_{2n}-x_{1})^{-1} & (y_{2n}-x_{2})^{-1} & \ldots \end{array}\right\} \sim \frac{1}{y_{2n}y_{2n-1}^2}\left\{\begin{array}{ccc}  x_{1} & x_{2} & \ldots \\ 1& 1 & \ldots \end{array}\right\}\, .
\eeq

To perform the integrations, we start by writing the Cauchy determinant as a sum over permutations,
\beq\label{Cnn}
C_{2n|2n}(\textbf{x}|\textbf{y}) = \frac{1}{2^{n}}\sum_{\sigma \in S_{2n}}\textrm{sign}(\sigma) \prod_{i=1}^{n} C_{2|2}(y_{2i-1}, y_{2i}| x_{\sigma(2i-1)}, x_{\sigma(2i)})\, ,
\eeq
and send $y_{2n}, y_{2n-1}$ to infinity, as in (\ref{firstLimiit}), leading to
\beq
C_{2n|2n-2}(\textbf{x}|\textbf{y})  = \frac{1}{2^{n}}\sum_{\sigma \in S_{2n}}\textrm{sign}(\sigma) (x_{\sigma(2n-1)}-x_{\sigma(2n)}) \prod_{i=1}^{n-1} C(y_{2i-1}, y_{2i}| x_{\sigma(2i-1)}, x_{\sigma(2i)})\, .
\eeq
The remaining $y$'s are then integrated pairwise using a factorized measure,
\beq\label{main}
\int \prod_{i=1}^{n-1}d\mu(y_{2i-1}, y_{2i}) C_{2n|2n-2}(\textbf{x}|\textbf{y}) = \frac{1}{2^{n}}\sum_{\sigma \in S_{2n}}\textrm{sign}(\sigma) (x_{\sigma(2n-1)}-x_{\sigma(2n)}) \prod_{i=1}^{n-1} A_{\sigma(2i-1), \sigma(2i)}\, ,
\eeq
where the elements $A_{ij} = -A_{ji}$ defines a $2n\times 2n$ antisymmetric matrix $A$, obtained by integrating the $2\times 2$ Cauchy determinant,
\beq\label{Aij}
\begin{aligned}
A_{ij} &= \int d\mu(y_{1}, y_{2}) C_{2|2}(y_{1}, y_{2}| x_{i}, x_{j}) \\
&=  \int \frac{d\mu(y_{1}, y_{2})(x_{i}-x_{j})(y_{2}-y_{1})}{(y_{1}-x_{i})(y_{2}-x_{i})(y_{1}-x_{j})(y_{2}-x_{j})}\, .
\end{aligned}
\eeq
Notice that the result of integrating the $2n$ $y$'s in $C_{2n|2n}$ in this manner using 
(\ref{Cnn}) is $n!\times \textrm{pf} \,(A)$, where $\textrm{pf}\, (A)$ is the Pfaffian of $A$. Equation (\ref{main}) is closely related to it and only differs in the presence of the `inhomogeneous' element $(x_{i}-x_{j})$.

Formula (\ref{main}) holds regardless of the measure chosen for integrating the $y$'s. In this appendix we work with
\beq
\int d\mu(v^{[+b]}, v^{[-b]}) = \sum_{b = 1}^{M}b\, \dashint \frac{dv}{2\pi}\, ,
\eeq
where a cut off $M$ was introduced to regularise the logarithmic divergences of the individual integrals, when $M\rightarrow \infty$. The individual integral $A_{ij}$ is obtained by closing the contour of integration at $\infty$ in (\ref{Aij}) and summing over the residues. It yields
\beq\label{Apm}
A_{ij} = A(u^{[p_{i}]}_{i}, u^{[p_{j}]}_{j}) = \textrm{div} +\tfrac{1}{2}i(u_{i}^{[p_{i}]}-u^{[p_{j}]}_{j}) K_{a_{i}a_{j}}^{(\textrm{sign}(p_{i}p_{j}))}(u_{i}-u_{j})\, ,
\eeq
where $p_{i, j} = \pm a_{i, j}$, and where
\beq\label{Kpm}
K^{(\pm)}_{ab}(u) = \sum_{k=0, 1} (H(k-1+\tfrac{1}{2}|a\pm b|+iu) + H(k-1+\tfrac{1}{2}|a\pm b|-iu))
\eeq
is such that $K^{(\pm)}_{ab}(u) = K^{(\pm)}_{ba}(u) = K^{(\pm)}_{ab}(-u)$ and $K^{(-)}_{aa}(0) = 0$. The regularisation dependent part in (\ref{Apm}) is given by
\beq
\textrm{div} = -2i(u_{i}^{[p_{i}]}-u^{[p_{j}]}_{j}) \log{(Me^{\gamma_{E}})} -(p_{i}-p_{j})\, .
\eeq
Neither the logarithm nor the subleading constants in the divergent part contribute to the final result in (\ref{main}) and the limit $M\rightarrow \infty$ can be safely taken in the end. We checked it explicitly for the two particular cases discussed below.

We can now specialise to the two examples met in Section \ref{Sect4}. Namely, the integral (\ref{int12}) is obtained by setting $n=2$ in the general formula, which gives
\beq
\begin{aligned}
&\sum_{b\geqslant 1}\dashint \frac{dv}{2\pi} \frac{b^2a_{1}a_{2}\Delta_{a_{1}a_{2}}(u_{1}-u_{2})}{\Delta_{ba_{1}}(u_{1}-v)\Delta_{ba_{2}}(u_{2}-v)} \\
& = ((u_{1}-u_{2})^2 +\tfrac{1}{4}(a_{1}+a_{2})^2)K^{(-)}_{a_{1}a_{2}}(u_{1}-u_{2}) - ((u_{1}-u_{2})^2 +\tfrac{1}{4}(a_{1}-a_{2})^2)K^{(+)}_{a_{1}a_{2}}(u_{1}-u_{2}) \\
& = a_{1}a_{2}K'_{a_{1}a_{2}}(u_{1}-u_{2}) \, , 
\end{aligned}
\eeq
where $K'_{ab}(u-v) = K_{ab}(u, v)-K_{aa}(u, u)$ is the subtracted scattering kernel, see (\ref{subtractedkernel}).  The next case, $n=3$, is more bulky, and corresponds to the integral in (\ref{double}).  
Averaging over the permutations, and setting $a_{1} = a_{2} = a_{3} = 1$, we obtain
\beq
\begin{aligned}\label{n3}
&\sum_{b_{1}, b_{2}\geqslant 1}\dashint \frac{dv_{1}dv_{2}}{(2\pi)^2} \frac{b_{1}^2b_{2}^2 \Delta_{b_{1}b_{2}}(v_{1}-v_{2})\prod_{i<j}^{3}\Delta_{11}(u_{i}-u_{j})}{\prod_{i, j}\Delta_{1b_{j}}(u_{i}-v_{j})} \\
&\qquad \qquad = -\frac{1}{2}((K'_{12})^2+(K'_{13})^2+(K'_{23})^2) +(K'_{13}K'_{12}+K'_{12}K'_{23}+K'_{13}K'_{23}) \\
&\qquad \qquad\quad\, -6\prod_{1 \leqslant i<j\leqslant 3}\frac{(u_{i}-u_{j})^2}{((u_{i}-u_{j})^2+1)}\, ,
\end{aligned}
\eeq
where $K'_{ij} = K'_{ji} = K'(u_{i}-u_{j})$ and where we used
\beq
K^{(+)}(u) = K'(u)+2\, , \qquad K^{(-)}(u) = K'(u) + \frac{2u^2}{u^2+1}\, .
\eeq
Note that the rational bit in (\ref{n3}) relates to the fact that $K^{(\pm)} - K' \neq 0$. If these two identities were observed, we would immediately obtain the formulae in (\ref{Bseries}), with no need for the matrix part.


\pdfbookmark[1]{\refname}{references}
\bibliographystyle{nb}
\bibliography{references}

\end{document}